%% file: main.tex
\RequirePackage{etex}
\documentclass[a4paper,twocolumn,10pt,unpublished]{quantumarticle}
\pdfoutput=1
\usepackage{array}
\input{sty/config}

\title{Quantum circuit compression using qubit logic on qudits}

\author{Liam Lysaght}
\email{liam.lysaght.external@quandela.com}
\affiliation{Quandela, 7 Rue Leonard de Vinci, 91300 Massy, France}

\author{Timothée Goubault}
\affiliation{Quandela, 7 Rue Leonard de Vinci, 91300 Massy, France}

\author{Patrick Sinnott}
\affiliation{Quandela, 7 Rue Leonard de Vinci, 91300 Massy, France}

\author{Shane Mansfield}
\affiliation{Quandela, 7 Rue Leonard de Vinci, 91300 Massy, France}

\author{Pierre-Emmanuel Emeriau}
\email{pe.emeriau@quandela.com}
\affiliation{Quandela, 7 Rue Leonard de Vinci, 91300 Massy, France}

\begin{document}

\maketitle

\begin{abstract}
  We present qubit logic on qudits (QLOQ), a compression scheme in which the qubits from a hardware agnostic circuit are divided into groups of various sizes, and each group is mapped to a physical qudit for computation. 
  QLOQ circuits have qubit-logic inputs, outputs, and gates, making them compatible with existing qubit-based algorithms and Hamiltonians.
  We show that arbitrary qubit-logic unitaries can in principle be implemented with significantly fewer two-level (qubit) physical entangling gates in QLOQ than in qubit encoding. We achieve this advantage in practice for two applications: variational quantum algorithms, and unitary decomposition.  
  The variational quantum eigensolver (VQE) for LiH took 5 hours using QLOQ on one of Quandela's cloud-accessible photonic quantum computers, whereas it would have taken 4.39 years in qubit encoding. 
  We also provide a QLOQ version of the Quantum Shannon Decomposition, which not only outperforms previous qudit-based proposals, but also beats the theoretical lower bound on the CNOT cost of unitary decomposition in qubit encoding. 
\end{abstract}

\input{sections/introduction}

\input{sections/qloqcore}

\input{sections/ansatz}

\input{sections/VQE}

\input{sections/qsd}

\input{sections/conclusion}

\section*{Acknowledgements}
The authors warmly thank Arno Ricou, Benjamin Stott, and Marc Bollée for their feedback and fruitful discussions. 
This work has been co-funded by the European Commission as part of the EIC accelerator program project SEPOQC under the grant agreement 190188855, by the Horizon-CL4 program project EPIQUE under the grant agreement 101135288, by the QuantERA project ResourceQ under the grant agreement ANR-24-QUA2-007-003, and by the EIC Pathfinder program project QUONDENSATE under the grant agreement 101130384.

\bibliographystyle{quantum}
\bibliography{mainbibliography}

\input{sections/appendices}

\end{document}

%% file: sty/config.tex
\usepackage[singlelinecheck=false]{caption}
\usepackage[singlelinecheck=false]{subcaption}
\usepackage{amsfonts}
\usepackage{amsthm}
\usepackage{amsmath}
\usepackage[english]{babel}
\usepackage{braket}
\usepackage{bm}
\usepackage{geometry}
\usepackage{amssymb}
\usepackage{lettrine}
\usepackage{tikz}
\usetikzlibrary{
	backgrounds,
	quantikz,
	arrows,
	quotes,
	angles,
	positioning,
	decorations.pathreplacing,
	shapes.symbols}
\usepackage{tikz-network}
\usepackage{graphicx}

\usepackage[classfont=sanserif,langfont=sanserif,funcfont=sanserif]{complexity}
\usepackage{enumitem}
\usepackage{multirow}
\usepackage[createShortEnv]{proof-at-the-end}
\usepackage{csquotes}
\usepackage{epigraph}
\usepackage{anyfontsize}
\usepackage{pgfplots}
\pgfplotsset{compat=newest}
\usepgfplotslibrary{groupplots}
\usepgfplotslibrary{dateplot}
\usepackage{refcount}
\usepackage[ruled,lined]{algorithm2e}

\usepackage{mathrsfs}
\usepackage{booktabs}

\PassOptionsToPackage{numbers}{natbib}
\usepackage{natbib}

\usepackage{bookmark}
\newclass{\PostBPP}{PostBPP}
\newclass{\SampP}{SampP}
\newclass{\SampBQP}{SampBQP}
\newclass{\FBPP}{FBPP}
\newclass{\BosonFP}{BosonFP}
\newclass{\BosonP}{BosonP}
\newclass{\CompClassA}{A}
\newclass{\CompClassB}{B}

\newtheorem*{proposition*}{Proposition}

\newtheorem*{lemma*}{Lemma}

\let\epsilon\varepsilon

\usepackage{apptools}
\AtAppendix{\captionsetup{width=0.9\textwidth}}
\definecolor{Navy}{HTML}{2874A6}
\definecolor{DarkGreen}{HTML}{117A65}
\definecolor{Green}{RGB}{115, 198, 182}
\definecolor{ultramarine}{RGB}{84,189,220}
\definecolor{violet}{RGB}{
	145,106,184
	}
\definecolor{darkviolet}{RGB}{
	99, 56, 142 
	}
\definecolor{lightviolet}{RGB}{
	158,136,180
	}

\definecolor{darkred}{RGB}{
	192, 57, 43
	}

\definecolor{darkgreen}{RGB}{
	39, 174, 96
	}

\hypersetup{
	colorlinks = true,
	allcolors = {darkviolet},
}

\usepackage[nameinlink]{cleveref} 
\crefname{chapter}{Chapter}{Chapters}
\crefname{section}{Section}{Sections}
\crefname{equation}{Equation}{Equations}
\crefname{observation}{Observation}{Observations}
\crefname{figure}{Figure}{Figures}
\crefname{table}{Table}{Tables}
\crefname{appendix}{Appendix}{Appendices}
\crefname{theorem}{Theorem}{Theorems}
\crefname{corollary}{Corollary}{Corollaries}
\crefname{lemma}{Lemma}{Lemmas}
\crefname{proposition}{Proposition}{Propositions}
\crefname{definition}{Definition}{Definitions}
\crefname{algorithm}{Algorithm}{Algorithms}

\crefformat{footnote}{#2\footnotemark[#1]#3}

\let\autoref\cref

\newcommand{\pr}[2][]{
	\mathop{
		\ifx &#1&
		\mathrm{Pr}
		\else
			\underset{#1}{\mathrm{Pr}}
		\fi
		\left[#2\right]}
}

\newcommand{\e}[2][]{
	\mathop{
		\ifx &#1&
			\mathbb{E}
		\else
			\underset{#1}{\mathbb{E}}
		\fi
		\left[#2\right]}
}

%% file: sections/introduction.tex
\section{Introduction}
Quantum computers have traditionally used two-level quantum systems called qubits to encode information. However, physical quantum systems often have more than two levels available for computation, and are artificially constrained to the two-level format, which resembles the binary logic of classical computers. In spatially encoded linear optical quantum computing (LOQC), for example, a photon can easily represent any $L$-dimensional qudit by occupying $L$ spatial modes \cite{reck1994experimental, clements2016optimal, da2021path}.
Multi-level qudits have been demonstrated on all major quantum computing platforms, including photonic \cite{lanyon2009simplifying, li2022chip}, superconducting \cite{galda2021implementing}, and ion trap quantum computers \cite{ringbauer2022universal}. 
Qudit encoding schemes have been examined extensively as a means of reducing the resource requirements of various quantum algorithms and operations \cite{wang2020qudits}. This includes qubit-to-qudit compression techniques, where hardware  

\begin{figure}[!ht]
    \centering
    \includegraphics[width=\columnwidth]{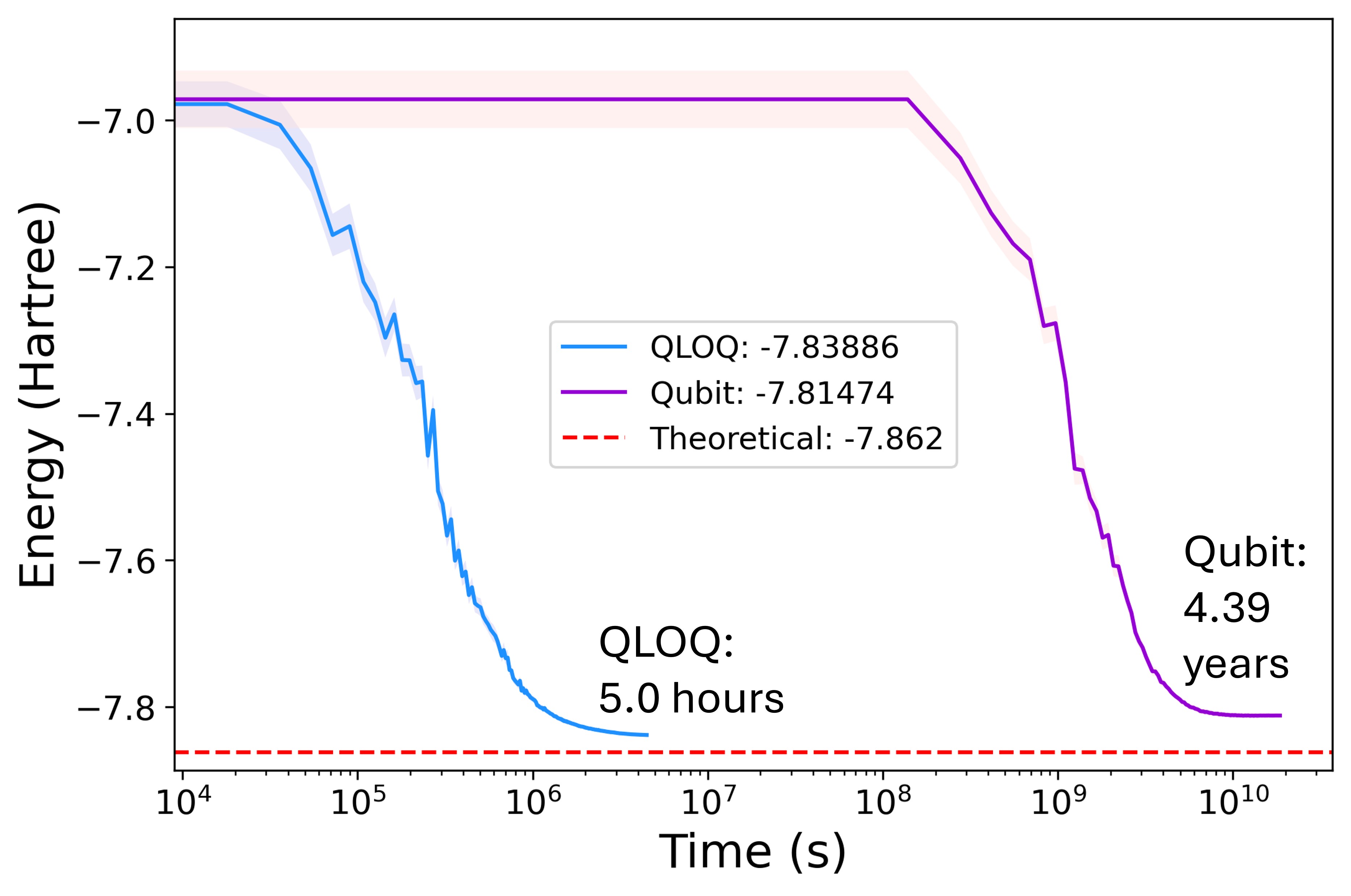}
    \caption{The average ground state energy estimate from 100 QLOQ and 100 qubit-based LiH VQE simulations, using COBYLA, converging over time. Shaded regions represent the standard error. An actual QPU run (see Figure \ref{fig:QPU_plot}) on one of Quandela's cloud-accessible photonic QPUs was used as the reference runtime for the QLOQ ansatz. The qubit-based runtime estimate was based on the difference in photon generation rate, success probability, and the average number of iterations in simulation between the ansatze. 
    }
    \label{fig:time_plots}
\end{figure}
\noindent agnostic qubit-based circuits are mapped to physical qudits for more efficient computation \cite{gao2022role,kiktenko2023realization, litteken2023qompress, mato2023compression}. 

Quantum systems are extremely sensitive to disturbances from their environment (noise), and may lose the quantum properties we rely on to perform computations through decoherence \cite{preskill2018quantum, zurek2003decoherence}.
This severely restricts the number of qubits and gate operations that can be deployed on currently available noisy intermediate scale quantum (NISQ) devices \cite{preskill2018quantum}.
Entangling operations between physical qubits or qudits are typically much more susceptible to noise and require more hardware resources to implement than local operations on a single qubit or qudit \cite{gao2022role}. 
In NISQ LOQC, for example, physical entangling gates are probabilistic, with the post-selected CNOT gate having a best-case success probability of 11.11\% \cite{ralph2002linear}. 
Entangling gates also require additional constrained resources, namely extra spatial modes and photons. Local (i.e.\ single-qudit) gates in LOQC meanwhile are theoretically deterministic, require no additional resources, and are composed of fewer optical components than entangling gates. 
Reducing the need for costly entangling gates is essential on all platforms to maximise the utility and performance of NISQ devices.

\subsection*{Our Contributions}

In this paper we introduce qubit logic on qudits (abbreviated QLOQ, pronounced ``clock"), a qubit-to-qudit compression scheme in which groups of qubits from a hardware agnostic circuit are mapped to physical qudits for computation, reducing the total number of physical entangling gates needed to implement the circuit.
QLOQ circuits have qubit-logic inputs, outputs and gates, so they are compatible with existing qubit-based algorithms and Hamiltonian formulations.  

Previous qubit-to-qudit compression schemes have proposed mapping the gates from qubit-based circuits onto qudits \textit{individually} \cite{kiktenko2023realization, litteken2023qompress, mato2023compression, nikolaeva2024efficient}. 
We show that this is only effective for extremely small or shallow circuits, or those with heavily restricted structures, and that the algorithms presented in previous works to find an optimal qubit-to-qudit map for qubit-based circuits do not work under realistic assumptions.  
However, we also demonstrate that arbitrary qubit-logic unitary operations can be implemented with far fewer physical entangling gates in QLOQ than qubit encoding. 
An advantage is therefore available where new circuits are designed from scratch for QLOQ, rather than directly translating existing qubit-native circuits. We provide circuit structures that achieve this advantage in practice for two applications: variational quantum algorithms (VQAs), and unitary decomposition. 

VQAs are hybrid quantum-classical algorithms in which a classical optimiser trains a parameterised quantum circuit (the ansatz) to solve a problem \cite{cerezo2021variational}.
We show that QLOQ ansatze for VQAs require significantly fewer physical entangling gates than qubit-based ansatze in a benchmark set to achieve a given level of expressibility. We also demonstrate the effectiveness of QLOQ ansatze by using one to run a variational quantum eigensolver (VQE) for Lithium Hydride (LiH) on Quandela's linear optical quantum computer. 

QLOQ can be deployed on a wide variety of quantum computing architectures, but provides a particularly strong advantage in photonics, where qudits are easy to deploy. This is illustrated by Figure \ref{fig:time_plots}, which shows the actual time required for QLOQ VQE (5 hours) and estimated time for qubit-based VQE (4.39 years) to converge on one of Quandela's photonic quantum processing units (QPU). 
QLOQ VQE also converged to a more accurate final ground state energy estimate in simulations than qubit-based VQE.  

Unitary decomposition is a process that maps the unitary matrix representing a quantum operation to a set of gates that can implement it on hardware. 
We show that the theoretical lower bound on the number of CNOTs required to decompose an arbitrary unitary is significantly lower in QLOQ than in qubit encoding. In fact, it decreases exponentially as the number of qubits mapped to each qudit increases. 
We also provide a QLOQ version of the Quantum Shannon Decomposition (QSD) \cite{shende2005synthesis} technique for decomposing unitaries. QSD is currently the state of the art for decomposing large unitaries \cite{krol2022efficient}. Numerical optimisation techniques can achieve lower CNOT counts, but this 
does not scale up to unitaries with more than a handful of qubits \cite{goubault2020methods, madden2022best, rakyta2022approaching}. 
QLOQ QSD not only outperforms previous proposals for decomposing unitaries using qudits \cite{di2013synthesis, li2013efficient}, it also beats the theoretical lower bound for the CNOT cost of unitary decomposition in qubit encoding \cite{shende2004minimal}. 
If sufficient unused qudit levels are available, QLOQ QSD can be deployed inside circuits that are otherwise qubit-based.

\subsection*{Paper Overview}
The paper is divided into sections as follows: in Section \ref{sec:QLOQ_section} the key concepts and gate decompositions of QLOQ are explained, along with theoretical evidence that arbitrary unitary operations can be implemented more efficiently in QLOQ than in qubit encoding. In Section \ref{{QLOQ_Ansatze}}, the structure of QLOQ ansatze for VQAs are presented and their performance metrics are compared to qubit-based ansatze from a benchmark set. In Section \ref{Experimental_VQE} we compare the results of LiH VQE simulations using QLOQ and qubit-based ansatze on a linear optical quantum computer. Section \ref{QSD Section} details the QLOQ Quantum Shannon Decomposition technique, and in Section \ref{Conclusion} we summarise our conclusions. 

\begin{table*}[!ht]
    \centering
    \begin{tabular}{c|l}
        $q$ & The number of physical qudits in a circuit \\  
        $i$ & The index of a particular qudit  \\ 
        $L$ & The number of levels in a qudit \\  
        $g$ & The number of qubits encoded on a qudit \\
        $G$ & The maximum number qubits that can be encoded on the qudits in a circuit \\ 
        $N$ & The total number of qubits encoded in a circuit \\ 
        $n$ & The number of encoded qubits to which a particular gate is applied\\  
        $k$ & The number of physical two-level entangling gates (e.g.\ CNOTs) required for a decomposition  \\   
    \end{tabular}
    \caption{\centering\label{notation_table} The most commonly used notation in this paper.}
\end{table*}

%% file: sections/qloqcore.tex
\section{Applying Qubit Logic to Qudits}
\label{sec:QLOQ_section}
\subsection{Core Concepts} 

The core concept of QLOQ is to divide the qubits of a hardware agnostic circuit into groups of various sizes, and to encode each group on a single physical qudit. Gates from the original qubit-based circuit may be referred to as `qubit-logical' or just `logical' gates in this paper, but this does not imply that error correction is being used. 
QLOQ has two key features which may significantly reduce the number of physical entangling gates required to implement a quantum circuit:

\begin{itemize}
\item Where multiple qubits are mapped to the same qudit, any entangling gates between them become local operations, which can generally be implemented with fewer resources on hardware than entangling operations \cite{gao2022role}. 
\item Multi-controlled unitary gates can be applied between all the qubits mapped to a pair of qudits with a single physical two-level (qubit) entangling gate (see Section \ref{key_decompositions} for details).
\end{itemize}

\newpage

There are also disadvantages associated with QLOQ, particularly:

\begin{itemize}
\item Entangling gates applied between some but not all of the qubits mapped to a pair of qudits may require more resources in a given QLOQ map than they would in qubit encoding. 
\item Multi-level qudits may be more difficult to deploy and control in certain hardware systems than two-level qubits. For example, in an ion trap quantum computer, the ions have a finite number of practically accessible energy levels, which imposes a limit on how large qudits can be. 
\end{itemize}

\subsection{Criteria For Implementing QLOQ}
The following criteria must be met for QLOQ to be deployed on a given quantum computer:

\begin{itemize}
    \item[1.] Qudits with at least $2^{g_i}$ levels are available, where $g_i$ is the number of qubits that are mapped to the physical qudit $i$ (at least one qudit with $4$ or more levels is required for a non-trivial encoding).
    \item[2.] Single-qudit operations are available that form a universal set on the $2^{g_i}$ levels in which qubit information is encoded.  
    \item[3.] All qudit levels in use can be distinguished at measurement.
    \item[4.] An entangling operation is available between the qudits in use. 
\end{itemize}

These criteria can be met on a variety of quantum computing platforms, including the Quandela photonic QPU used for the demonstration in this paper. On devices where criteria 2 and 4 are both met, any qubit-logical operation can be applied to the mapped qubits, because universal single-qudit gates combined with any entangling gate form a universal set \cite{bremner2002practical, brylinski2002universal}. For our analysis of QLOQ we assume that CNOT and CZ are the only natively available physical entangling gates.  
Restricting ourselves to two-level (qubit) entangling gates allows for a direct comparison of entangling gate cost with qubit encoding, and removes the need to develop qudit-specific entangling gates to implement QLOQ on existing hardware.  
Additionally, certain multi-level entangling gates would remove the primary drawback of directly translating qubit-based circuits into QLOQ, i.e.\ that logical entangling operations between qubits on different qudits may become more expensive. This would provide an overly favourable outlook for QLOQ, because these gates are not generally available on modern QPUs.

\begin{figure*}[!ht]
    \centering
    \includegraphics[width=0.7\textwidth]{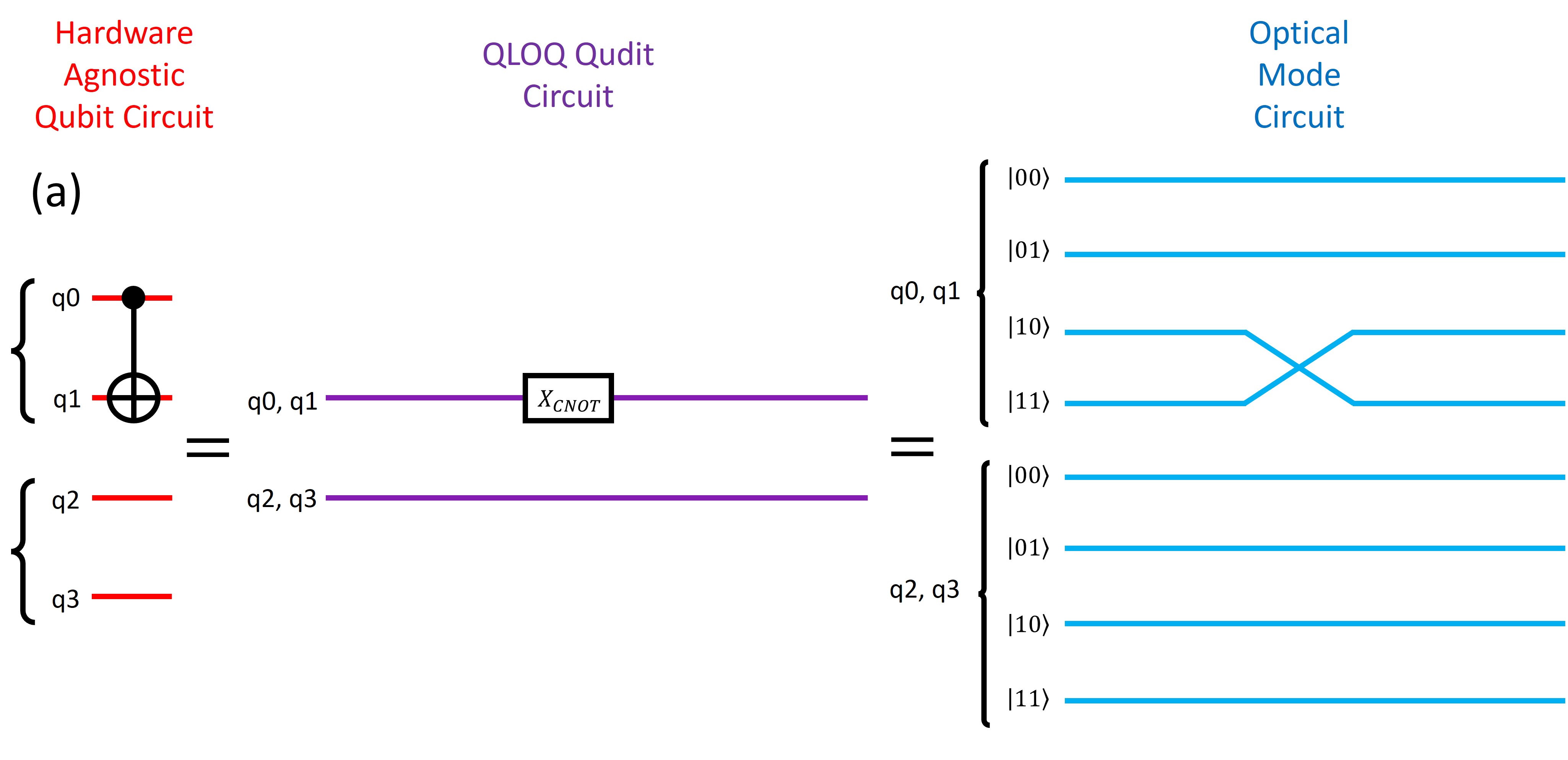}
    \includegraphics[width=0.7\textwidth]{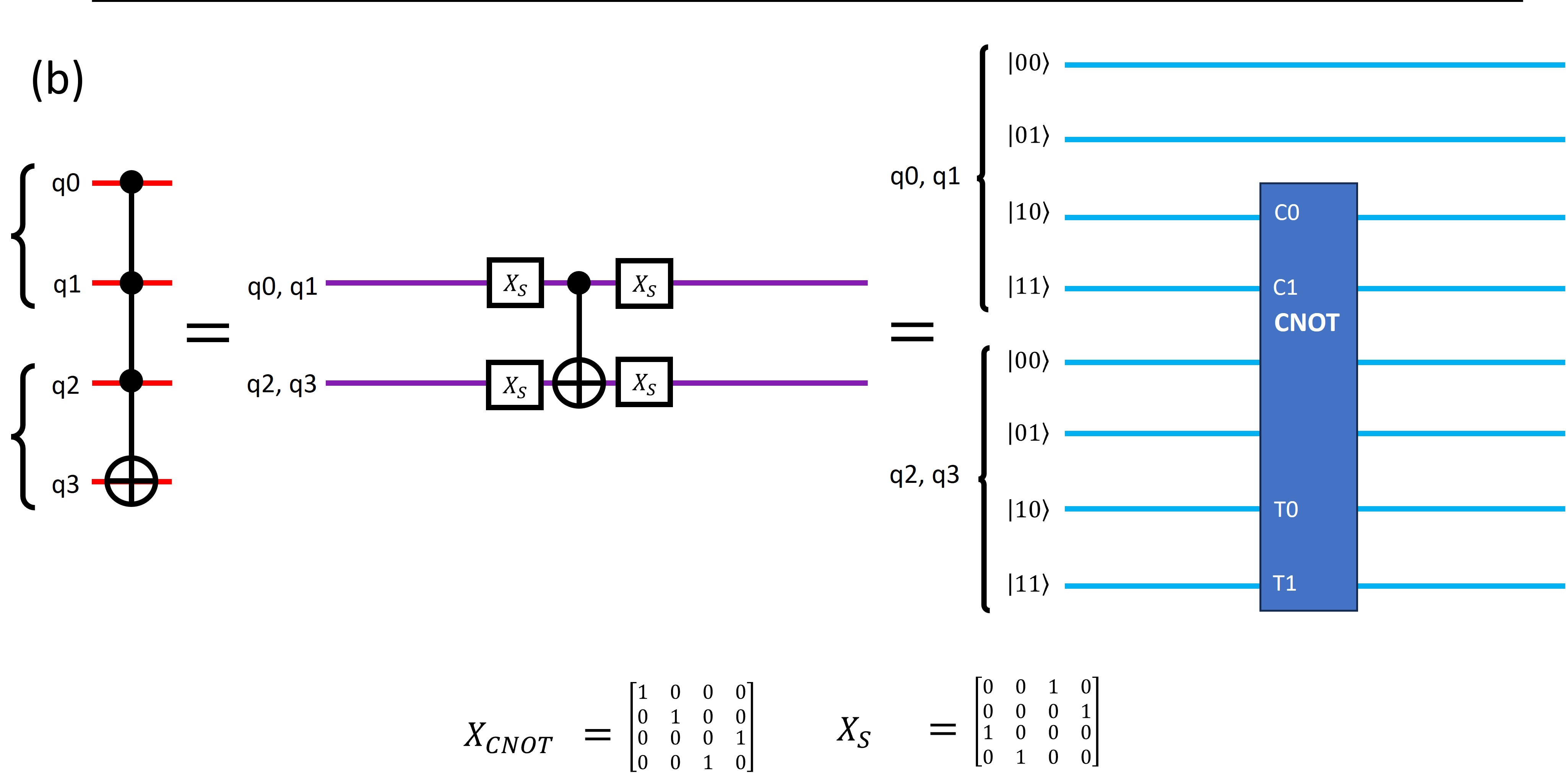}
    \caption{Key gate decompositions represented as hardware agnostic qubit-based circuits (red lines, left), QLOQ qudit-based circuits (purple lines, centre), and optical mode circuits (blue lines, right). The black brackets to the left of the diagrams denote qubits that are mapped to the same qudit. (a) A CNOT between two qubits encoded on the same qudit. (b) A CCCX (4-qubit Toffoli) gate between four qubits mapped to two qudits.}
    \label{fig:QLOQ Decompositions}
\end{figure*}

\subsection{Key Gate Decompositions in QLOQ}
\label{key_decompositions}
\begin{figure*}[!ht]
    \centering
    \includegraphics[width=0.9\textwidth]{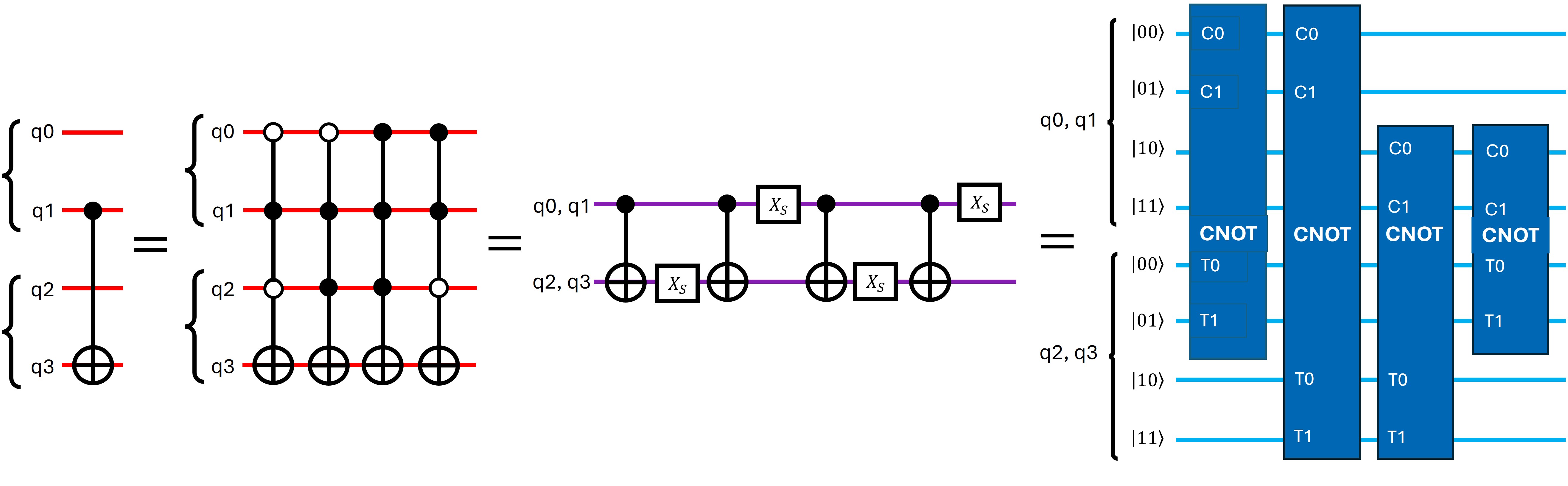}
    \caption{The QLOQ decomposition of a CNOT between two qubits mapped to different 4-level qudits. White control circles on a CCCX indicate that the control state for that qubit is $|0\rangle_{L}$ instead of $|1\rangle_{L}$. 
    } 
    \label{fig:QLOQ CNOT Between Qudits}
\end{figure*}

We present three important QLOQ gate decompositions in Figures \ref{fig:QLOQ Decompositions} and \ref{fig:QLOQ CNOT Between Qudits}. In these diagrams the gates are represented (from left to right) as hardware agnostic qubit-based circuits, QLOQ qudit-based circuits, and spatially encoded linear optical circuits. The qudit levels in a linear optical circuit are the spatial modes (blue lines) in which photons may travel, which makes these diagrams helpful for visualising qudit operations. QLOQ is particularly well suited to linear optical hardware, as discussed in Section \ref{sec:scaling_advanatges}.

The notation QLOQ(a, b)(c, d, e)...(y, z) is used throughout this paper to indicate that the qubits within each pair of brackets have been mapped to the same physical qudit. All qubits not specified are encoded individually as physical qubits. 
The QLOQ(0,1)(2,3) map is used in Figures \ref{fig:QLOQ Decompositions} and \ref{fig:QLOQ CNOT Between Qudits}. There are two 4-level qudits in this map, which each encode two qubits. Qubits mapped to the same qudit are enclosed by black brackets on the left hand side of circuit diagrams in this paper. For clarity we may use $|.\rangle_{P}$ when referring to physical qudit states, and $|.\rangle_{L}$ for logical qubit states. The symbol $\leftrightarrow$ below indicates the exchange of two states e.g.\ the qubit NOT operation is applied as $|0\rangle_{P} \leftrightarrow |1\rangle_{P}$.

For our qudit circuit diagrams (purple rails in Figures \ref{fig:QLOQ Decompositions} and \ref{fig:QLOQ CNOT Between Qudits}) we use the convention that physical CNOTs can only be applied directly to the qudit levels $|0\rangle_{P}$ and $|1\rangle_{P}$. If the control or target qudit for any two-level entangling gate is in another state ($|2\rangle_{P}, |3\rangle_{P}, |4\rangle_{P}$, etc.), neither qudit is affected by the gate. 
We do not restrict CNOTs in the optical circuit diagrams (blue rails in Figures \ref{fig:QLOQ Decompositions} and \ref{fig:QLOQ CNOT Between Qudits}) to the qudit levels $|0\rangle_{P}$ and $|1\rangle_{P}$, since it is visually obvious which qudit levels we are applying them to, as indicated by the control ($C0$, $C1$) and target ($T0$,$T1$) labels.  

Figure \ref{fig:QLOQ Decompositions} (a) shows a logical CNOT between two qubits mapped to the same qudit, which becomes an entirely local physical operation. 
A CNOT with qubit 0 as its control and qubit 1 as its target can be implemented by applying $|2\rangle_{P} \leftrightarrow |3\rangle_{P}$ to the qudit on which they are mapped (corresponding to the qubit-logical operation $|10\rangle_{L} \leftrightarrow |11\rangle_{L}$), as shown in the optical mode diagram.

Figure \ref{fig:QLOQ Decompositions} (b) shows a CCCX (4-qubit Toffoli) gate implemented with a single physical CNOT between two qudits.
The $X_s$ gates perform the operation $|0 \rangle_{P} \leftrightarrow |2 \rangle_{P}$, $|1 \rangle_{P} \leftrightarrow |3 \rangle_{P}$.
If the control qudit was originally in the state $|3 \rangle_{P}$ ($|11 \rangle_{L}$), the first $X_s$ gate will transfer the state to $|1 \rangle_{P}$, which is the control level for the CNOT. The CNOT will then apply $|0 \rangle_{P} \leftrightarrow |1 \rangle_{P}$ to the target qudit. Together with the single qudit gates, this has the effect of flipping the logical state of the target qubit only if all three other qubits were originally in the state $|1 \rangle_{L}$, which is the operation of a CCCX gate. 
This can be generalised to apply any $n$-qubit Toffoli with a single physical CNOT by mapping more qubits to each qudit, and adjusting the single qudit gates accordingly. 
Replacing the physical CNOT in this construction with another two-qubit gate will create a multi-controlled version of that gate. Using a CZ for example will create a CC...CZ. 
Internal SWAP operations can be used to rearrange the order of qubits on a qudit as necessary. See Appendix Section \ref{sec:single_qubit_ops} for details on how to apply internal SWAP and single-qubit logical operations in QLOQ.   

The QLOQ CCCX decomposition represents a significant improvement over qubit encoding, where 14 CNOTs are required to implement a CCCX without feed-forward techniques or ancillas \cite{nakanishi2021quantum}. The theoretical lower bound for $n$-qubit Toffoli gates decomposed with these conditions is $2n$ CNOTs (8 CNOTs for a CCCX) \cite{shende2008cnot}, compared to 1 CNOT in QLOQ. Previous decompositions that use auxiliary qudit levels can apply an $n$-qubit Toffoli with $2n-3$ entangling gates (e.g.\ 5 CNOTs for a CCCX) while maintaining qubit logic at the gate's inputs and outputs, \cite{lanyon2009simplifying, kiktenko2020scalable, nikolaeva2022decomposing, wang2011improved, baker2020improved}. Such techniques can also be applied in QLOQ to create $n$-qubit Toffolis between $q$ qudits with $2q-3$ CNOTs, reducing the size of the qudits needed to compress certain circuits.  

\begin{figure*}[!ht]
    \centering
    \includegraphics[width=0.9\textwidth]{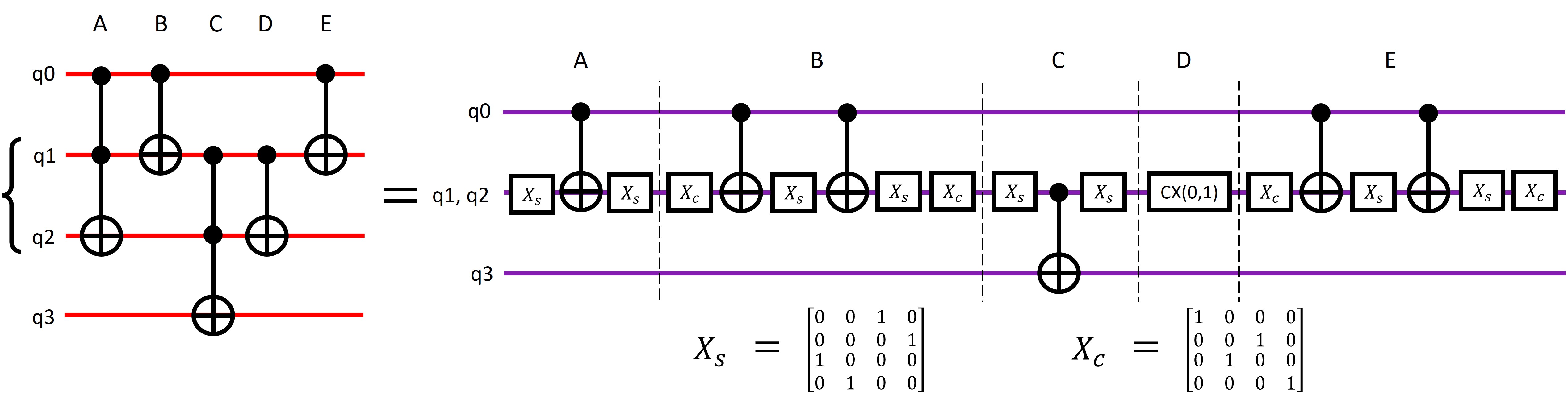}
    \caption{\centering The quantum full adder circuit \cite{feynman2023quantum} in qubit encoding and QLOQ(1,2).}
    \label{fig:QFA}
\end{figure*}

A control qubit can be removed from a multi-controlled gate by applying the original gate followed by a second instance of the gate with the control qubit in question flipped: i.e.\ once with the control state for that qubit as $|0\rangle_{L}$, and again with the control state as $|1\rangle_{L}$. Each control qubit removed doubles the gate's physical cost. Figure \ref{fig:QLOQ CNOT Between Qudits} shows a CNOT between qubits mapped to different qudits, constructed by combining CCCXs gates (removing controls). Extending this to the general case, the number of physical two-qubit gates $k$ required to implement an $n$-qubit multi-controlled gate between two qudits to which $g_a$ and $g_b$ qubits have been mapped is given by Equation \eqref{eq:abk}:

\begin{equation}
k = 2^{g_a+g_b-n} \; .
\label{eq:abk}
\end{equation}

This formula was previously presented in Ref \cite{nikolaeva2024efficient}. The resulting potential increase in physical gate cost depends on the structure of the uncompressed circuit and the QLOQ map chosen, and is the main drawback directly converting existing qubit-based circuits into QLOQ.

We present the quantum full adder (QFA) circuit \cite{feynman2023quantum} in the map QLOQ(1,2) in Figure \ref{fig:QFA}. This circuit is the quantum equivalent of the classical full adder, and will serve as a simple example to illustrate the potential benefits of translating qubit-based circuits into QLOQ.

Implementing each of the gates in the QFA circuit individually in qubit encoding would require 15 CNOTs, but the circuit as a whole can be optimised to 11 CNOTs \cite{qiskit2024}.
If the aforementioned auxiliary qudit level techniques are used for the Toffoli gates, the total cost drops to 9 CNOTs. In QLOQ(1,2) however, which requires one 4-level (4L) qudit, the quantum full adder only costs 6 physical CNOTs. If qudits with 8 or more levels are available, the map QLOQ(0,1,2) can be used instead, resulting in a QFA circuit costing only 2 physical CNOTs, shown in Section \ref{sec:8L_max} of the Appendix.

\subsection{Related Work}
\label{related_work}
Previous qubit-to-qudit compression schemes include Refs \cite{gao2022role, kiktenko2023realization, litteken2023qompress, mato2023compression, nikolaeva2024efficient, nikolaeva2023generalized}.

Refs \cite{gao2022role} and \cite{mato2023compression} assume that qubit-logic entangling gates between qubits mapped to different qudits (multi-level entangling gates) are available for the same resource cost as two-level entangling gates. This avoids the possible increase in physical gate count from Equation \eqref{eq:abk}, and allows any qubit-based circuit to be compressed by any non-trivial qubit-to-qudit map, since the cost of each gate would either decrease or stay the same. These multi-level entangling gates would remove the primary drawback of QLOQ and drastically improve its performance relative to qubit encoding. 
Ref \cite{gao2022role} acknowledges that these multi-level gates may be difficult to implement in practice (to our knowledge they do not exist in spatial LOQC for example). Nevertheless, Ref \cite{gao2022role}'s upper bound on the number of entangling operations in a compressed circuit assumes that these gates are available or does not account for the increase in physical gate cost when only two-level entangling gates are available. 

Ref \cite{litteken2023qompress} examines qubit-to-qudit compression with 4L superconducting transmon qudits in simulation, using pulse duration as their main gate cost metric. They find that logical CNOT gates where at least one of the qubits involved is mapped to a 4L qudit have a pulse duration at least twice as high (and often higher) as in qubit encoding i.e.\ multi-level entangling gates cost more than two-level entangling gates. 

Ref \cite{nikolaeva2024efficient} examines qubit-to-qudit compression where qubit-logic entangling gates between qudits are not natively available. Ref \cite{kiktenko2023realization} compares the impact of having two-level or multi-level entanglers (specifically the qudit Mølmer-Sørensen gate \cite{molmer1999multiparticle}) available for qubit-to-qubit compression. 

Refs \cite{kiktenko2023realization, litteken2023qompress, mato2023compression, nikolaeva2024efficient} suggest schemes in which each gate in a qubit-based circuit is converted into qudit form individually, and present algorithms for finding the optimal qubit-to-qudit mapping. 
However, we have found that this strategy of translating gates from qubit to qudit form individually, which we will call \textit{gate-by-gate compression}, is ineffective for the vast majority of circuits. The optimal qubit-to-qudit map for circuits with many entangling gates will almost always be qubit encoding. Note that we assume, as in Ref \cite{nikolaeva2024efficient} and in line with usual modern hardware constraints, that only two-level (qubit) physical entangling gates (e.g.\ CNOT/CZs) are available between qudits. 

\begin{figure}[!ht]
    \centering
    \includegraphics[width=0.475\textwidth]{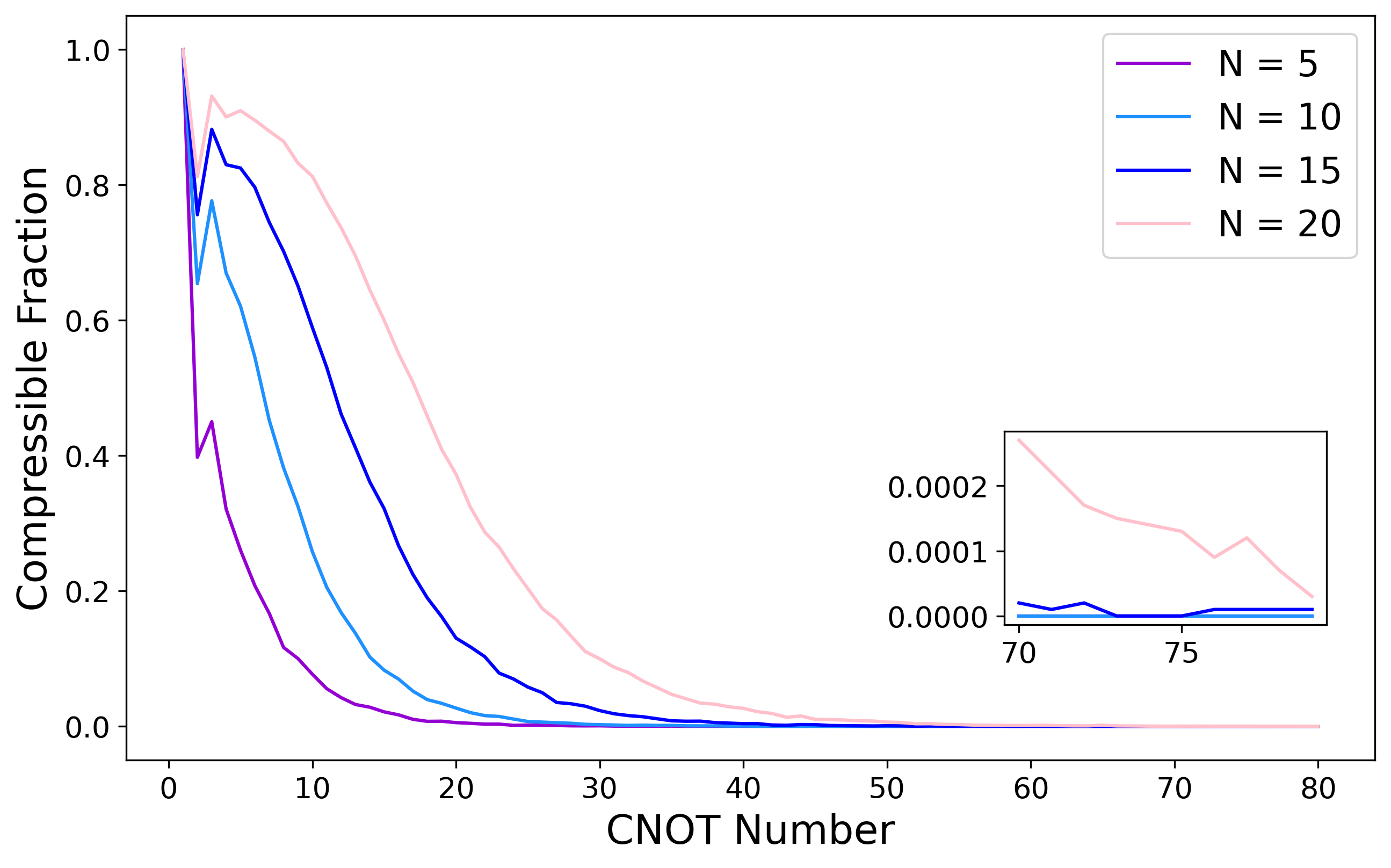}
    \caption{The fraction of randomly generated CNOT circuits where previously proposed gate-by-gate compression schemes could provide an advantage. Section \ref{lower_bound} describes how QLOQ can avoid the performance decay observed in previous schemes as more CNOTs are added to the circuits. $N$ is the number of qubits in the circuits. 10,000 circuits were generated for each data point in the plot, except for the final 10 values for each $N$ (shown in the inset), where 100,000 circuits were used per point.}
    \label{fig:FractionComp}
\end{figure}

If we map a pair of qubits to the same qudit, with the rest of the circuit encoded as qubits, the pair's internal connections become local operations (Cost: -1) and by Equation \eqref{eq:abk} their external connections become twice as expensive (Cost: +1). Assuming that at most two qubits can be mapped to each qudit ($G=2$), if a circuit contains any pair of qubits with more internal than external connections, then at least one qubit-to-qudit map exists for which gate-by-gate compression provides an advantage. Otherwise, no advantageous maps exist for that circuit.

We randomly generated connectivity graphs for qubit-based circuits composed entirely of equal cost two-qubit gates (e.g.\ CNOTs), with single qubit rotations being ignored. Figure \ref{fig:FractionComp} shows the fraction of these circuits for which at least one qubit-to-qudit map existed where gate-by-gate compression could reduce the physical CNOT cost. 10,000 circuits were generated and tested per data point, except for the final 10 iterations, where 100,000 circuits were used per data point.

It is clear from Figure \ref{fig:FractionComp} that the utility of gate-by-gate compression rapidly decays as more gates are added to the qubit-based target circuit. 
In Section \ref{gate_by_gate} of the Appendix we provide a detailed analysis showing that this also occurs for circuits containing multi-qubit gates, and that using larger qudits actually makes the problem worse. 

While it may reduce the entangling gate count for shallow circuits or those with heavily restricted structures, in general gate-by-gate compression does not provide any advantage. 
Ref \cite{litteken2023qompress} benchmark their compression scheme on circuits with ``localised groups of qubits that interact together", while Ref \cite{nikolaeva2023generalized} examines the scaling of their scheme on Grover's algorithm circuits where the only entanglement is generated by circuit-spanning Toffoli gates. 

\subsection{Unitary Decomposition Lower Bounds}
\label{lower_bound}

Qubit-to-qudit compression is generally not effective when individually translating each gate from a qubit-based circuit into qudit form (gate-by-gate). 
However, a strong advantage is available when quantum circuits are created from scratch with the design constraints of qudit encoding schemes like QLOQ in mind. In this section, we investigate how QLOQ can implement arbitrary $n$-qubit unitaries with significantly fewer physical entangling gates than is possible in qubit encoding.

In qubit encoding, the worst case number of CNOT gates needed to implement an $n$-qubit unitary is lower bounded by \cite{shende2004minimal}:
\begin{equation}
k = \Bigl\lceil\frac{4^n-3n-1}{4}\Bigr\rceil \; .
\label{eq:qubit_decomp}
\end{equation}

Whether or not this is a tight lower bound in practice for all $n$ is an open research question.
The CNOT operation itself is an example of a specific two-qubit unitary that requires only 1 physical CNOT gate to implement. However, 3 or more CNOTs (with single qubit rotations) can implement \textit{any} two-qubit unitary i.e.\ 3 CNOTs is the tight lower bound for two-qubit unitaries \cite{shende2004minimal}. No practical algorithm has been found that can achieve the lower bound CNOT cost for arbitrary unitaries for all $n$.

\begin{figure}[!ht]
    \centering
    \includegraphics[width=0.48\textwidth]{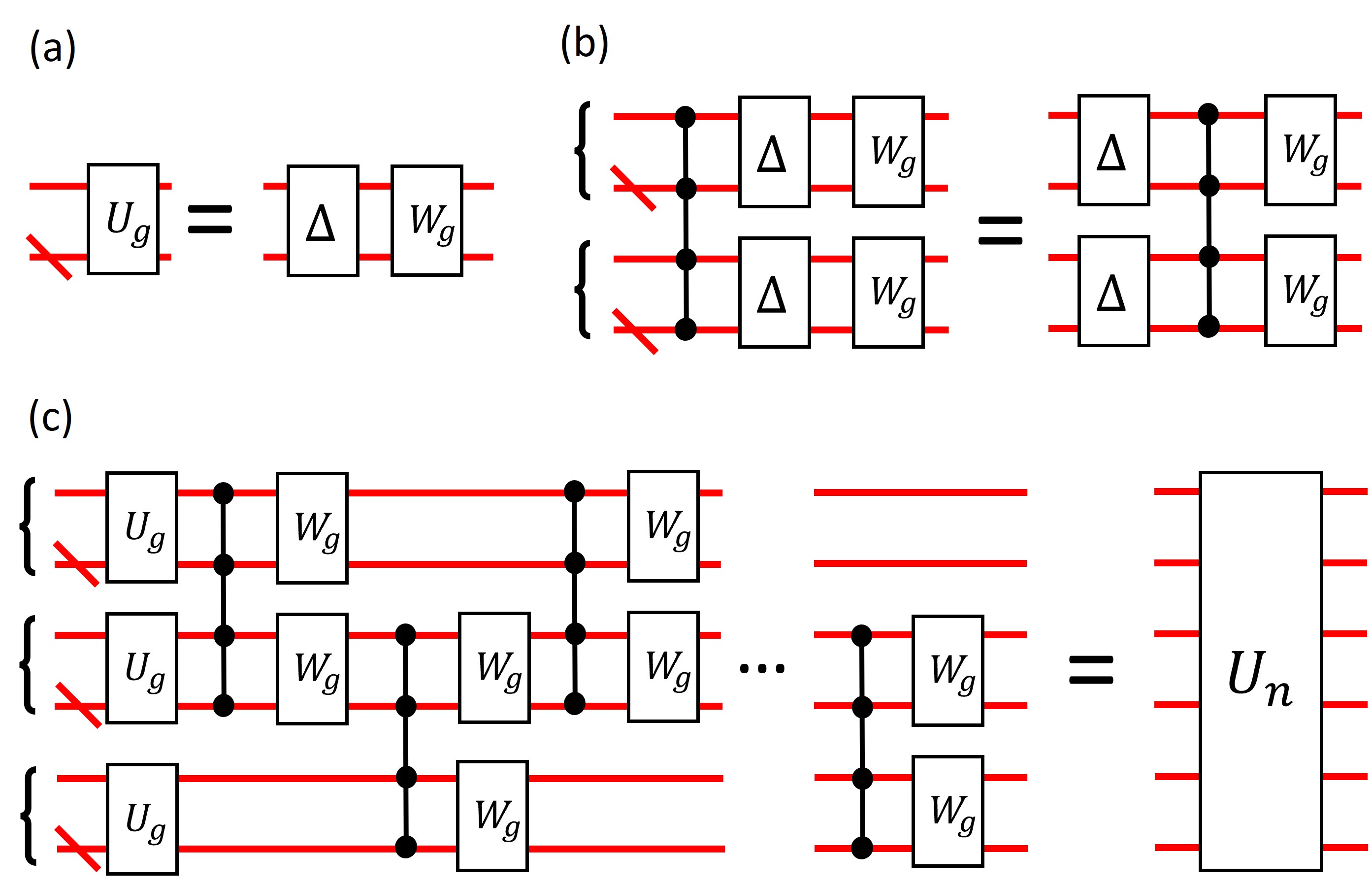}
    \caption{This figure shows how an arbitrary $n$-qubit unitary may be decomposed into CC...CZ gates and $g$-qubit unitaries, which can be efficiently implemented in QLOQ. (a) A generic $g$-qubit unitary $U_g$ is decomposed into a diagonal gate $\Delta$ and a remaining unitary $W_g$. 
    (b) Diagonal gates commuting through a CC...CZ (multi-controlled sign shift) gate. (c) A circuit structure for the decomposition of an $n$-qubit unitary into CC...CZ gates and local $g$-qubit unitaries in QLOQ. A diagonal red line on a qubit rail indicates that the rail represents arbitrarily many qubits. One qubit is separated out from each multi-qubit rail to clearly illustrate that the entangling gates are multi-controlled Z gates, not two-qubit CZs.} 
    \label{fig:unitary_decomp}
\end{figure}

An $n$-qubit unitary can be implemented with $4^{n}-1$ free parameters or less \cite{shende2004minimal}. 
Consider the QLOQ unitary decomposition of an $n$-qubit unitary shown in Figure \ref{fig:unitary_decomp} (c), 
composed of CC...CZ (multi-controlled sign shift) gates and local (single-qudit) unitaries. The construction starts with a layer of $q$ local unitaries $U_i$, each applied to $g_i$ qubits. Each of these unitaries contributes $4^{g_i}-1$ parameters to the circuit, with global phase being disregarded. For every CC...CZ gate added to the circuit, two additional local unitaries can be added as well. However, a diagonal gate with $2^{g_i}$ parameters (including global phase) can commute through the CC...CZ gate, as shown in Figures \ref{fig:unitary_decomp} (a) and (b), whereupon it will be absorbed into the previous local unitary. Therefore, each CC...CZ gate applied between qudits $i$ and $i+1$ only allows $4^{g_i}-2^{g_i}+4^{g_{i+1}}-2^{g_{i+1}}$ additional parameters to be added to the circuit.
Since CC...CZ gates between all of the qubits mapped to a pair of qudits can be implemented with a single physical CNOT in QLOQ (ignoring single qubit rotations), the number of physical CNOTs $k$ needed to implement an arbitrary $n$-qubit unitary in QLOQ is subject to the following inequality:

\begin{equation}
\sum_{i=0}^{q-1}4^{g_i}-1+\sum_{a=0}^{k-1} 4^{g_{a_c}}-2^{g_{a_c}}+4^{g_{a_t}}-2^{g_{a_t}} \geq 4^n-1 \; .
\label{eq:general_inequality}
\end{equation}

\noindent $a_c$ and $a_t$ are the qudit indexes for the control and target qudits to which the CC...CZ gate $a$ is applied. $i$ is the index for the qudit to which each $g_i$-qubit unitary in the first layer is applied. When the same number of qubits ($g$) are encoded on all qudits, every $g_{a_c}$ = $g_{a_t}$ = $g_i$ = $g$, and $q = n/g$. Equation \eqref{eq:general_inequality} then simplifies to:

\begin{equation}
(4^g-1)\frac{n}{g}+2k(4^g-2^g) \geq 4^n-1 \; ,
\label{eq:Nmodn_inequality}
\end{equation}

\noindent from which we derive the following expression for the minimum CNOT cost to implement an arbitrary unitary in QLOQ:

\begin{equation}
k_{min} = \Bigl\lceil\frac{4^n-1-(4^g-1)\frac{n}{g}}{2(4^g-2^g)}\Bigr\rceil \; .
\label{eq:Nmodn_equation}
\end{equation}

At $g = 1$ we recover Equation \eqref{eq:qubit_decomp}, the theoretical lower bound for qubit encoding. Equation \eqref{eq:Nmodn_equation} shows that as the number of qubits encoded on each qudit $g$ increases, the lower bound on the number of physical CNOTs required to implement an arbitrary $n$-qubit unitary decreases exponentially. Where the number of qubits mapped to every qudit is not equal, $k_{min}$ can still be found numerically from Inequality (\ref{eq:general_inequality}). 

Table \ref{Lower_bound_table} shows the lower bound for QLOQ with max qubits-per-qudit of $G$ = 1 (qubit encoding), 2 and 3. These values were derived from Equation \eqref{eq:Nmodn_equation} where the same number of qubits are mapped to each qudit ($n$ $mod$ $G = 0$) and Equation \eqref{eq:general_inequality} otherwise. 
No assumptions were made about the structure of the decomposed circuits, so the lower bounds were found by applying CNOTs only between the largest qudits. For the case where $G = 2$ and $n = 5$ (marked by * in Table \ref{Lower_bound_table}) this results in one of the physical qubits being completely unentangled, so we consider the corresponding value of 42 CNOTs to be an estimate rather than a true lower bound.

\begin{table}[!ht]
    \centering
    \begin{tabular}{|l|l|l|l|}
    \hline
        n & $G =1$ & $G =2$ & $G =3$ \\ \hline
        2 & 3 & 0 & 0 \\ \hline
        3 & 14 & 4 & 0 \\ \hline
        4 & 61 & 10 & 4 \\ \hline
        5 & 252 & 42* & 14 \\ \hline
        6 & 1020 & 169 & 36 \\ \hline
    \end{tabular}
    \caption{\label{Lower_bound_table} The lower bounds on the number of physical CNOT gates required to decompose an arbitrary $n$-qubit unitary with various max qubits-per-qudit values $G$, found using Equation \eqref{eq:Nmodn_equation} and Inequality \eqref{eq:general_inequality}.
    }
\end{table}

In practice, whole circuits with many qubits are not implemented directly through unitary decomposition because of the exponential increase in CNOT cost with qubit number (Equation \eqref{eq:qubit_decomp}). Unitary decomposition is however a widely used tool in quantum computing, with many algorithms utilising the technique \cite{krol2022efficient}. We present a practical method for decomposing unitaries more efficiently than the theoretical lower bound of qubit encoding in Section \ref{QSD Section} (QLOQ QSD). 

The improved theoretical lower bound supports the conclusion that QLOQ is generally more efficient and expressive per physical CNOT than qubit encoding. 
This makes intuitive sense, since each physical CNOT in QLOQ can entangle $2G$ qubits, and allows two additional $G$-qubit unitaries to be added to the circuit. Qubit encoding corresponds to the case $G =1$, so each CNOT can only entangle two qubits, and add two single-qubit unitaries to the circuit.

\subsection{Circuit Design}

As in qubit encoding, circuit structures can be found in QLOQ that perform specific operations with fewer CNOTs than the corresponding theoretical lower bound for unitary decomposition. Finding such structures for specific applications is a time-consuming manual process in both qubit encoding \cite{krol2022efficient} and QLOQ. 

\begin{figure}[!ht]
    \centering
    \includegraphics[width=0.45\textwidth]{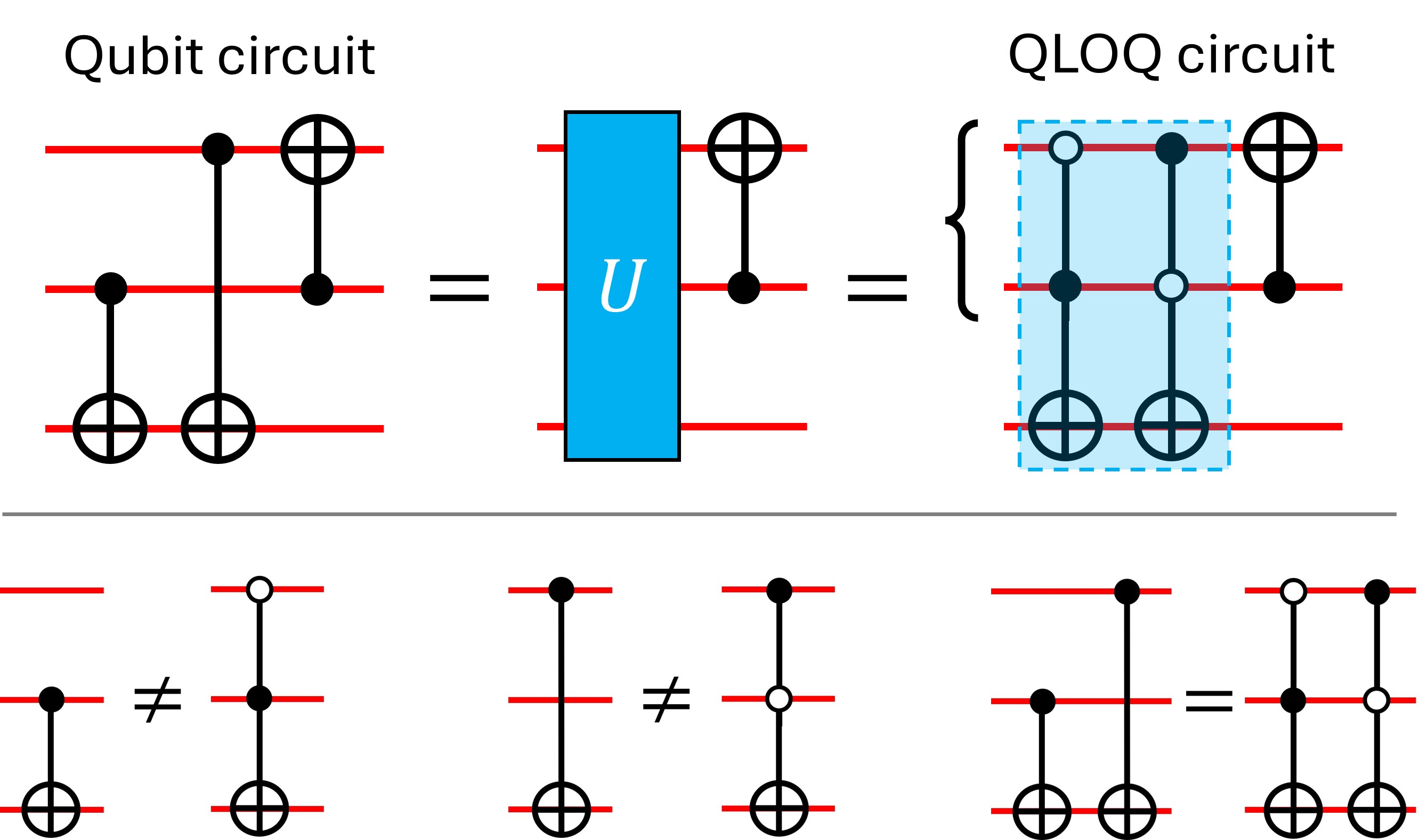}
    \caption{The qubit-based circuit shown costs at least 4 physical CNOTs in any 4L QLOQ map using gate-by-gate compression. However, the first two CNOTs can be combined into a single unitary, which is then decomposed into CCX gates, for a total QLOQ circuit cost of 2 physical CNOTs. The CCX gates are not equivalent to the CNOTs individually, only when combined into one unitary.}
    \label{fig:combined_gates}
\end{figure}

For applications where it is difficult to design circuits from scratch to suit the constraints of QLOQ, it may be desirable to employ an improved version of direct gate-by-gate compression.
This can be achieved by combining adjacent gates in the qubit-based circuit into a single unitary. These unitaries can then be decomposed into gates that are suitable for QLOQ through QLOQ QSD, numerical optimisation, or by hand. An example is shown in Figure \ref{fig:combined_gates}, where a circuit with 3 CNOTs in qubit encoding would cost at least 4 physical CNOTs in any 4L QLOQ map when translating each gate individually. However, by combining two of the CNOTs into a single unitary and manually decomposing that in QLOQ, the overall cost is reduced to 2 physical CNOTs.
This technique still won't scale well for unstructured circuits, but may be useful in specific circumstances. 

QLOQ and qubit encoding apply different constraints to circuit design. Multi-qubit gates for example are an efficient way to generate entanglement in QLOQ, while they are generally avoided wherever possible in qubit encoding due to their high cost when decomposed into physical two-qubit gates. Entangling operations between qubits mapped to the same qudit have a negligible cost in QLOQ, whereas in qubit encoding physical entangling gates are always relatively expensive. Converting circuits designed for QLOQ to qubit encoding will generally increase the number of physical entangling gates required, and vice versa. 
We now present a circuit structure designed specifically for QLOQ with a strong advantage over its qubit-based equivalents: the QLOQ layer-based ansatze for variational quantum algorithms.

%% file: sections/ansatz.tex
\section{QLOQ Ansatze for VQAs}
\label{{QLOQ_Ansatze}}
\subsection{Layer-Based Structures}

Many existing VQA ansatz designs contain repeated layers of two-qubit entangling and single-qubit rotation gates, including the hardware efficient ansatz \cite{kandala2017hardware} and the variational quantum classifier \cite{havlivcek2019supervised}. This simple structure can be sufficient to solve important practical problems, such as finding the ground state energy of molecules with VQE.
We will focus our discussion on layer-based QLOQ ansatze, but note that many other structures are possible. For instance, a QLOQ quantum neural network structure is presented briefly in Section \ref{alt_structures} of the Appendix.

We define two categories of QLOQ entangling layers for convenience: global layers and local layers. Global layers are composed of entangling gates between different qudits, while local layers are composed of operations on a single qudit (which are generally of negligible cost). 
QLOQ ansatze are composed of a combination of global and local layers. 

The minimum number of physical entangling gates $k$ required to entangle $N$ qubits with a maximum qubits-per-qudit of $G$ is given by Equation \eqref{eq:minimal}: 

\begin{equation}
k = \Bigl\lceil\frac{N}{G}\Bigr\rceil-1 \; .
\label{eq:minimal}
\end{equation}

\begin{figure}[!ht]
    \centering
    \includegraphics[width=0.45\textwidth]{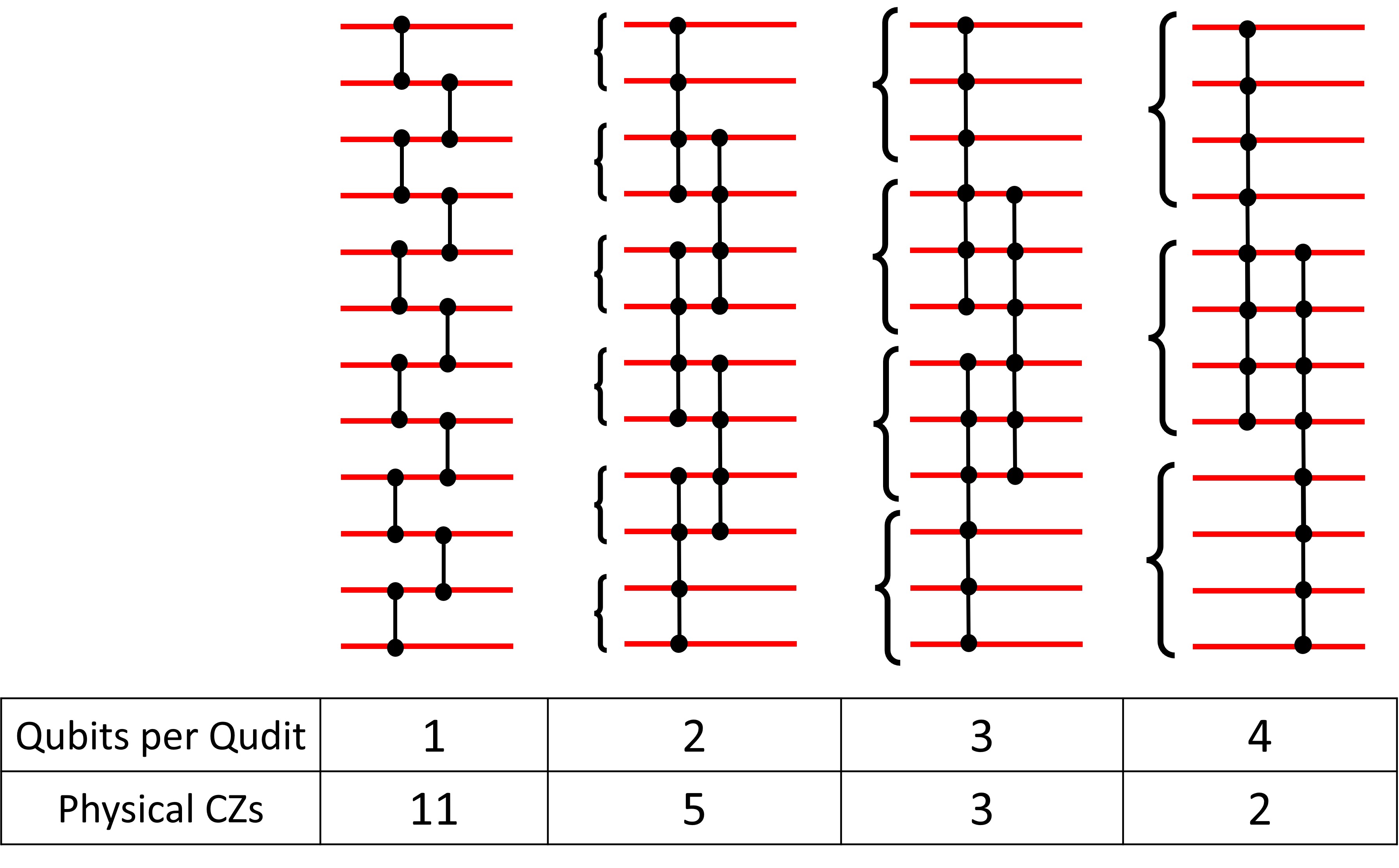}
    \caption{Global entangling layers that use the fewest possible physical entangling gates to entangle 12 qubits in various QLOQ encodings. The qubits per qudit and number of physical CZs for each layer is shown in the table column directly below them. The brackets to the left of each layer indicate which qubits are mapped to the same qudits.}
    \label{fig:minimal layer}
\end{figure}

We derive this from the fact that $q-1$ two-qudit entangling gates are needed to connect $q$ qudits, and there are at least $\lceil \frac{N}{G} \rceil$ qudits and qubits in a circuit. 
In Figure \ref{fig:minimal layer} we show global layers that can entangle 12 qubits using the fewest possible entangling gates in various QLOQ encodings. Such layers are useful for designing resource-efficient ansatze for VQAs. Each qubit-logical multi-controlled gate is implemented by a single physical entangling gate. 
The more qubits are mapped to each qudit, the fewer physical entangling gates are needed for each global layer. 

\begin{figure}[ht]
    \centering
    \includegraphics[width=0.4\textwidth]{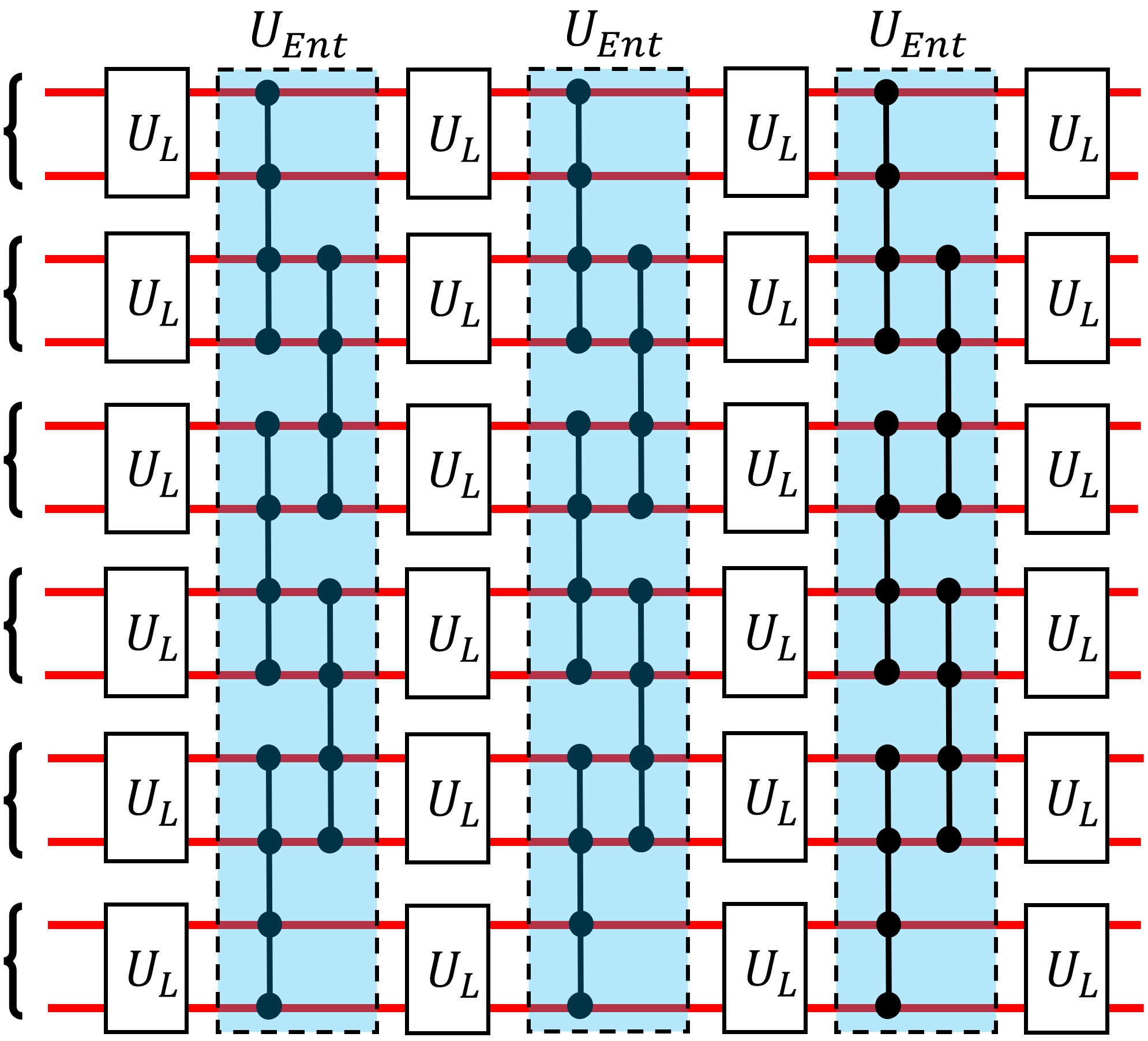}
    \caption{A layer-based QLOQ ansatz in an encoding scheme where qudits have 4 levels. Each global layer $U_{Ent}$ entangles 12 qubits with 5 physical CZs. Any two-qubit unitary can be used for each local $U_L$ gate, adding qubit-logical entanglement and expressibility without requiring additional physical entangling gates.}
    \label{fig:10}
\end{figure}

A QLOQ ansatz that combines these optimal global entangling layers with local layers is shown in Figure \ref{fig:10}. Since the physical entangling gates in the global layers have a nearest-neighbour structure, this ansatz is as effective on qudits with a linear topology as on qudits with all-to-all connectivity. 
There are no qubit-logical entangling operations between pairs of qubits mapped to different qudits in this structure, which avoids one of the primary drawbacks of QLOQ. 
This leaves hardware limitations as the only constraint on qudit size.

QLOQ ansatze have two primary advantages over qubit-based ansatze: 

\begin{itemize}
\item QLOQ global layers can entangle $N$ qubits with fewer physical entangling gates ($\lceil \frac{N}{G} \rceil-1$) than qubit encoded layers ($N-1$). 
\item Local layers consisting of only single-qudit physical operations can increase the expressibility and qubit-logical entanglement of an ansatz without using any physical entangling gates. 
\end{itemize}

\subsection{Expressibility \& Entanglement }
We will examine two metrics of particular importance in comparing the performance of ansatze for VQAs: expressibility (Expr) and entangling capability (Ent) \cite{sim2019expressibility}. The expressibility of an ansatz is its ability generate pure states that are representative of the Hilbert space, 
while entangling capability is a measure of its ability to generate entangled
states. 
In general, ansatze that are more expressive and have higher entangling capability will arrive at more accurate answers when used in VQAs, although an ansatz that is too expressive may require more optimisation iterations to converge \cite{sim2019expressibility} or become stuck in a barren plateau \cite{holmes2022connecting}. 

\begin{figure*}[!ht]
    \centering
    \includegraphics[width=0.9\textwidth]{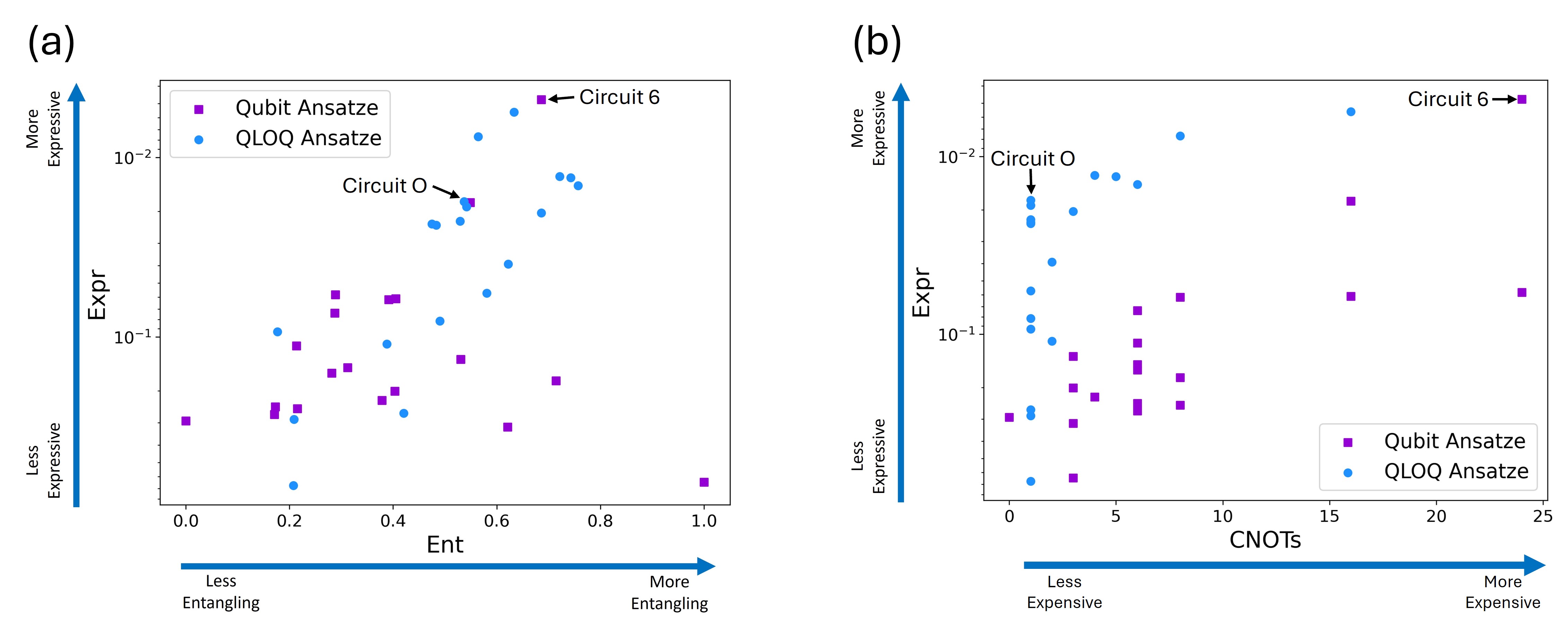}
    \caption{(a) Entangling capability vs expressibility for the ansatze tested in the qubit-based Sim et al.\ (purple) and QLOQ (blue) benchmark sets. (b) Expressibility vs physical CNOT cost for the same ansatze. Smaller Expr values (higher on the $y$-axis) indicate a more expressible circuit.}
    \label{fig:expr_plots}
\end{figure*}

\begin{figure*}[!ht]
    \centering
    \includegraphics[width=0.9\textwidth]{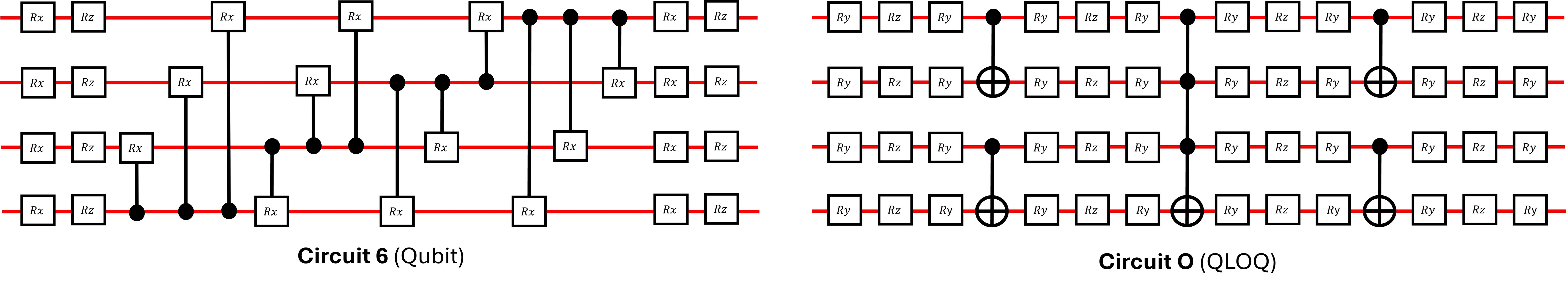}
    \caption{Example ansatze from the Sim et al.\ (Circuit 6) and QLOQ (Circuit O) benchmark sets.}
    \label{fig:ansatze_qubit_v_qloq}
\end{figure*}

We note that expressibility and entangling capability are not perfect metrics for assessing VQA ansatze. Hamiltonians of molecular systems for example may contain symmetries that can be exploited by problem-specific ansatze, which is not accounted for by these metrics \cite{rasmussen2020reducing}. However, they remain a useful tool for judging the problem-agnostic effectiveness of ansatze.
 
A fidelity distribution is a histogram with the fidelity between pairs of randomly sampled states from a given set on the $x$-axis, and the corresponding probability of each fidelity occurring on the $y$-axis. 
As a measure of expressibility, Sim et al.\ 2019 \cite{sim2019expressibility} used the standard KL divergence $D_{KL}$ \cite{kullback1951information} between the fidelity distribution generated by an ansatz $\hat{P}_{Ansatz}$ and the fidelity distribution of Haar random states $P_{Haar}$ \cite{sim2019expressibility}:

\begin{equation}
Expr = D_{KL}(\hat{P}_{Ansatz} ||P_{Haar} ) \; . 
\end{equation}

\noindent $P_{Haar}$ for a d-dimensional Hilbert space is found analytically as \cite{sim2019expressibility}:

\begin{equation}
P_{Haar}(F) = (d-1)(1-F)^{d-2} \; .
\end{equation}
\noindent The fidelity distribution of an ansatz $\hat{P}_{Ansatz}$ is constructed by randomly sampling pairs of parameter vectors $\theta_0$ and $\theta_1$, applying the ansatz with these parameters to the input state 
$|0 \rangle ^{\otimes n}$ to produce the output states 
$|\psi_{\theta_{0}} \rangle$ and $|\psi_{\theta_{1}} \rangle$, 
computing the fidelity $F$ between these states as $F = |\langle \psi_{\theta_{0}} | \psi_{\theta_{1}} \rangle | ^{2}$, and plotting these fidelities as a histogram. 
Smaller Expr values indicate a more expressible circuit.

The entangling capability of an ansatz is given by the average of its Meyer-Wallach measure $Q(|\psi\rangle)$ \cite{meyer2002global} for a set of parameter vectors $S = \{\theta_{m}\}$ \cite{sim2019expressibility}:

\begin{equation}
 Ent = \frac{1}{|S|} \sum\limits_{\theta_{m} \in S} Q(|\psi_{\theta_{m}}) \rangle \; .
\end{equation}

\noindent The Meyer-Wallach measure for an ansatz can be found from the sum of the purity of each of its $N$ qubits \cite{brennen2003observable}:

\begin{equation}
 Q(|\psi\rangle) = 2 (1 - \frac{1}{N} \sum\limits_{j = 0}^{N-1} \text{Tr} \left[\rho_{j}^{2}\right] )  \; ,
\end{equation}

\noindent where $\rho_{j}$ is the reduced state of qubit $j$. 

Sim et al.\ compare the entangling capability and expressibility of a selection of ansatze used in various VQA applications, such as data classification, quantum autoencoding, and VQE \cite{kandala2017hardware, schuld2020circuit, romero2017quantum}. We re-evaluated these metrics for the Sim et al.\ benchmark set in simulation, and compared them to a set of 19 QLOQ ansatze in the QLOQ(0,1)(2,3) map. A circuit from each of these sets is shown in Figure \ref{fig:ansatze_qubit_v_qloq}: Circuit O (QLOQ) and Circuit 6 (qubit). The rest of the QLOQ ansatze are presented in Figure \ref{fig:all_qloq_ansatze} of the Appendix. 
As in Sim et al.\ a bin size of 75 was taken for the KL divergence histograms \cite{sim2019expressibility}. 5000 sample pairs were taken per ansatz, compared to 1000 in Sim et al. 

Figure \ref{fig:expr_plots} (a) shows a scatter plot of the expressibility and entanglement results for both sets of ansatze. It can be seen that the QLOQ ansatze have similar or better ratios of expressibility to entangling capability than the qubit-based ansatze. The exact descriptor values measured are shown in Table \ref{expr table} of the Appendix. 

Figure \ref{fig:expr_plots} (b) shows a plot of the expressibility of each ansatz against its entangling gate cost in terms of physical CNOTs. CZs are equivalent to CNOTs plus single-qubit rotations, while controlled-rotation gates can be decomposed into two CNOTs plus single-qubit rotations. QLOQ ansatze can thus achieve much greater expressibility (lower Expr values) per physical entangling gate than the qubit-based ansatze from the benchmark set. QLOQ Circuit O for example exceeds the expressibility of all the qubit-based circuits in the Sim et al.\ benchmark set except for Circuit 6, despite only using 1 physical CNOT. Circuit 6 meanwhile employs the equivalent of 24 physical CNOTs (assuming all-to-all connectivity). 
As physical entangling gates are a constrained resource for quantum computers in the NISQ era (and beyond) this is a key result of our research. 

Figure \ref{fig:expr_plots} (b) supports our previous assertion that QLOQ circuits are generally more expressive per two-level entangling gate than qubit encoded circuits. This makes intuitive sense given that global layers use fewer entangling gates in QLOQ than in qubit encoding, and QLOQ local layers add expressibility and qubit-logical entanglement without requiring physical entangling gates at all. 

If the difficulty of implementing qudits with at least 4 levels on hardware does not outweigh this entangling gate reduction, QLOQ ansatze have the potential to significantly improve the performance of VQAs on NISQ devices. We will now examine this advantage in detail for spatially encoded LOQC, an architecture where qudits are easy to deploy and control.

%% file: sections/VQE.tex
\section{VQE} 
\label{Experimental_VQE}
\subsection{Linear Optical Quantum Computing}

Several routes have been proposed to develop fault-tolerant universal photonic quantum computers \cite{de2024spin, hilaire2024enhanced, Kok2007linear, nielsen2004optical, knill2001scheme, yoran2003deterministic, browne2005resource}. 
Current photonic NISQ devices \cite{maring2024versatile} rely on probabilistic gates to generate entanglement, typical examples of which are the heralded Knill CZ \cite{knill2002quantum} 
and post-selected Ralph et al.\ CNOT \cite{ralph2002linear}. These gates usually fail to apply the desired operation on their input photons, but we can identify which measurements correspond to gate failures and discard those results. When doing so using post-selection, the output states of gate failures are not valid qubit or qudit states e.g.\ two photons in the same mode in spatial encoding. When using heralding, ancilla modes will have a particular state when the gate has been applied successfully e.g.\ a photon in each of the ancillas for the Knill CZ. As a rule of thumb, post-selected gates can only be used at the end of a linear optical circuit, but they may have a higher success probability than heralded gates, which can be used anywhere in the circuit. Every Knill and Ralph et al.\ gate used in a circuit requires two additional ancilla modes.

The overall success probability of a probabilistic LOQC circuit is given by the product of the success probabilities of each gate in the circuit. The circuit's success probability approaches zero as the number of entangling gates increases. This challenge can be overcome in various ways by the aforementioned routes to fault-tolerant universal photonic quantum computing, but these are more demanding in terms of resources and hardware performance than probabilistic NISQ LOQC. This is also why there is such a strong benefit to removing entangling operations from probabilistic NISQ LOQC circuits. 

\subsection{VQE Explained}

The variational quantum eigensolver (VQE) is a hybrid quantum-classical algorithm first proposed by Peruzzo et al.\ \cite{peruzzo2014variational}. VQE can be used to compute an upper bound on the ground state-energy of a Hamiltonian, which has applications in material science, drug discovery, and chemical engineering \cite{tilly2022variational}. VQE is based on the variational principle, whereby the ground state energy $E_0$ associated with a Hamiltonian $\hat{\mathcal{H}}$ and an ansatz wavefunction $\ket {\psi (\vec \theta)}$ parameterised by $\vec \theta$ satisfies the following inequality \cite{tilly2022variational}: 

\begin{equation}
E_0 \leq \frac{\bra {\psi (\vec \theta)} \hat{\mathcal{H}} \ket {\psi (\vec \theta)}}{\bra {\psi (\vec \theta)} {\psi (\vec \theta)} \rangle} \; .
\label{eq:VQE}
\end{equation}
VQE seeks to estimate the ground state energy $E_0$ by varying the quantum circuit parameters $\vec \theta$ to minimise the expectation value of the circuit $\bra {\psi (\vec \theta)} \hat{\mathcal{H}} \ket {\psi (\vec \theta)}$. 

\begin{figure*}[t]
    \centering
    \includegraphics[width=0.9\textwidth]{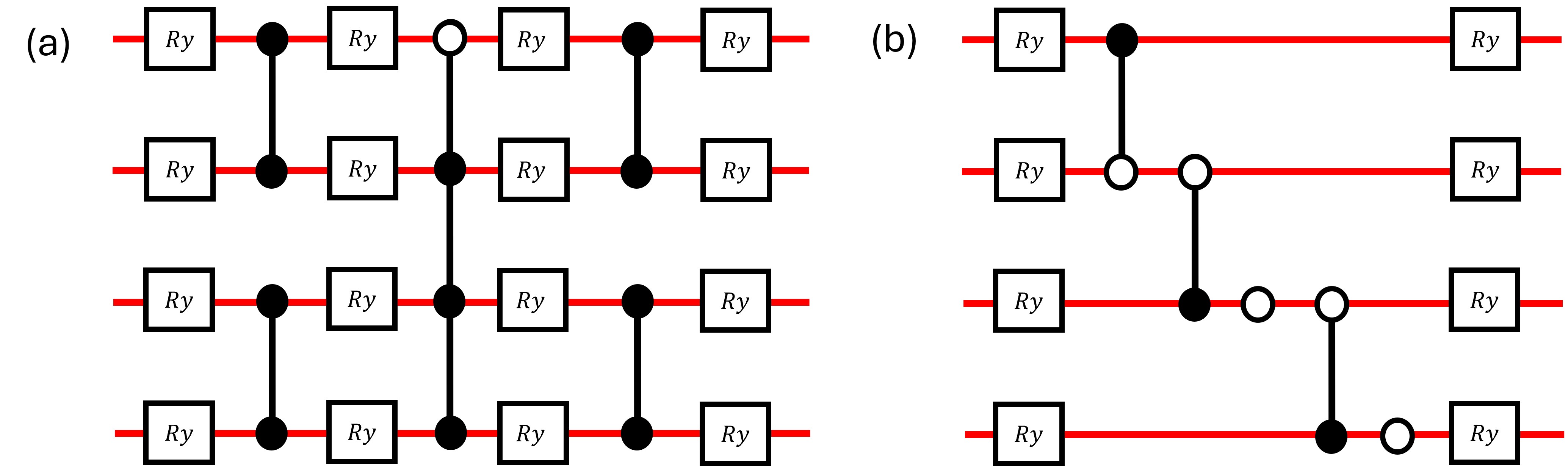}
    \includegraphics[width=0.9\textwidth]{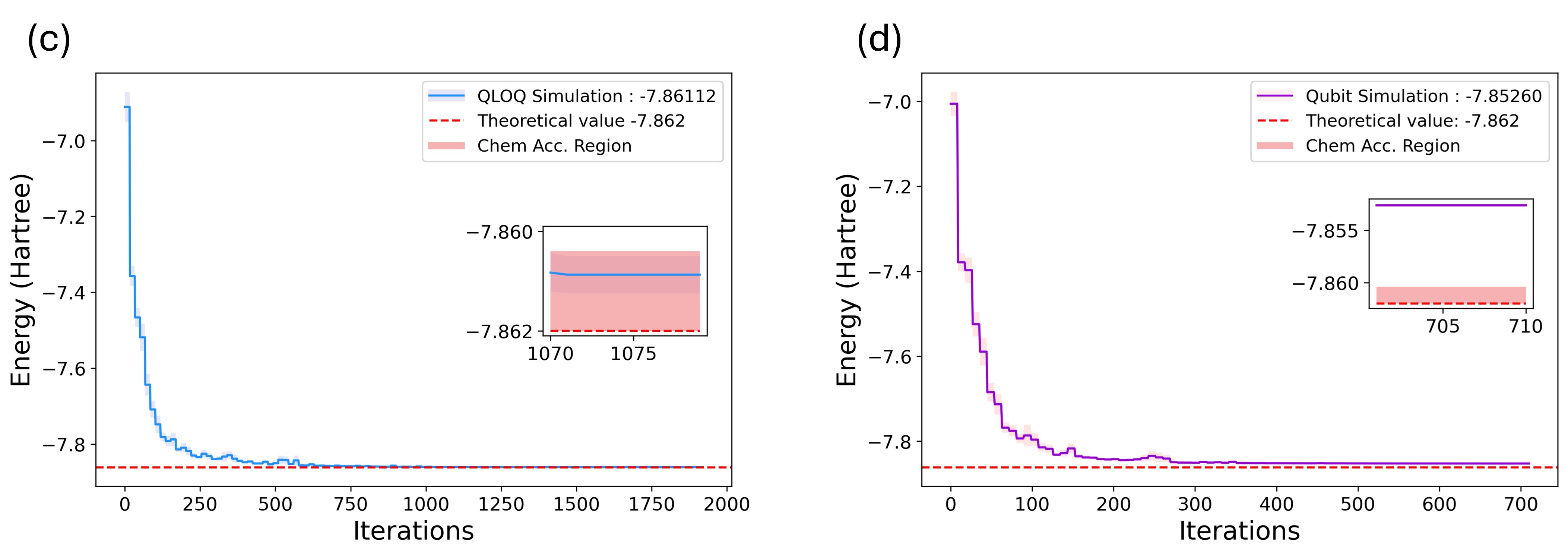}
    \caption{ (a) The QLOQ ansatz used for the LiH VQE. (b) The qubit-based ansatz used for the LiH VQE simulations, with entanglement generated by cascading CZs from Ralph et al.\ 2004 \cite{ralph2004scaling}. A white control circle on a CZ or CCCZ in these diagrams indicates that the phase shift is applied when that qubit is in the state $|0\rangle_{L}$ instead of $|1\rangle_{L}$. The other white circles are sign flip (Pauli-Z) gates. (c) The average of 100 QLOQ VQE runs using L-BFGS-B. (d) The average of 100 qubit-based VQE runs using L-BFGS-B.} 
    \label{fig:sim ansatze}
\end{figure*}

The VQE algorithm can be summarised as follows:
\begin{itemize}
    \item[1.] Find a Hamiltonian formulation of the problem to be minimised. For our experiment, we used the parity mapping \cite{seeley2012bravyi} available through the OpenFermion python package \cite{mcclean2020openfermion}. 
    This package provides the number of qubits required to run the VQE for a particular molecule, and a decomposition of the molecular Hamiltonian into Pauli terms, real-coefficients, and associated measurement bases. QLOQ allows qubit-based Hamiltonians such as those generated by the parity mapping to be implemented using qudits. 
    \item[2.] Select an ansatz for the quantum circuit with the correct number of qubits and sufficient expressibility to reach a good approximation of the ground state energy. For our experiment we used the QLOQ ansatz from Figure \ref{fig:sim ansatze} (a). 
    \item[3.] Estimate the expectation value of the Hamiltonian $\langle \hat{\mathcal{H}} \rangle$ for a set of $\vec \theta$ parameters. We used the formula $\langle \hat{\mathcal{H}} \rangle = \sum_a^T h_a \langle P_a \rangle$ for all $T$ Pauli terms $P_a$ with their real coefficients $h_a$ \cite{kandala2017hardware}. 
   $\langle P_a \rangle$ is found for each Pauli term by applying the corresponding basis rotations to the ansatz and taking many measurements (the required number depends on desired precision \cite{tilly2022variational}). LiH has 99 Pauli terms, some of which are grouped and measured in the same basis, for a total of 25 Pauli groups (tensor product basis sets \cite{kandala2017hardware}).  
   \item[4.] Optimise the ansatz parameters via a classical optimisation procedure (e.g.\ the L-BFGS-B algorithm \cite{zhu1997algorithm}) to reach a lower value of $\langle \hat{\mathcal{H}} \rangle$. 
    \item[5.] Repeat steps 3 and 4 until the ground state energy estimate (GSEE) given by $\langle \hat{\mathcal{H}} \rangle$ has converged (see Figures \ref{fig:sim ansatze} (c) and (d)).
\end{itemize}

\subsection{VQE Setup}

We ran noise-free LOQC simulations of an LiH VQE using the SLOS algorithm \cite{heurtel2023strong} in the software package Perceval \cite{heurtel2023perceval}. Two classical optimisers were tested: L-BFGS-B \cite{zhu1997algorithm} and COBYLA \cite{powell1994direct}. L-BFGS-B is more accurate than COBYLA in the absence of noise, but COBYLA performs better in a noisy environment (such as a QPU).
Three ansatze were tested: one in qubit encoding with no entanglement, one in qubit encoding with entanglement, and one in QLOQ with entanglement. 
For the qubit VQE simulations with no entanglement, each qubit was parameterised by a single rotation about the $y$-axis on the Bloch sphere (an $R_y$ gate). This ansatz provided a performance baseline against which the other ansatze could be compared. 100 simulations of each type were performed. 

The QLOQ ansatz chosen to perform the VQE is shown in Figure \ref{fig:sim ansatze} (a). It uses a standard Ralph et al.\ CZ \cite{ralph2002linear} as the physical entangling gate to create a CCCZ. The CCCZ phase shift is applied on the combined qubit state $|0111\rangle_{L}$ instead of $|1111\rangle_{L}$ to simplify the optical circuit, which does not affect the outcome of the VQE. Details on the hardware implementation are provided in Section \ref{sec:optical} of the Appendix.

Figure \ref{fig:sim ansatze} (b) shows the qubit-based ansatz (with entanglement) we selected for our VQE. 
We chose to use cascading Ralph CZs \cite{ralph2004scaling} to provide the entanglement because they provided the highest available success probability for a 4-qubit ansatz in qubit encoding (1/81). An ansatz with the same number of parameters using 3 standard Ralph et al.\ CZs for entanglement would have had a success probability of just 1/729. A photon is required for each qudit or qubit, so the QLOQ ansatz needed 2 photons and the qubit ansatz needed 4. The QLOQ ansatz would have had a success probability of 1/9 from its single physical Ralph et al.\ CZ, but further linear optical optimisations raised its average success probability to 0.448 (see Section \ref{sec:optical}). 

For VQE runs using the COBYLA classical optimiser, both simulated and on the QPU, 3 million shots were taken for each Pauli group in each iteration. A shot is counted if at least 1 photon was detected at the output of the circuit. We count shots instead of samples because NISQ LOQC circuits are probabilistic and the number of samples may vary. For simulations with the classical optimiser L-BFGS-B, it was assumed that infinitely many samples were taken i.e.\ the effects of sampling were ignored. L-BFGS-B performs poorly in the presence of noise, including sampling noise, but can reach a more accurate final answer than COBYLA.

It is standard practice when estimating the ground state energy of a molecule with VQE to use its Hartree-Fock wavefunction as the circuit's initial input state \cite{tilly2022variational, zhang2021mutual}. For LiH, this is $|0101\rangle_{L}$ \cite{choy2023molecular}, so that is the qubit-logical input state we used.

\subsection{VQE Results}
\label{vqe_results}

The results from our VQE simulations are presented in Figures \ref{fig:time_plots} and \ref{fig:sim ansatze}. Further details are provided in Table \ref{results table} in the Appendix. 
Only the QLOQ ansatz with the classical optimiser L-BFGS-B was able to achieve chemical accuracy for the ground state energy of LiH: $-7.862$ $\pm 0.0016$ Hartree \cite{kandala2017hardware, tilly2022variational, choy2023molecular, peterson2012chemical}. For both classical optimisers tested, the QLOQ ansatz outperformed the qubit-based ansatz on average and in terms of the minimum (most accurate) value reached across all runs. The qubit-based ansatz with no entanglement performed worst, as expected. Figures \ref{fig:sim ansatze} (c) and (d) show the average GSEE convergence of 100 VQE simulations in QLOQ and qubit-encoding 

\newpage

\begin{figure}[!ht]
    \centering
    \includegraphics[width=0.45\textwidth]{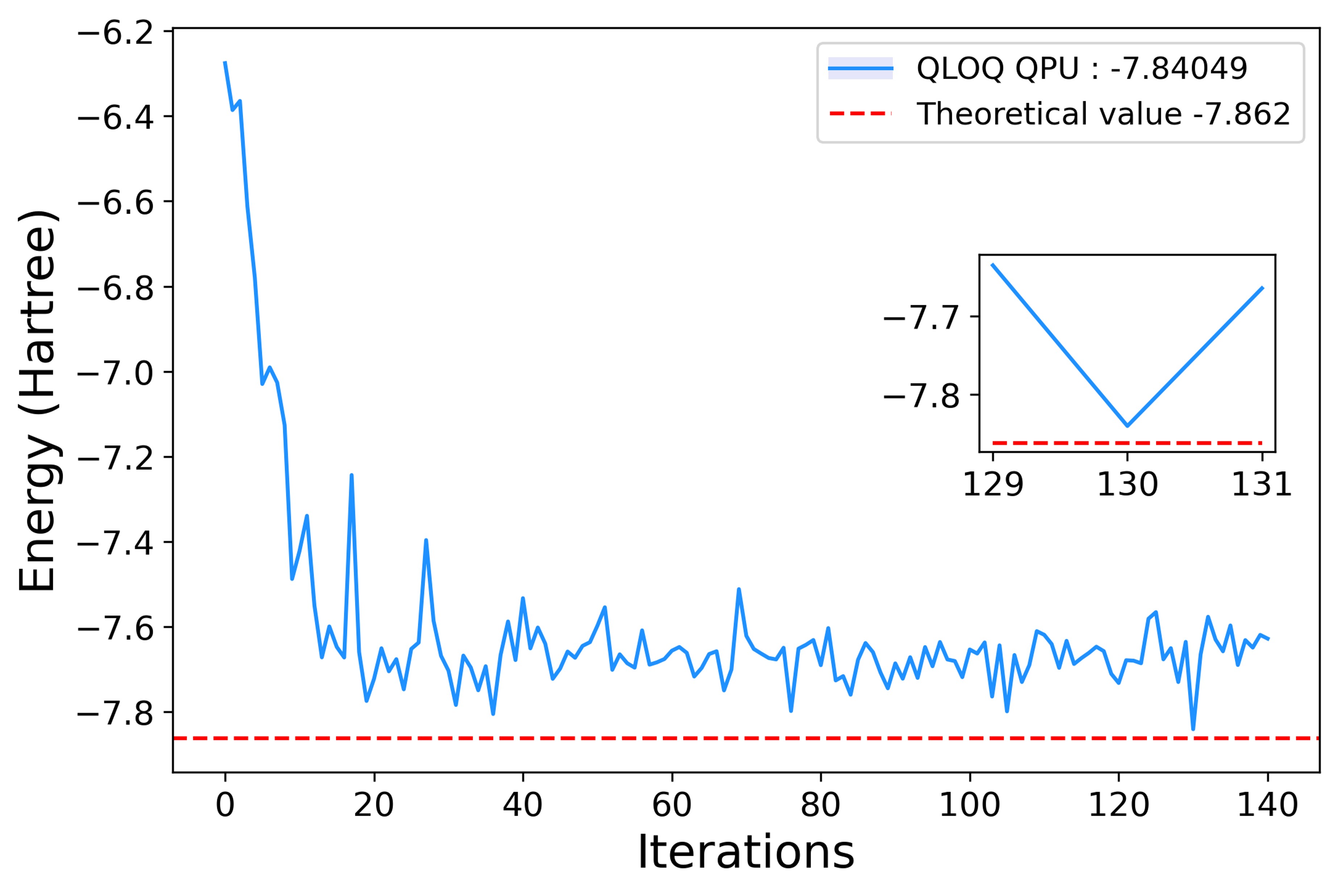}
    \caption{(a) The convergence plot from running the LiH VQE in QLOQ on Quandela's photonic QPU, using COBYLA as the classical optimiser. The error estimate is rendered as a shaded light blue region, but this may be too small to see at the current image resolution. The error estimate is plotted separately in Section \ref{vqe_error}.
    }
    \label{fig:QPU_plot}
\end{figure}

\noindent respectively, with L-BFGS-B as the classical optimiser. 

Figure \ref{fig:QPU_plot} shows the VQE convergence plot from the QLOQ ansatz on Quandela's linear optical QPU, with COBYLA as the classical optimiser. The QPU VQE run reached a similar final GSEE to the average of the noise-free COBYLA simulations. 
When the experiment was run, the QPU could generate and detect pairs of coincident photons at a rate of approximately 9 kHz, while the 4-photon coincidence rate was approximately 20 Hz. 
On average, the COBYLA simulations in QLOQ required approximately twice as many iterations to converge as those in qubit encoding (218 vs 103 terations), possibly because the qubit ansatz has fewer parameters to optimise. 
Ignoring the increase in certain errors, such as photon loss, that arise from having more photons in the circuit, we would expect the QLOQ ansatz to run approximately 7705 times faster than the qubit-based ansatz. This accounts for the differences between the two ansatze in success probability, photon coincidence rate, and average number of iterations required to converge. 
Since the QLOQ VQE took 5 hours on Quandela's QPU, we predict that the qubit VQE would take approximately 4.39 years on the same device. 
For obvious reasons, we did not confirm this experimentally.

Not only is the QLOQ VQE more accurate than its qubit-based counterpart, its lower resource requirement allows it to actually be performed on NISQ hardware in a realistic amount of time. 
The 4-qubit QLOQ ansatz on Quandela's QPU generated approximately 7800 successful samples per second during the experiment, compared to the reported approx.\ 4000 samples per second available with some cloud-accessible superconducting QPUs \cite{sachdeva2024quantum}.

\subsection{Scaling Advantages}

\label{sec:scaling_advanatges}
QLOQ ansatze significantly increase the size of problems that are tractable to VQAs on NISQ linear optical hardware. 
$\lceil\frac{N}{G}\rceil$  photons are required to encode $N$ qubits in QLOQ, so larger qudits also reduce a circuit's photon requirement. This greatly improves run times both on QPU and in classical simulations with software such as Perceval \cite{heurtel2023perceval}. Loss errors increase and shot rates decrease exponentially as more photons are added to a circuit, while using more spatial modes to implement larger qudits is essentially free, provided the modes are available on chip.

\begin{figure}[!ht]
    \centering
    \includegraphics[width=0.45\textwidth]{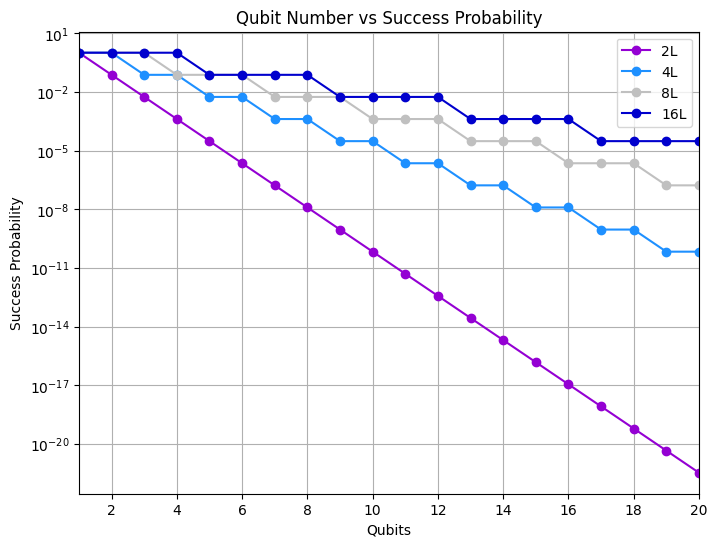}
    \caption{Best-case success probability vs qubit number for heralded LOQC entangling layers based on the Knill CZ. The legend shows the number of qudit levels $L$ (spatial modes) required by each qudit for a given encoding. Note that the exponential decrease in success probability shown applies specifically to the NISQ LOQC paradigm we are considering, as opposed to other scalable schemes with in-built loss and fault-tolerance. }
    \label{fig:Heralded_Layer_Probs}
\end{figure}

\begin{figure*}[!ht]
    \centering
    \includegraphics[width=0.95\textwidth]{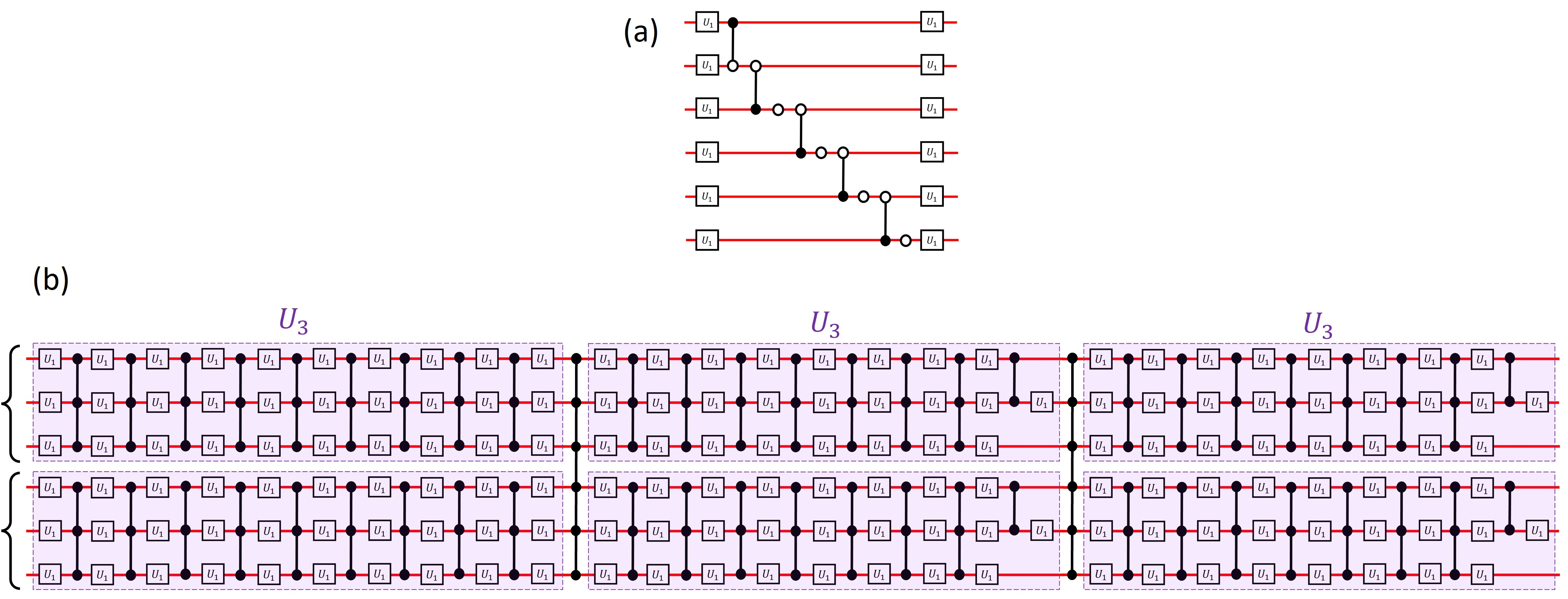}
    \caption{(a) The largest 6-qubit ansatze that will fit on Quandela's 6-photon, 20-mode QPU in qubit encoding. (b) An ansatz for the map QLOQ(0,1,2)(3,4,5) that runs 27,000 times faster than the ansatz in (a) on the same QPU, and has 7.6 times more free parameters.
    The purple $U_{3}$ boxes represent generic three-qubit unitary operations. The first $U_{1}$ gate applied to each qubit represents 3 free parameters, with the rest representing 2 free parameters. In practice, the $U_{3}$ structures shown are not sufficient to apply all possible three-qubit unitaries, they are only meant to illustrate the theoretical lower bound on the number of $U_{1}$ gates required.}
    \label{fig:ansatze comparison}
\end{figure*}
 
The number of available spatial modes and maximum optical component depth are the only constraints on qudit size in spatial LOQC. 
In this architecture, QLOQ is subject to a trade-off between the $2^{g_i}$ modes needed to encode $g_i$ qubits on qudit $i$, the $2x$ reduction in modes provided by the net $x$ physical CNOT/CZs converted into deterministic local operations, and the dump ports (extra spatial modes) required to balance the success probabilities of certain post-selected gates between qudits (see Appendix section \ref{sec:optical} for details. 

Figure \ref{fig:Heralded_Layer_Probs} shows the success probability of heralded global entangling layers composed of CC...CZs using the minimum number of physical entangling gates (as in Figure \ref{fig:minimal layer}).
Even with relatively few qubits, QLOQ layers have success probabilities many orders of magnitude higher than qubit encoded layers. This leads to proportionally faster computation, even before considering the reduced photon requirement. The heralded entangling layers with 2 or 3 qubits per qudit (4L or 8L) also require fewer spatial modes overall than qubit-based layers.
Post-selected layers based on the cascading Ralph CZ require more modes (but still fewer photons) as more qubits are mapped to each qudit, since they require dump ports to balance their success probabilities. 

Figure \ref{fig:ansatze comparison} (a) shows the largest 6-qubit ansatz that could be implemented on Quandela's 6-photon 20-mode linear optical quantum computer in qubit-encoding at the time of testing. Cascading Ralph CZs can entangle 6 qubits with a success probability of 1/729. The resulting ansatz has 30 free parameters. In QLOQ(0,1,2)(3,4,5), an ansatz with two CCCCCZs based on the Knill CZ will have 350 free parameters and a success probability of 4/729. Since they require the same number of photons, we would expect the QLOQ ansatz to be 4 times faster and dramatically more expressive than its qubit-based counterpart.

Alternatively, if only 1 Knill-based CCCCCZ is used, the circuit will have 230 free parameters, and a success probability of 2/27. This ansatz only requires 4 photons, compared to 6 photons for the ansatz in Figure \ref{fig:ansatze comparison} (a). At time of testing the 4-photon coincidence rate was approximately 500 times higher than the 6-photon coincidence rate on Quandela's hardware. The resulting QLOQ ansatz (shown in Figure \ref{fig:ansatze comparison} (b)) will therefore have 7.6 times more free parameters and would be expected to run approximately 27,000 times faster than the largest qubit-based ansatz on this device. These more powerful, faster QLOQ ansatze increase the size of problems that can be solved on NISQ devices. 

The two main drawbacks of QLOQ are that logical operations between qubits mapped to different qudits require more physical entangling gates, and qudits can be difficult to implement in hardware. The first drawback is avoided by designing circuit structures to suit QLOQ, such as layer-based QLOQ ansatze, and in LOQC encoding two qubits as one qudit actually requires fewer resources (4 modes, 1 photon) than encoding them separately as qubits (4 modes, 2 photons). Therefore QLOQ ansatze are generally superior to qubit-based ansatze for practical layer-based VQAs in spatially-encoded probabilistic LOQC.

%% file: sections/qsd.tex
\section{Quantum Shannon Decomposition}
\label{QSD Section}

In Section \ref{lower_bound} we showed that the lower bound on the number of physical entangling gates required to implement an arbitrary unitary is much lower in QLOQ than in qubit encoding. Here, we provide a scheme for accessing this advantage in practice by adapting the Quantum Shannon Decomposition (QSD) technique \cite{shende2005synthesis} to QLOQ.

\begin{figure}[ht]
    \centering
    \includegraphics[width=0.4\textwidth]{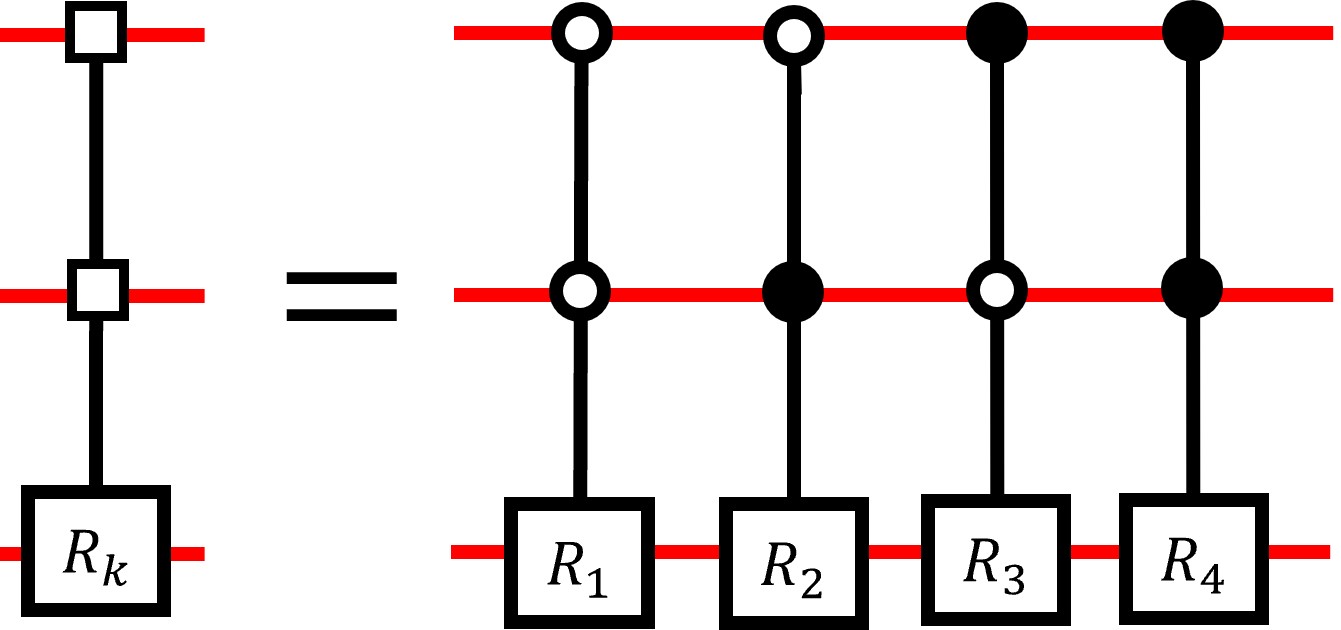}
    \caption{Decomposition of a 3-qubit rotational multiplexor into multi-controlled rotation gates. A white circle indicates that the control state of the gate for that qubit is $|0\rangle_{L}$. White squares indicate the multiplexor's select qubits.}
    \label{multiplexor_decomp}
\end{figure}

QSD relies on multiplexors, which are multi-qubit gates that apply a different unitary to the target qubit depending on the state of one or more `select' qubits. A rotational multiplexor and its decomposition into multi-controlled rotation gates are shown in Figure \ref{multiplexor_decomp}, with the select qubits marked by small white squares.

\begin{figure}[!ht]
    \centering
    \includegraphics[width=0.95\columnwidth]{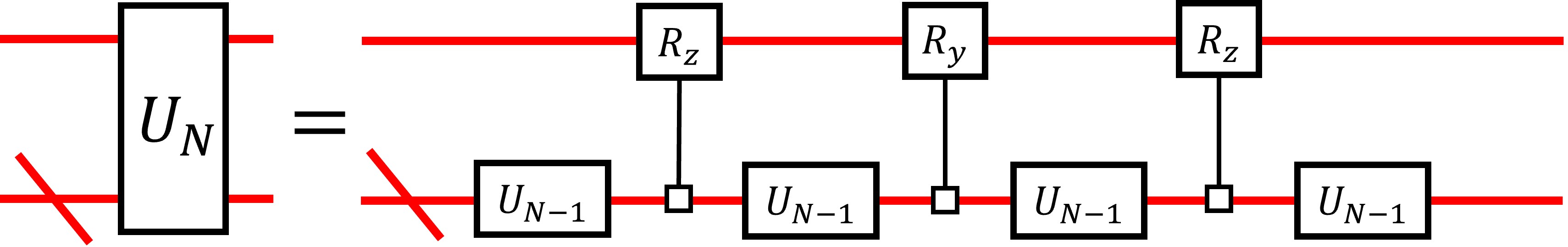}
    \caption{The Quantum Shannon Decomposition. 
    }
    \label{QSD}
\end{figure}

QSD is a recursive decomposition, as shown in Figure \ref{QSD}. An $n$-qubit unitary can be decomposed into 3 rotational multiplexors and four $[n-1]$-qubit unitaries, which can themselves be decomposed into multiplexors and $[n-2]$-qubit unitaries, and so on. In qubit encoding, when the recursion reaches 2-qubit unitaries, alternative decompositions are used that cost fewer CNOTs than continuing the recursion.

Our proposed version of the QSD for QLOQ uses a single qudit to which $g$ qubits have been mapped, with the rest of the circuit being encoded as qubits. The final $g$-qubit unitaries in the recursion are implemented locally on that qudit for 0 physical CNOTs. 

\begin{figure}[!ht]
    \centering
    \includegraphics[width=0.98\columnwidth]{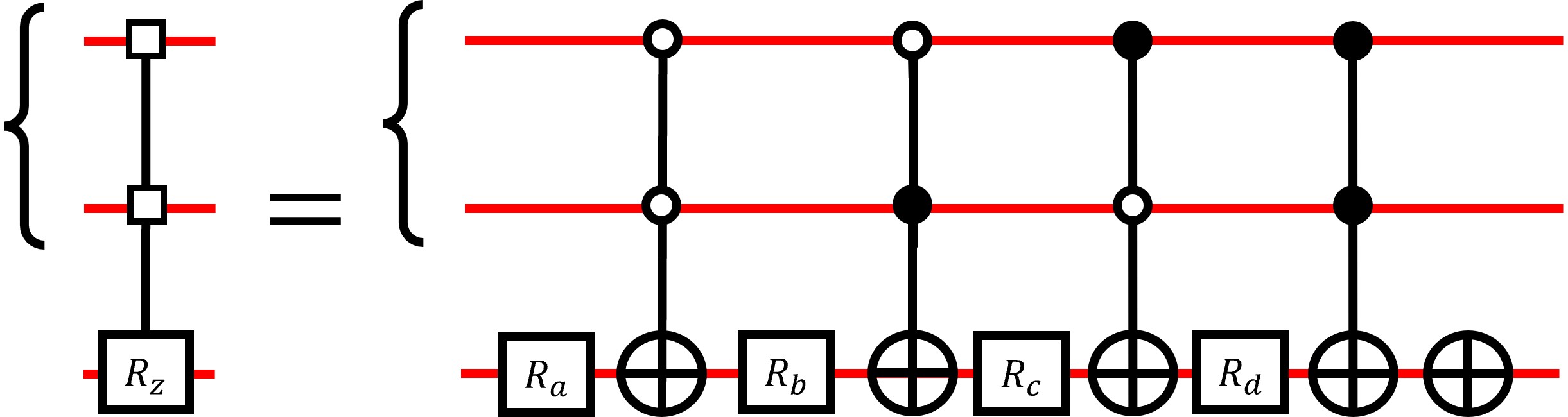}
    
    \vspace{0.4cm} 
    
    \includegraphics[width=0.7\columnwidth]{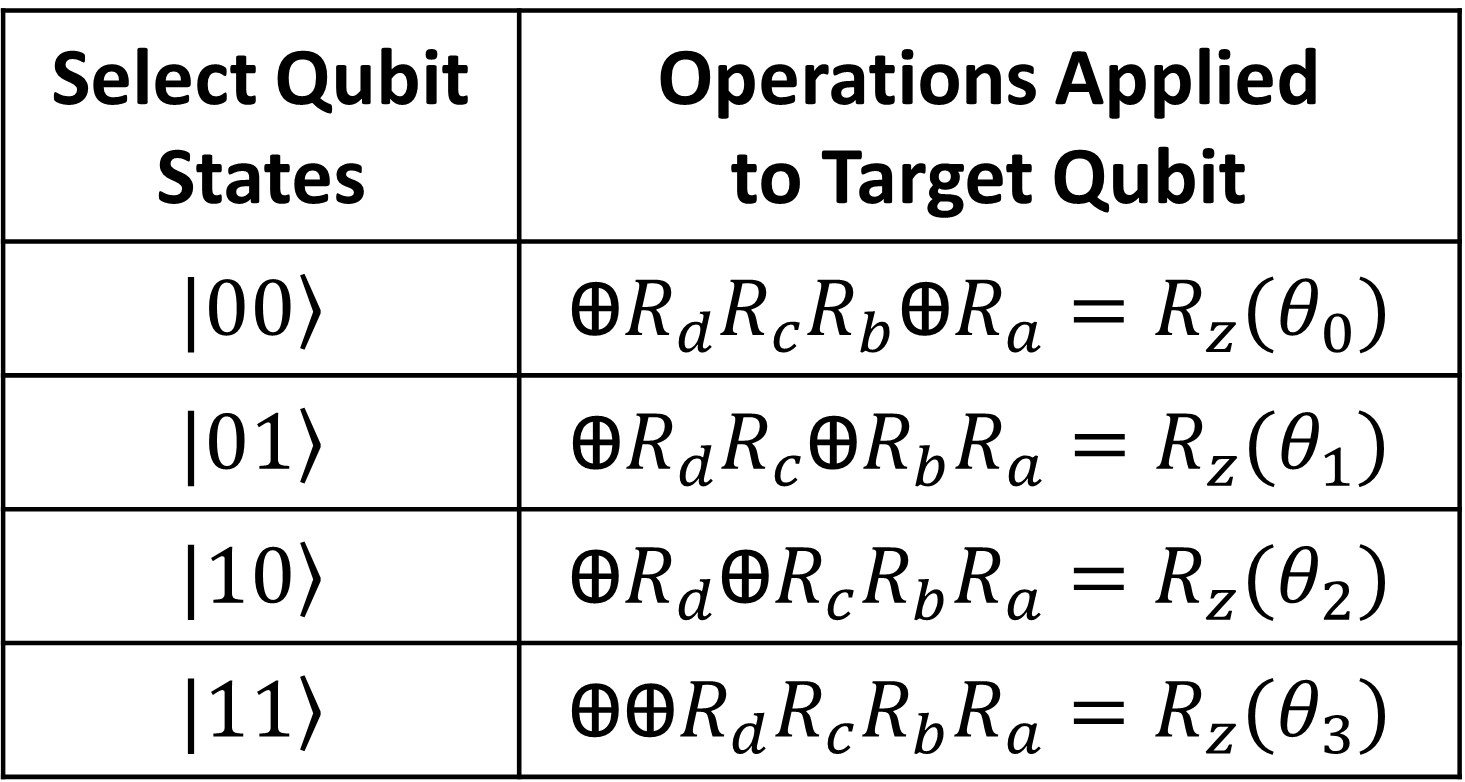}
    
    \vspace{0.3cm} 
    
    \includegraphics[width=0.7\columnwidth]{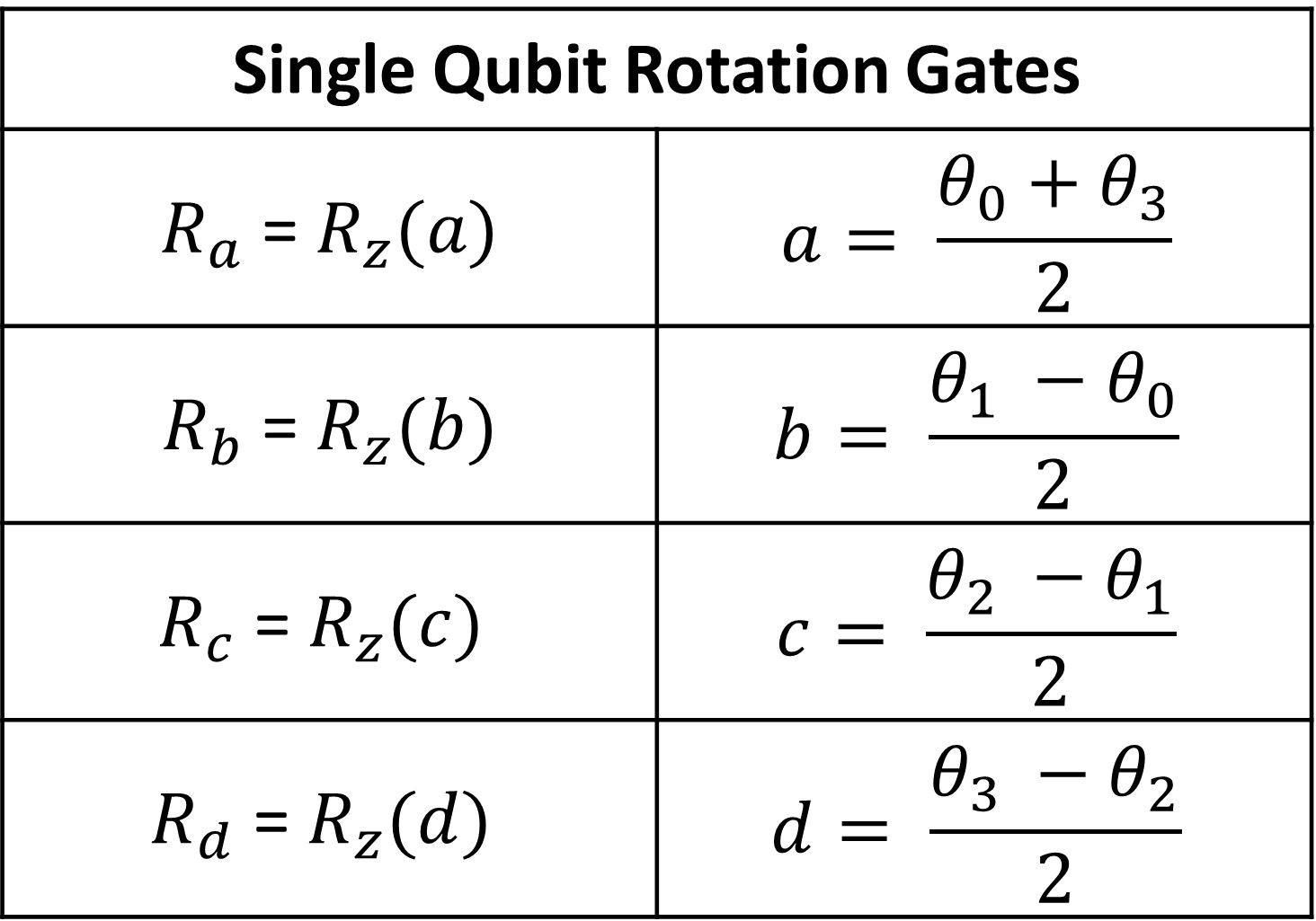}
    \caption{A decomposition of the three-qubit Z rotation multiplexor into single-qubit rotations and Toffoli gates. The upper table shows the operations applied to the target qubit depending on the state of the select qubits. $\oplus$ represents a NOT operation applied by a Toffoli gate. The lower table shows the relationships between the desired rotations \{$\theta_0,\theta_1,\theta_2,\theta_3$\} and the rotations applied by the single-qubit $R_z$ gates \{$a,b,c,d$\}. Y rotation multiplexors can be applied by replacing the single-qubit Z rotations with Y rotations.}
    \label{fig:QLOQ_multiplexor_decomp}
\end{figure}

The multiplexor decomposition from the original QSD proposal (Shende et al.\ 2005 \cite{shende2005synthesis}) performs worse in this QLOQ map than in qubit encoding, so an alternative is needed. We present a decomposition based on Toffoli gates that can be efficiently implemented in QLOQ. The QLOQ decomposition of a three-qubit Z rotation multiplexor is shown in Figure \ref{fig:QLOQ_multiplexor_decomp}. 
The upper table in this figure shows which single qubit operations are applied to the target qubit depending on the states of the select qubits, and the corresponding rotation $\theta$ that is applied by the multiplexor. This creates a set of simultaneous equations which can be solved to find the relationships between the single-qubit $R_z$ rotation angles {$a,b,c,d$}, and the multiplexor rotation angles {$\theta_0,\theta_1,\theta_2,\theta_3$} for each combination of select qubit states (shown in the lower table). $R_z$ gates rotate qubits about the $z$-axis of the Bloch sphere.

$2^{s}$ Toffoli gates are required to implement a multiplexor using the construction from Figure \ref{fig:QLOQ_multiplexor_decomp}: one for each of the $2^s$ combined computational basis states for the $s$ select qubits. A three qubit multiplexor for example has two select qubits, which have $2^2$ computational basis states, so 4 Toffoli gates are required.

\begin{figure*}[!ht]
    \centering
    \includegraphics[width=0.75\textwidth]{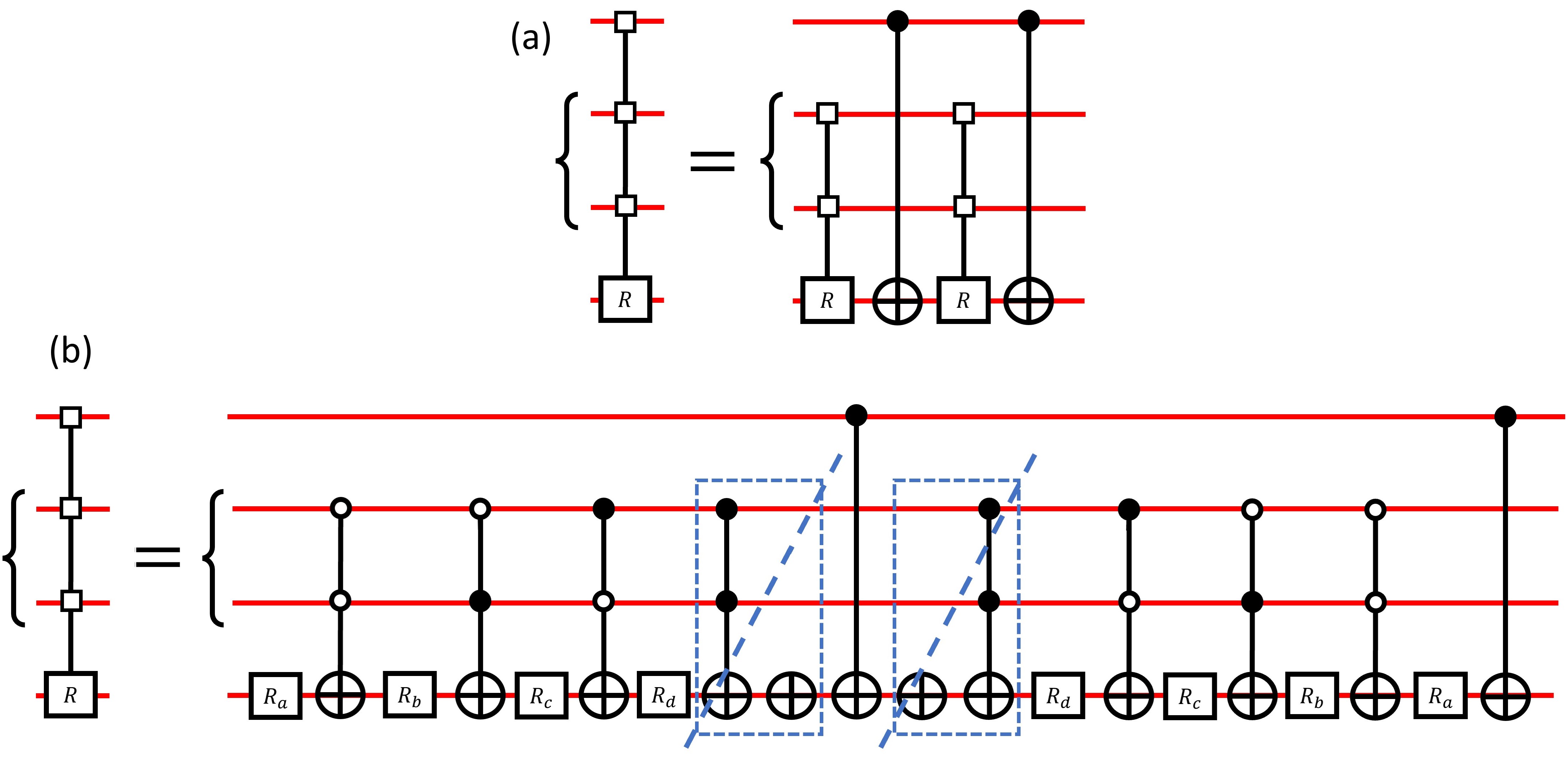}
    \caption{(a) A method for adding select qubits to multiplexors from Shende et al.\ 2005 \cite{shende2005synthesis}. In QLOQ QSD the CNOTs are applied between qubits, and do not interact directly with the qudit. (b) Two physical CNOTs can be cancelled for each select qubit added to the multiplexor circuit, so $2^s$ physical CNOTs are needed in total to implement a multiplexor with $s$ select qubits.}
    \label{fig:adding_select_qubits}
\end{figure*}

Select qubits can be added to a multiplexor using additional CNOTs \cite{shende2005synthesis}, as shown in Figure \ref{fig:adding_select_qubits} (a).
The additional CNOTs are always applied between physical qubits, and do not interact directly with the only qudit in this QLOQ map. This multiplexor decomposition applies the same operation if the order of gates is reversed, since that would be equivalent to transposing the circuit, and the $R_z$-multiplexor is invariant under transposition. 
This allows us to cancel a pair of physical CNOTs for each select bit added to a multiplexor, as shown in Figure \ref{fig:adding_select_qubits} (b). A multiplexor with $s$ select bits can therefore be implemented with $2^s$ physical CNOTs in this QLOQ map, the same amount as required by the qubit-based decomposition from Shende et al.\ 2005 \cite{shende2005synthesis}.

The number of CNOTs $c_n$ required to decompose an arbitrary $n$-qubit unitary with standard QSD is given by Shende et al.\ 2005 \cite[Equation (18)]{shende2005synthesis}:

\begin{equation}
c_n \leq 4^{n-l}(c_l + 3 \times 2^{l-1})-3 \times 2^{n-1} \; ,
\label{eq:shende_QSD}
\end{equation}

\noindent where $c_l$ is the cost of implementing the final $l$-qubit operator in the QSD recursion. For QLOQ QSD, $l=g$ and the final $g$-qubit operators are local single-qudit gates ($c_l = 0$). The CNOT cost $k$ of QLOQ QSD is therefore:

\begin{equation}
k = 4^{n-g} \times 3 \times 2^{g-1}-3 \times 2^{n-1} \; ,
\label{eq:QLOQ_QSD}
\end{equation}

\noindent which we can rearrange to show that $k$ decreases exponentially as $g$ increases: 

\begin{equation}
k = \tfrac{3}{2} 4^{n} (\tfrac{1}{2})^{g} - 3\times 2^{n-1} \; .
\label{eq:QLOQ_QSD_rearranged}
\end{equation}

A QLOQ map where many qubits are mapped to one qudit and the rest are encoded individually will increase the entangling gate cost of some circuit structures, which is a problem when deploying unitary decomposition as a subroutine within a larger algorithm. Fortunately, it is possible to change which qubits are mapped to which physical qudits mid-circuit (remapping) using CNOTs. This lets us start with a circuit where all the qubits are encoded separately, remap several qubits onto one qudit to deploy QLOQ QSD, and then remap again to return the circuit to its original encoding. 

\begin{table*}[!ht]
    \centering
    \begin{tabular}{|l|*{8}{>{\centering\arraybackslash}m{1cm}|}}
    \hline
        $n$ & Qubit Lower Bound & Qubit QSD & Li et al. & QLOQ $g=2$ & QLOQ $g=3$ & QLOQ $g=4$ \\ \hline
        3 & 14 & 20 & 64 & 20 & 24 & 56 \\ \hline
        4 & 61 & 100 & 272 & 80 & 48 & 56 \\ \hline
        5 & 252 & 444 & 1184 & 344 & 168 & 104 \\ \hline
        6 & 1020 & 1868 & 4848 & 1448 & 696 & 344 \\ \hline
        7 & 4091 & 7660 & 20016 & 5960 & 2904 & 1400 \\ \hline
    \end{tabular}
    \caption{The number of physical CNOTs required to decompose an arbitrary $n$-qubit unitary using QLOQ QSD, qubit QSD, and a qudit-based decomposition from Li et al.\ \cite{li2013efficient}. QLOQ QSD requires one qudit with at least $2^{g}$ levels, with the rest of the circuit encoded as qubits. We assume that only two-level physical CNOTs are available.}
    \label{QSD_comparison}
\end{table*}

\begin{figure}[!ht]
    \centering    
    \includegraphics[width=0.48\textwidth]{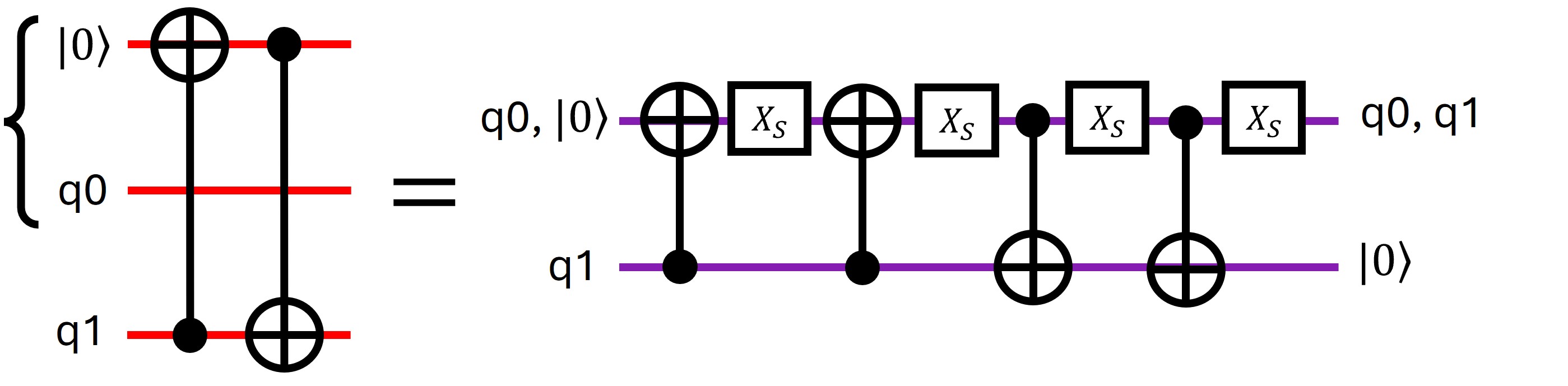}
    \caption{A circuit that will SWAP the state of a physical 2L qubit with a $|0\rangle_{L}$ ancilla encoded on a 4L qudit. The matrix for $X_S$ is provided in Figure \ref{fig:QLOQ Decompositions}. }
    \label{fig:map_change}
\end{figure}

Let's say we want to remap qubit 1 from a physical 2L qubit onto a qudit with a max qubits-per-qudit of $G$, on which $G-1$ qubits are currently encoded. We can treat the available space on that qudit as an ancilla qubit in the state $|0\rangle_{L}$, and use two logical CNOTs to SWAP it for qubit 1, as shown in Figure \ref{fig:map_change}. From Equation \eqref{eq:abk}, this will cost $2^{G}$ physical CNOTs.
If we want to move qubit 1 onto a qudit that is already carrying $G$ qubits, we need to use a full SWAP gate (3 logical CNOTs) costing $3 \times 2^{G-1}$ physical CNOTs.

Equation \eqref{eq:remap_cost} gives the cost to map $f$ qubits onto a single qudit, and then separate them onto physical qubits again (useful for QLOQ QSD). It is found as the sum of the costs to map one qubit on top of another to occupy 4 levels of a qudit, then adding another to occupy 8 levels, and so on until $f$ qubits have been mapped to the same qudit. The cost is then doubled to return all qubits to their original positions, for a CNOT cost $k_{\text{remap}}$ of:

\begin{equation}
k_{\text{remap}} = 2\sum^{f}_{a=2}2^a = 2^{f+2}-8\; .
\label{eq:remap_cost}
\end{equation}

This allows QLOQ QSD to be inserted into a qubit-based circuit without affecting the performance of other gates, provided that at least one qudit is available with sufficiently many unused levels. With remapping, the total CNOT cost $k$ of QLOQ QSD is:

\begin{equation}
k = \tfrac{3}{2} 4^{n} (\tfrac{1}{2})^{g} - 3\times2^{n-1} + 2^{g+2}-8\; .
\label{eq:total_qsd_cost_with_remap}
\end{equation}

\noindent This compares favourably to qubit encoding, where QSD costs:

\begin{equation}
k = \tfrac{23}{48} \times 4^{n} - \tfrac{3}{2}\times2^{n} + \tfrac{4}{3} \; .
\label{eq:optimise_shende}
\end{equation}

Note that the improved performance of Equation \eqref{eq:optimise_shende} compared to Equation \eqref{eq:shende_QSD} relies on additional optimisations provided in the appendix of Shende et al.\ 2005, which we did not implement for QLOQ QSD.  

Table \ref{QSD_comparison}, shows the cost of decomposing $n$-qubit unitaries for $3\leq n\leq7$ using qubit QSD, QLOQ QSD, a previous qudit-based method, and the theoretical lower bound for qubit encoding. 
Where a 4L qudit ($g = 2$) is available, QLOQ QSD with remapping beats qubit QSD for all $n > 3$. Where an 8L qudit ($g = 3$) is available, QLOQ QSD with remapping beats the theoretical lower bound of qubit encoding for all $n > 3$. This is shown analytically in Appendix Section \ref{sec:QSD_cost}, and can be seen for $ n\leq7$ in Table \ref{QSD_comparison}. 
Related works are discussed in Section \ref{QSD_related_works} of the Appendix. The performance of a previous qudit-assisted unitary decomposition technique from Li et al.\ is included in Table \ref{QSD_comparison} for comparison. 

The only drawback of using QLOQ QSD with remapping is the potential hardware constraints on deploying qudits. In architectures where this is not a concern, such as spatial LOQC, QLOQ QSD outperforms the theoretical lower bound on gate-based unitary decomposition for qubit encoding. QLOQ QSD also outperforms all previous qudit-assisted unitary decomposition techniques of which we are aware. 

Further improvement is possible for relatively small unitaries by applying numerical optimisation techniques, such as those found in Refs \cite{goubault2020methods, madden2022best, rakyta2022approaching}, to QLOQ ansatze. This would approach the theoretical lower bound CNOT cost for decomposing unitaries in QLOQ (presented in Table \ref{Lower_bound_table}). The lower bound for a 6-qubit unitary for example is 1020 CNOTs in qubit encoding, but only 36 CNOTs in QLOQ with 8L qudits. Remapping 6 qubits to a pair of 8L qudits and back would require 48 CNOTs (from Equation \eqref{eq:remap_cost}), so any 6-qubit unitary could potentially be inserted into an otherwise qubit-based circuit for only $\geq$ 84 CNOTs.

%% file: sections/conclusion.tex
\section{Conclusion}
\label{Conclusion}
We have shown that QLOQ can significantly increase the size of circuits that are tractable on NISQ devices. The lower bound on the number of CNOTs required to implement an arbitrary unitary operation is much lower in QLOQ than in qubit encoding, and we have provided practical schemes to achieve this advantage for unitary decomposition and variational quantum algorithms. 

The QLOQ Quantum Shannon Decomposition can implement unitaries within a qubit-based circuit with fewer CNOTs than the lower bound for qubit encoding, provided a qudit with sufficient extra levels is available. As far as we are aware, QLOQ QSD outperforms the current state of the art for qudit-assisted qubit-logic unitary decomposition.
QLOQ ansatze for VQAs meanwhile require fewer physical entangling gates to achieve a given level of expressibility than qubit-based ansatze.

The case for QLOQ is particularly strong in spatial LOQC, where qudits actually require fewer resources (in particular photons) than qubits to encode a given amount of information. 
QLOQ can be used in LOQC to optimise a circuit's photon or mode requirement, ansatz expressibility, optical component depth, success probability, or a combination of these. The encoding scheme offers a great deal of flexibility, with different QLOQ maps and circuit structures providing trade-offs that suit particular algorithms and hardware constraints. 

Even relatively small QLOQ ansatze for VQAs can have many times more free parameters and run orders of magnitude faster in LOQC than their qubit-based equivalents. We performed a QLOQ VQE calculation in hours on Quandela's photonic QPU that would have taken years in qubit encoding. We conclude that QLOQ is generally superior to qubit-encoding for layer-based VQAs in spatially encoded probabilistic LOQC. 

Future research will be required to assess the usefulness of QLOQ for fault-tolerant quantum computing schemes (FTQC). For example, QLOQ error correcting codes may be able to encode logical qubits onto physical qudits more efficiently than existing qubit-based schemes. Although it can be deployed on many architectures, QLOQ could be particularly useful in photonic FTQC, e.g.\ for preparing entangled resource states.

%% file: sections/appendices.tex
\onecolumn\newpage
\appendix

\section{Gate-by-Gate Compression}
\label{gate_by_gate}
Gate-by-gate qubit-to-qudit compression is presented in several previous works \cite{kiktenko2023realization, litteken2023qompress, mato2023compression, nikolaeva2024efficient}. In such schemes, each gate in a qubit-based circuit is translated into qudit form individually. Several algorithms have been proposed for finding the optimal qubit-to-qudit map for gate-by-gate compression \cite{litteken2023qompress, mato2023compression, nikolaeva2024efficient}. 
However, if we assume that only two-level (qubit) physical entangling gates are available between qudits (e.g.\ CNOT/CZ), as in Ref \cite{nikolaeva2024efficient} and in line with modern hardware restrictions, we find that the fraction of circuits for which \textit{any} advantageous map exists approaches 0 as the circuit size increases. 

Figure \ref{fig:FractionComp} shows the fraction of randomly generated circuits composed of equal-cost two-qubit gates that could be compressed with 4L-qudits. The figure shows that the utility of gate-by-gate compression diminishes rapidly as circuits get larger. 
We now show that this is also true for circuits containing multi-qubit gates, and for those where bigger qudits are available. 

For a circuit to be compressible it must contain at least one group of qubits such that encoding them on the same qudit would reduce the circuit's physical gate count. An optimal map may be composed of several such qubit-to-qudit groups. Each of these groups must provide an advantage individually for the map to be optimal. 
This is because, according to Equation \eqref{eq:abk}, the cost of a qudit's external gates is minimised when the rest of the circuit is encoded as qubits. If a map contains one qudit, and you add another qudit on which $b$ qubits are encoded, the cost of every external gate between the two qudits will increase by a factor of $2^{b-1}$, while the reduction in cost from converting entangling gates to internal local operations is fixed for each qudit. 

At first glance, multi-qubit gates appear to create circumstances where a map containing two qudits could provide an advantage where neither qudit would individually. For example, a CCCZ can be decomposed into 14 CZs in qubit encoding \cite{nakanishi2021quantum}. Mapping any pair of qubits to the same qudit to create a 4L-2L-2L map would not provide an advantage for the 14 CZ circuit (and for some possible qubit pairs it would actually increase the overall cost). In a 4L-4L map meanwhile, the CCCZ can be implemented with only 1 CZ. 

\begin{figure}[!ht]
    \centering
    \includegraphics[width=0.7\textwidth]{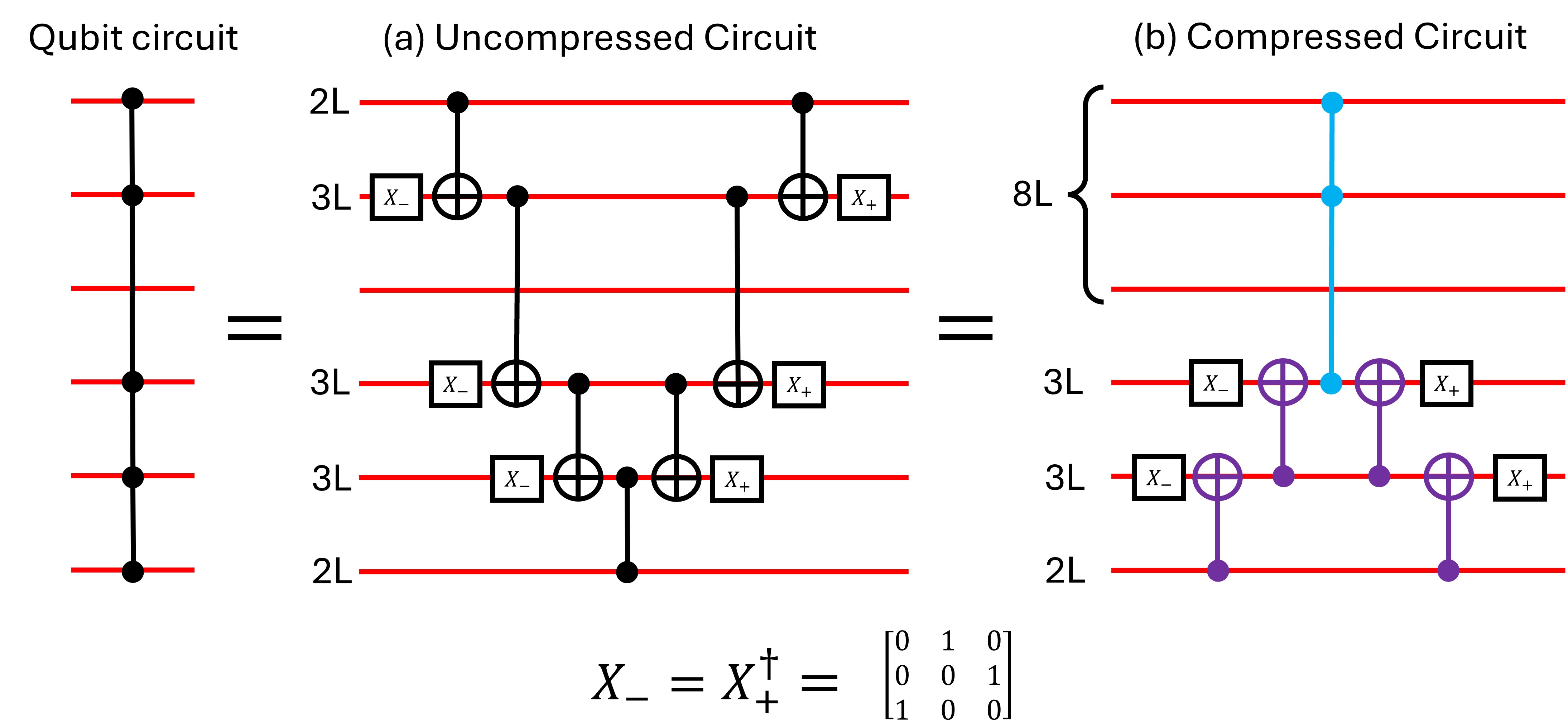}
    \caption{Decompositions of a CCCCZ gate using (a) auxiliary qudit levels (b) auxiliary qudit levels and qubit-to-qudit compression. The decomposition in (a) costs 7 physical CNOTs and in (b) costs 6 physical CNOTs. The cost of the decomposition in (b) can be calculated using Equation \eqref{external_cost}, with the matching colours showing how each gate contributes to the overall cost.}
    \label{fig:compressed_CCCCZ}
\end{figure}

However, since qubit-to-qudit compression requires qudits with at least 4 levels, its performance when compressing multi-qubit gates should be compared with auxiliary qudit level techniques \cite{lanyon2009simplifying, kiktenko2020scalable, nikolaeva2022decomposing, wang2011improved, baker2020improved}, not pure qubit encoding. These decompositions have qubit-logic inputs and outputs, so they do not incur a performance trade-off with other gates for a mapping algorithm to optimise. The technique presented in Figure \ref{fig:compressed_CCCCZ} (a) uses 3L-qudits and $2n-3$ physical CNOTs to create an $n$-qubit CC...CZ. An equivalent technique (ignoring single qudit rotations) is presented in Ref \cite{kiktenko2023realization}. 

Where multiple qubits have been mapped to the same qudit, they can be added to the construction (\textcolor[RGB]{160,48,200}{purple}) using a QLOQ CC...CZ (\textcolor[RGB]{0,176,240}{blue}), as shown in Figure \ref{fig:compressed_CCCCZ} (b). 
If an $n$-qubit CC...CZ gate is applied between some of the $g$ qubits mapped to a single qudit, and $x$ independent (external) qubits (which have an auxiliary qudit level available as needed) the physical CNOT cost $k$ is given by:

\begin{equation}
k = \textcolor[RGB]{160,48,200}{2x-2}+\textcolor[RGB]{0,176,240}{2^{g-n+x}} \; ,
\label{external_cost}
\end{equation}

\noindent The colours in this equation show how the gates in Figure \ref{fig:compressed_CCCCZ} (b) contribute to the decomposition's overall cost. Note that this formula is not valid for gates that are entirely internal to a qudit ($x=0$), which have a cost of $k=0$. We structure our results below (Equation \eqref{R_eq}) such that $x > 0$ and this detail can be omitted for simplicity. 

By Equation \eqref{external_cost}, a CCCZ costs 5 CNOTs without compression, and 3 CNOTs if any two qubits are paired onto a single qudit. Both of the qudits in an optimal map (0,1)(2,3) would provide an advantage individually with the rest of the circuit encoded as qubits (with auxiliary levels available). 
Equation \eqref{external_cost} gives the lowest physical CNOT cost of which we are aware for a CC...CZ gate in such a map. The condition where each qudit in an optimal map must provide an advantage individually therefore also holds for CC...CZ and equivalent multi-qubit gates.

We define the expected cost ratio $R$ of an $n$-qubit gate as its average cost in a given map, assuming all possible gate placements are equally likely (no restrictions on circuit structure), divided by its original cost without qubit-to-qudit compression. We examine an $N$-qubit circuit where $g$ qubits are mapped to one qudit and the rest are encoded independently (each with an auxiliary level available). We will exclude gates that do not interact with the size $g$ qudit from our cost ratio calculation, since they are not being compressed.
The cost of each possible gate arrangement depends on its number of external qubits, as specified by Equation \eqref{external_cost}.  
For an $n$-qubit CC...CZ gate $R$ can therefore be found as: 

\begin{equation}
R = \sum \frac{\text{\#arrangements with \textit{x} external qubits}}{\text{\#arrangements involving the qudit}} \times \frac{\text{Cost of each arrangement}}{\text{Cost of the uncompressed gate}}\; .
\end{equation}

The number of possible arrangements with $x$ external qubits is found as the product of the number of arrangements of internal qubits 
$\binom{g}{n-x}$ and the number of arrangements of external qubits $\binom{N-g}{x}$. Internal and external qubits refer to the qubits to which the gate is applied, on or outside of the single qudit in this map respectively. The cost of the gate for each arrangement is given by Equation \eqref{external_cost}, while the cost of the uncompressed gate is given by $2n-3$ \cite{kiktenko2023realization}. The number of arrangements involving the qudit is given by the total number of arrangements $\binom{N}{n}$ minus the number of arrangements that don't involve the qudit $\binom{N-g}{n}$. This gives us the final equation for $R$:

\begin{equation}
R = \sum^{n-1}_{x = s} \frac{\binom{g}{n-x}\binom{N-g}{x}[2x-2+2^{g-n+x}]}{[\binom{N}{n}-\binom{N-g}{n}][2n-3]} \; .
\label{R_eq}
\end{equation}

\noindent where $s = n - g$ if $n > g$ and $s = 1$ otherwise. 

When the expected cost ratio $R$ of a gate in a particular map is greater than 1, adding that gate to a circuit is likely to make the compressed version more expensive relative to the uncompressed version (assuming no restrictions on placement). Since an optimal map is composed of single qudit maps, each of which must provide an advantage individually, by showing that $R>1$ for the single qudit case, we have shown that $R>1$ for \textit{every possible} map composed of qudits of size $g$ and qubits with auxiliary levels.
This is why we see a steep decrease in the fraction of circuits that are compressible in Figure \ref{fig:FractionComp} as more CNOT gates are added to the $N$-qubit circuits: 
when $g = n = 2$, Equation \eqref{R_eq} simplifies to $R = \frac{4N-8}{2N-3}$, and $R > 1 \forall N \geq 3$.

Table \ref{N_for_R} shows the lowest values of $N$ for which $R>1$. From this we see that gate-by-gate compression is ineffective even for relatively small circuits. 
Table \ref{R_at_Ninf}  shows the values of $\lim_{N \to \infty} R$ at particular $g$ and $n$, estimated by solving Equation \eqref{R_eq} for $N=1e12$. We see that gate-by-gate compression does not become effective for larger circuits or where larger qudits are available: in fact, as qubits per qudit $g$ increases, there is a steep increase in the expected cost ratio $R$ of all gates. 
The results in both Tables were verified by randomly generating circuits composed of $n$-qubit gates and calculating the cost of each gate under Equation \eqref{external_cost}. 

\begin{table}[!ht]
    \centering
    \begin{tabular}{|l|l|l|l|l|l|l|l|l|}
    \hline
        ~ & \textbf{n = 2} & \textbf{n = 3} & \textbf{n = 4} & \textbf{n = 5} & \textbf{n = 6} & \textbf{n = 7} & \textbf{n = 8} & \textbf{n = 9} \\ \hline
        \textbf{g = 2} & 3 & 6 & 8 & 10 & 12 & 14 & 16 & 18 \\ \hline
        \textbf{g = 3} & 4 & 6 & 7 & 9 & 11 & 13 & 15 & 17 \\ \hline
        \textbf{g = 4} & 5 & 6 & 7 & 9 & 11 & 13 & 14 & 16 \\ \hline
        \textbf{g = 5} & 6 & 6 & 7 & 9 & 11 & 12 & 14 & 16 \\ \hline
        \textbf{g = 6} & 7 & 7 & 8 & 9 & 11 & 12 & 14 & 15 \\ \hline
        \textbf{g = 7} & 8 & 8 & 8 & 9 & 10 & 12 & 13 & 15 \\ \hline
    \end{tabular}
    \caption{\label{N_for_R} The lowest values of $N$ for which $R>1$. $R$ is the expected cost of a randomly applied $n$-qubit gate in a map with one qudit on which $g$ qubits are encoded, divided by the uncompressed cost of that gate. Gate-by-gate compression is likely to be effective for circuits with arbitrarily many $n$-qubit gates, provided the circuits have fewer qubits than shown in this table.
    }
\end{table}

\begin{table}[!ht]
    \centering
    \begin{tabular}{|l|l|l|l|l|l|l|l|l|}
    \hline
        ~ & \textbf{n = 2} & \textbf{n = 3} & \textbf{n = 4} & \textbf{n = 5} & \textbf{n = 6} & \textbf{n = 7} & \textbf{n = 8} & \textbf{n = 9} \\ \hline
        \textbf{g = 1} & 1 & 1 & 1 & 1 & 1 & 1 & 1 & 1 \\ \hline
        \textbf{g = 2} & 2 & 1.333 & 1.2 & 1.143 & 1.111 & 1.091 & 1.077 & 1.067 \\ \hline
        \textbf{g = 3} & 4 & 2 & 1.6 & 1.429 & 1.333 & 1.273 & 1.231 & 1.2 \\ \hline
        \textbf{g = 4} & 8 & 3.333 & 2.4 & 2 & 1.778 & 1.636 & 1.538 & 1.467 \\ \hline
        \textbf{g = 5} & 16 & 6 & 4 & 3.143 & 2.667 & 2.364 & 2.154 & 2 \\ \hline
        \textbf{g = 6} & 32 & 11.333 & 7.2 & 5.429 & 4.444 & 3.818 & 3.385 & 3.067 \\ \hline
        \textbf{g = 7} & 64 & 22 & 13.6 & 10 & 8 & 6.727 & 5.846 & 5.2 \\ \hline
    \end{tabular}
    \caption{\label{R_at_Ninf} The values of $\lim_{N \to \infty} R$, estimated by solving Equation \eqref{R_eq} for $N=1e12$. $R$ is the expected cost of a randomly applied $n$-qubit gate in a map with one qudit on which $g$ qubits are encoded, divided by the uncompressed cost of that gate. 
    }
\end{table}

Based on these results, we conclude that gate-by-gate compression is not effective, except for extremely small or shallow circuits. 
It is of course possible to intentionally construct circuits that perform much better in qudit encoding than qubit encoding: that is the primary result of our paper. Several such QLOQ circuit structures are provided for specific applications in the main text. However, we have shown here that the gate-by-gate mapping algorithms presented in previous works are not useful, because for the overwhelming majority of qubit-based circuits the optimal map will be qubit-encoding.

\section{Single Qubit Operations in QLOQ}
\label{sec:single_qubit_ops}
\begin{figure*}[!ht]
    \centering
    \includegraphics[width=0.4\textwidth]{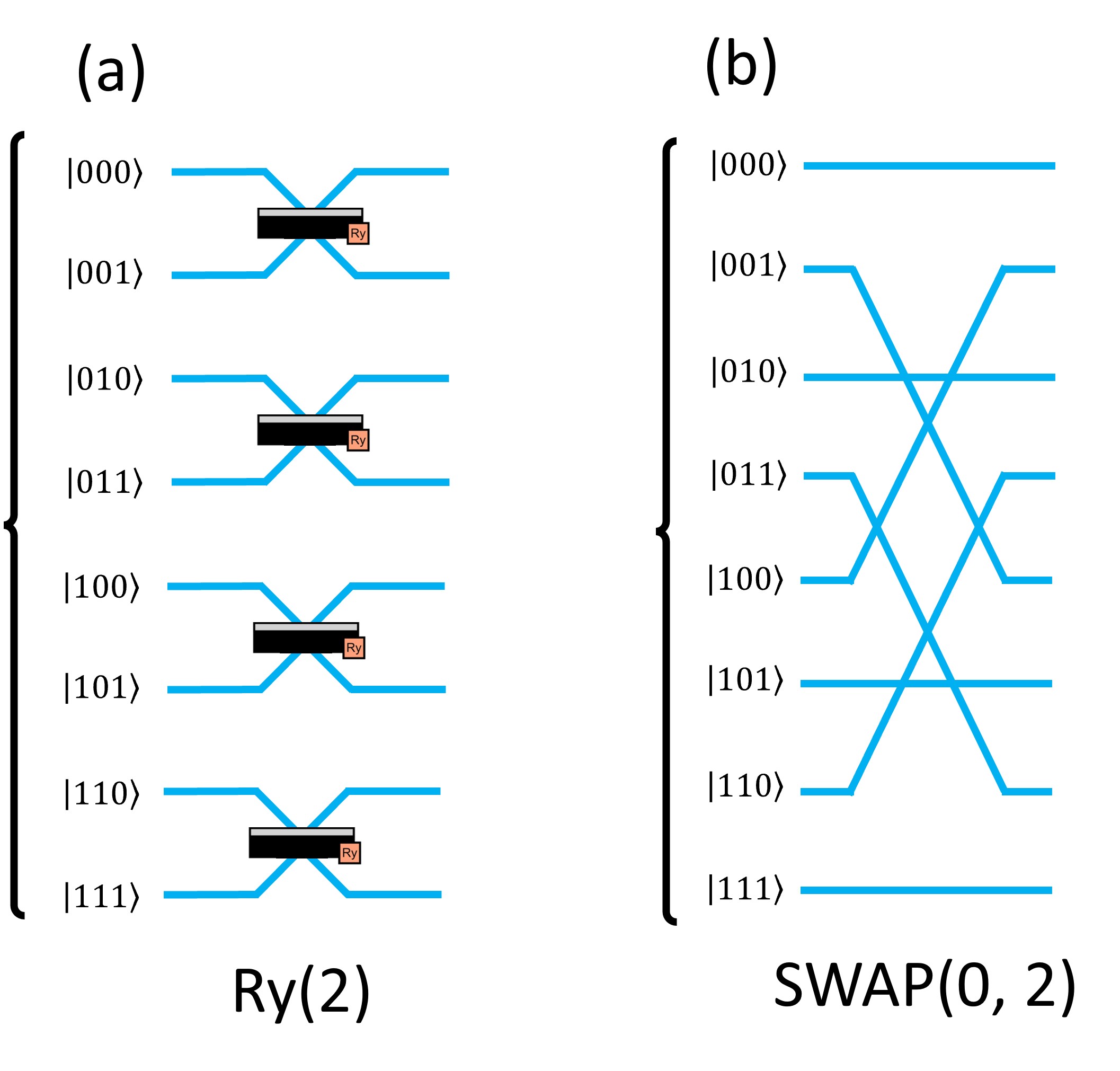}
    \caption{Optical component diagrams for an 8-level qudit to which 3 qubits have been mapped. (a) Applying an $R_y$ gate to qubit 2 (b) Applying a SWAP gate between qubits 0 and 2.}
    \label{fig:Single_qubit_operations}
\end{figure*}

To apply a single-qubit operation on the least significant qubit mapped to a qudit, one can simply apply that operation to each consecutive pair of qudit levels. For example, for an 8-level qudit to which three qubits have been mapped (q0, q1, q2), a logical $R_y$ rotation can be applied to q2 by applying physical $R_y$ rotations to the qudit level pairs ($|0\rangle_{P}$,$|1\rangle_{P}$), ($|2\rangle_{P}$,$|3\rangle_{P}$), ($|4\rangle_{P}$,$|5\rangle_{P}$), and ($|6\rangle_{P}$,$|7\rangle_{P}$). This is shown in Figure \ref{fig:Single_qubit_operations} (a) for spatial LOQC. 

The significance/order of qubits mapped to a qudit can be rearranged with internal swap operations. On the same 3 level qudit, q0 and q2 can be swapped by exchanging the information of particular qudit levels as shown in Figure \ref{fig:Single_qubit_operations} (b). 
To apply a single-qubit operation to a qubit that is not the least significant qubit mapped to a given qudit, simply swap it with the least significant qubit, apply the operation, and swap it back. 

\newpage
\section{Ansatz Benchmark Results}

\begin{figure*}[!ht]
    \centering
    \includegraphics[width=0.8\textwidth]{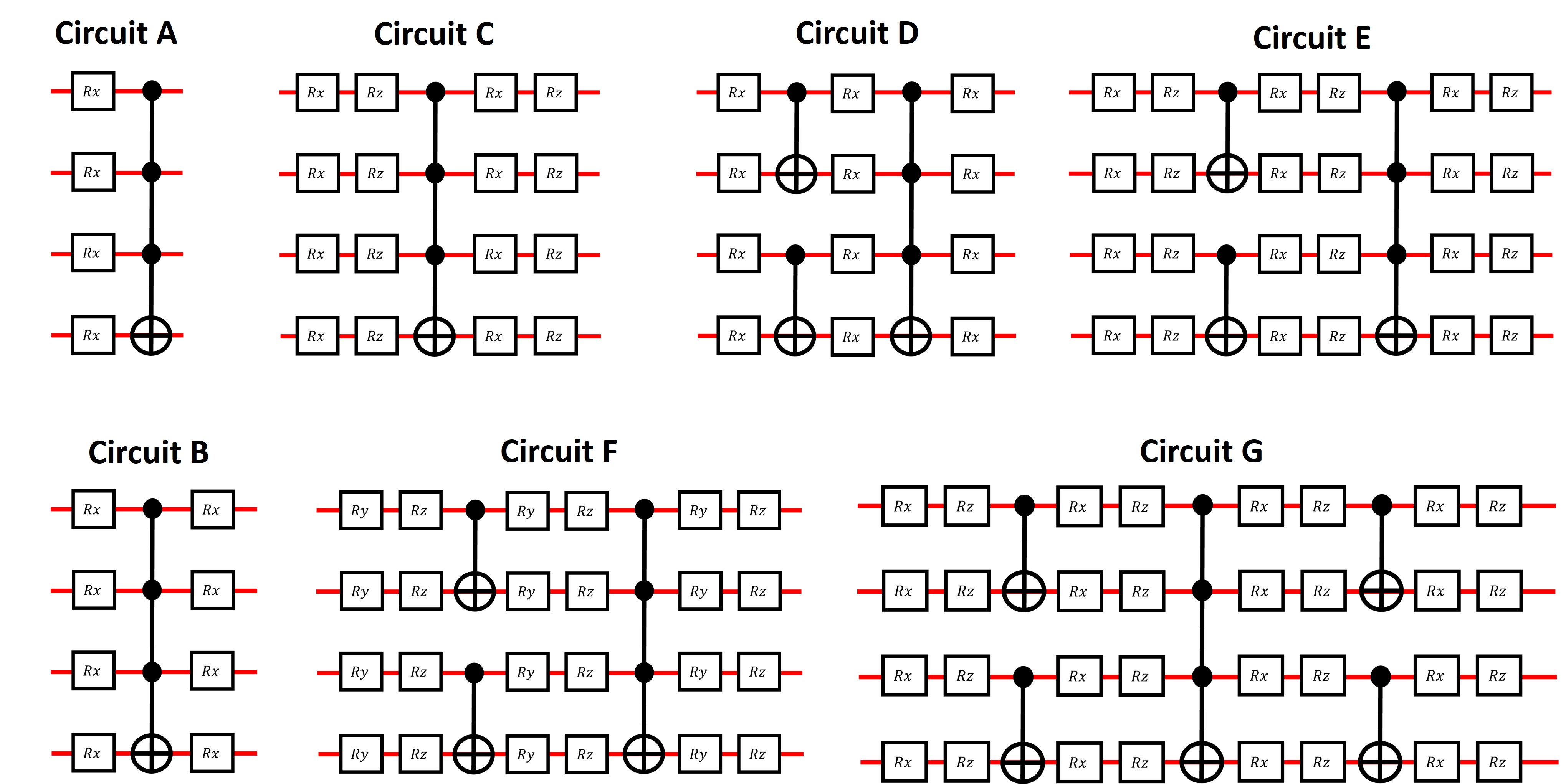}
    \vspace{0.5cm} 
    
    \includegraphics[width=0.8\textwidth]{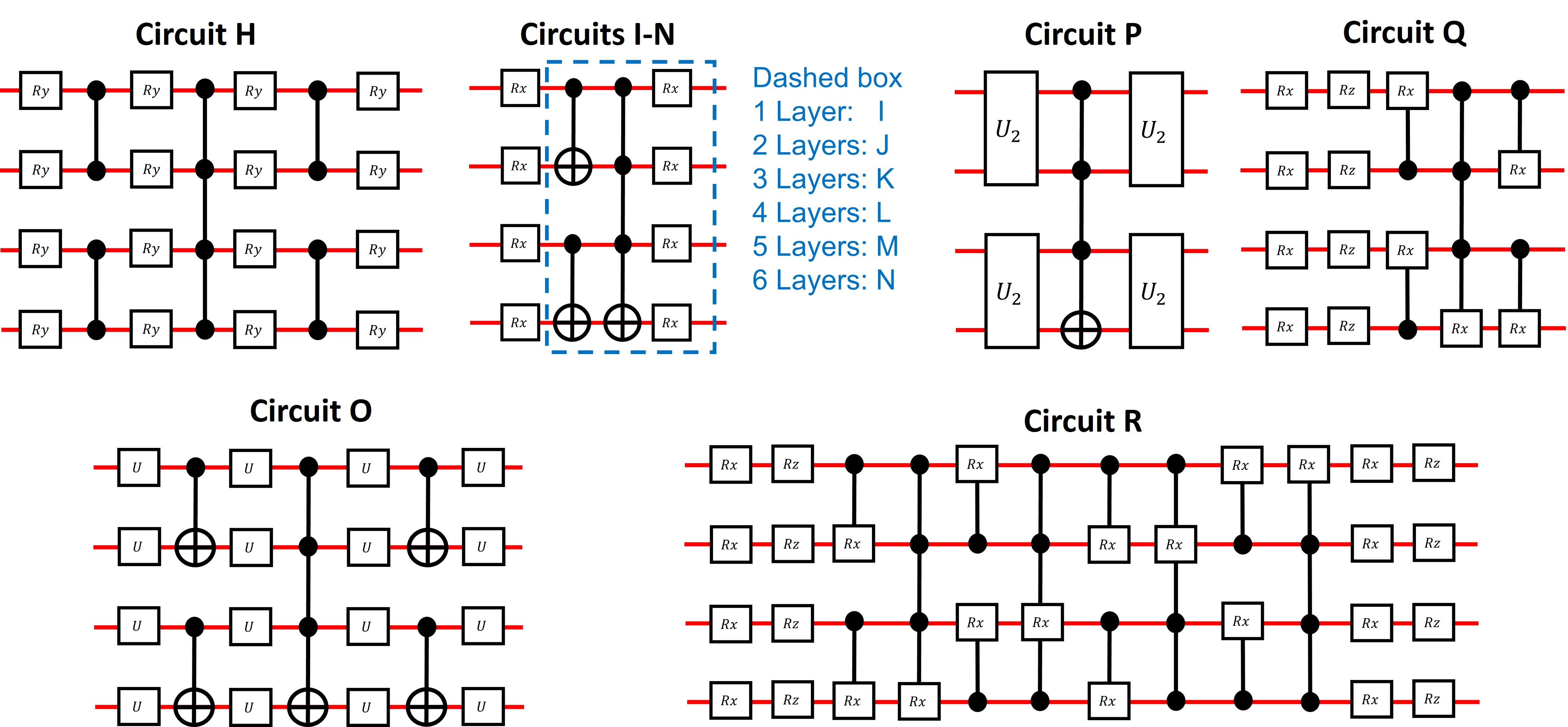}
    
    \vspace{0.5cm} 
    \includegraphics[width=0.7\textwidth]{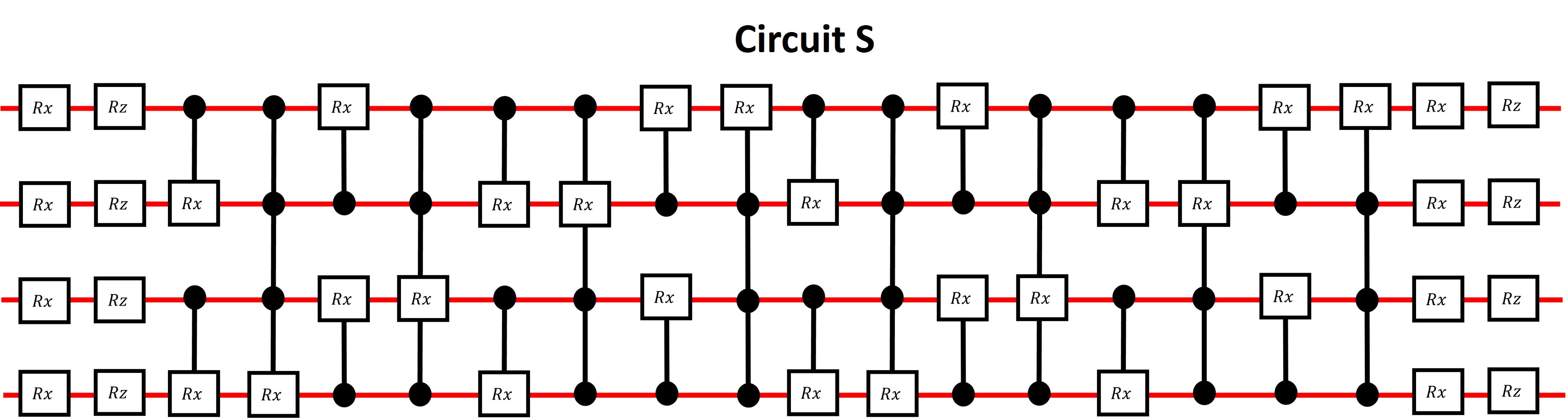}
    \caption{The 19 ansatze in our QLOQ benchmark set. The blue box on Circuits I, J, K, L, M, N contains a repeated layer, with I having one layer, J having two, and so on.}
    \label{fig:all_qloq_ansatze}
\end{figure*}

Figure \ref{fig:all_qloq_ansatze} shows the ansatze used in the QLOQ benchmark set. 
Circuits I, J, K, L, M, and N are all instances of the same circuit structure with repeated layers. As was found for most of the circuits in the benchmark set in Sim et al.\ 2019 \cite{sim2019expressibility}, the expressibility of this circuit saturates after a certain number of layers (4 in this case), beyond which adding more layers does not improve performance.

The specific expressibility and entangling capability values found for the QLOQ and Sim et al.\ benchmark sets of 4-qubit ansatze are presented in Table \ref{expr table}. 
The cost of implementing these ansatze is shown in CNOT equivalents and parameters. The maximum qudit size was assumed to be 4 levels, which is the minimum required for non-trivial QLOQ. 

\begin{table*}[!ht]
    \centering
    \begin{tabular}{|l|l|l|l|l|l|l|l|l|l|l|}
    \hline
        ID & CNOTs & Params & Expr & Ent & ~ & ID & CNOTs & Params & Expr & Ent \\ \hline
        1 & 0 & 8 & 0.2930 & 0.0000 & ~ & A & 1 & 4 & 0.673069874 & 0.207796485 \\ \hline
        2 & 3 & 8 & 0.3176 & 0.6210 & ~ & B & 1 & 8 & 0.28741122 & 0.20855039 \\ \hline
        3 & 6 & 11 & 0.2454 & 0.1727 & ~ & C & 1 & 8 & 0.093357187 & 0.176728067 \\ \hline
        4 & 6 & 11 & 0.1479 & 0.3123 & ~ & D & 1 & 12 & 0.056974447 & 0.580449973 \\ \hline
        5 & 24 & 28 & 0.0581 & 0.2883 & ~ & E & 1 & 24 & 0.023855159 & 0.48322078 \\ \hline
        6 & 24 & 28 & 0.0048 & 0.6857 & ~ & F & 1 & 24 & 0.023445757 & 0.475109623 \\ \hline
        7 & 6 & 19 & 0.1118 & 0.2131 & ~ & G & 1 & 32 & 0.022675845 & 0.529186039 \\ \hline
        8 & 6 & 19 & 0.0737 & 0.2871 & ~ & H & 1 & 16 & 0.265845891 & 0.420539123 \\ \hline
        9 & 3 & 4 & 0.6450 & 1.0000 & ~ & I & 1 & 8 & 0.081421924 & 0.49003053 \\ \hline
        10 & 4 & 8 & 0.2260 & 0.3784 & ~ & J & 2 & 12 & 0.039202524 & 0.622025717 \\ \hline
        11 & 3 & 12 & 0.1332 & 0.5306 & ~ & K & 3 & 16 & 0.020372673 & 0.68568794 \\ \hline
        12 & 3 & 12 & 0.2001 & 0.4033 & ~ & L & 4 & 20 & 0.012759938 & 0.721151532 \\ \hline
        13 & 16 & 16 & 0.0611 & 0.4051 & ~ & M & 5 & 24 & 0.012978008 & 0.742729297 \\ \hline
        14 & 16 & 16 & 0.0178 & 0.5490 & ~ & N & 6 & 28 & 0.014358822 & 0.756693472 \\ \hline
        15 & 8 & 8 & 0.1756 & 0.7143 & ~ & O & 1 & 36 & 0.017601636 & 0.536901877 \\ \hline
        16 & 6 & 11 & 0.2703 & 0.1707 & ~ & P & 1 & 96 & 0.018824992 & 0.541514054 \\ \hline
        17 & 6 & 11 & 0.1589 & 0.2817 & ~ & Q & 2 & 13 & 0.109578906 & 0.387939208 \\ \hline
        18 & 8 & 12 & 0.2505 & 0.2154 & ~ & R & 8 & 28 & 0.007640917 & 0.564235526 \\ \hline
        19 & 8 & 12 & 0.0620 & 0.3914 & ~ & S & 16 & 40 & 0.005586178 & 0.633387543 \\ \hline
    \end{tabular}
    \caption{\label{expr table}The cost metrics (physical CNOTs and parameters) and descriptors (Expr and Ent) for the qubit-based circuits from the Sim et al.\ benchmark set (left, numerical IDs) and a selection of QLOQ ansatze (right, alphabetical IDs). Smaller Expr values indicate a more expressible circuit. Several ansatze have single-qubit gates which could commute through entangling gates and be absorbed by other single qubit gates, so the parameter counts shown are not optimal. 
    }
\end{table*}

\newpage

\section{QFA in QLOQ(1,2)}
\label{sec:8L_max} 

\begin{figure*}[!ht]
    \centering
    \includegraphics[width=0.9\textwidth]{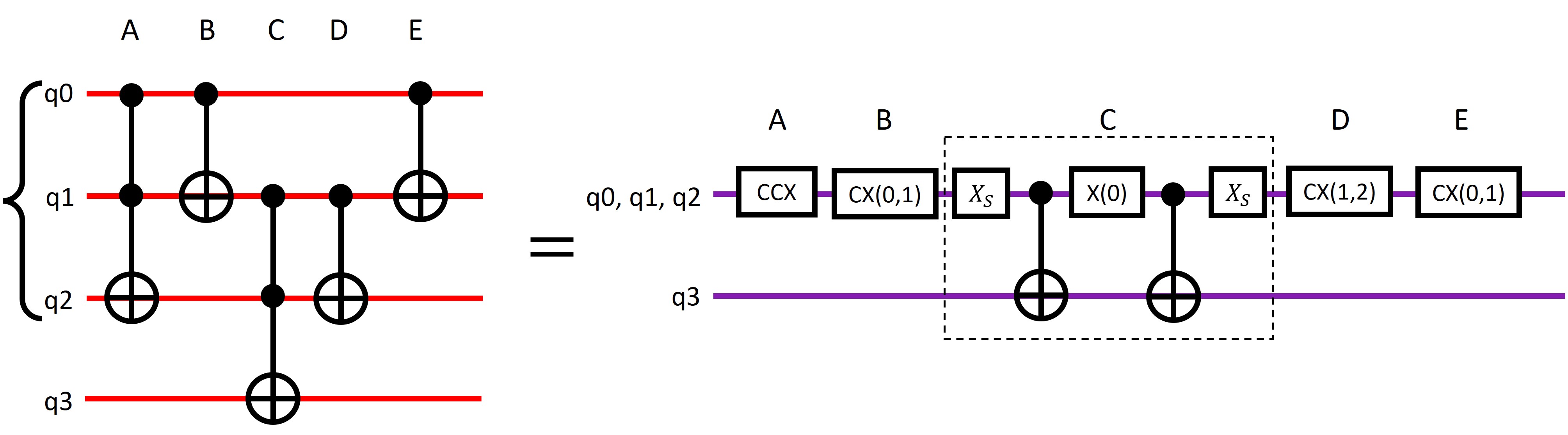}
    \caption{The quantum full adder circuit mapped to QLOQ(0,1,2). X(0) represents a logical NOT operation on qubit 0, while $X_S$ is the 8-level-qudit extension of the $X_s$ gate defined in Figure \ref{fig:QLOQ Decompositions}.}
    \label{fig:8L_max}
\end{figure*}

Figure \ref{fig:8L_max} shows the quantum full adder circuit \cite{feynman2023quantum} in QLOQ(0,1,2)

\newpage

\section{Other Ansatz Structures}
\label{alt_structures} 

\begin{figure*}[!ht]
    \centering
    \includegraphics[width=0.3\textwidth]{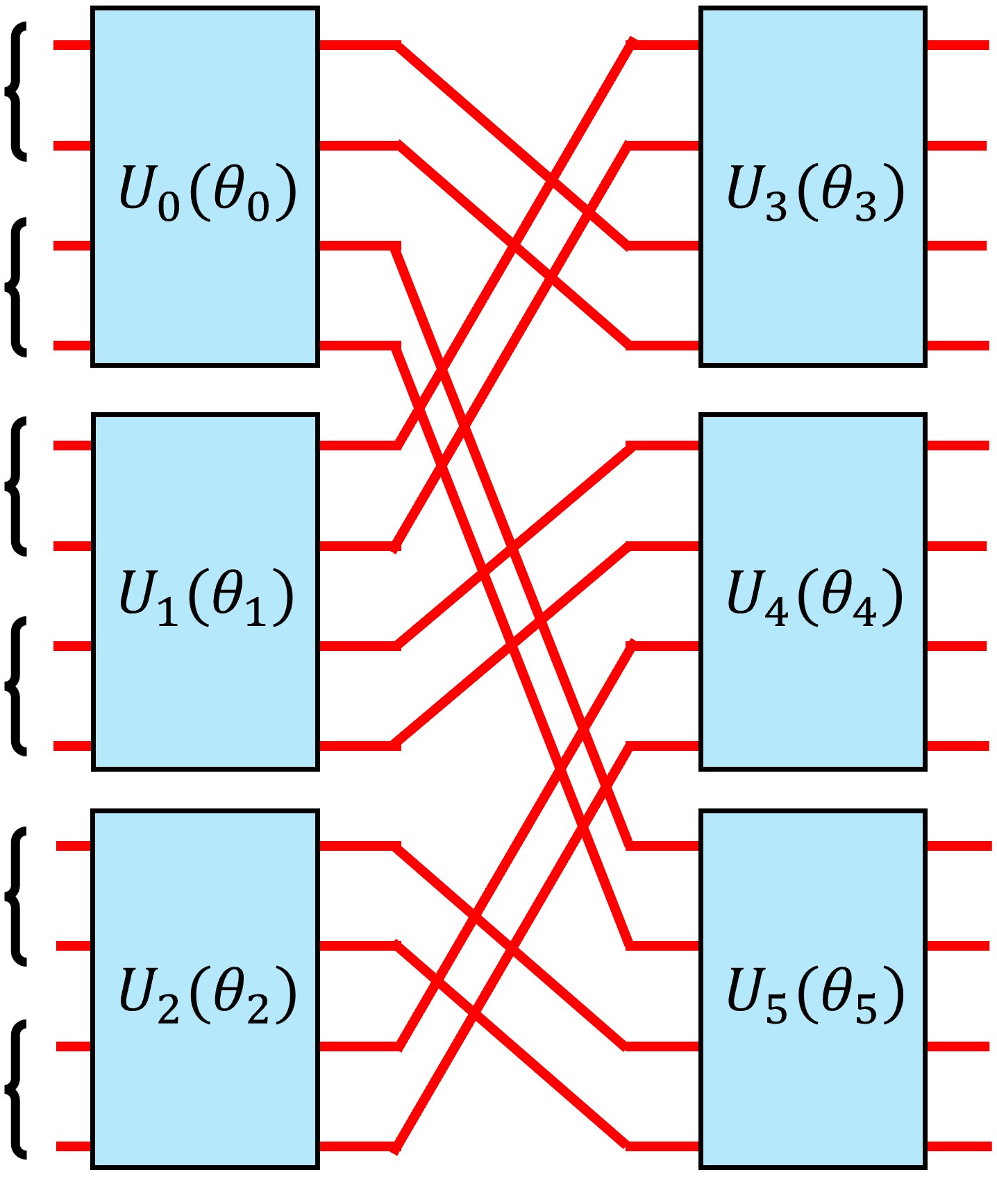}
    \includegraphics[width=0.4\textwidth]{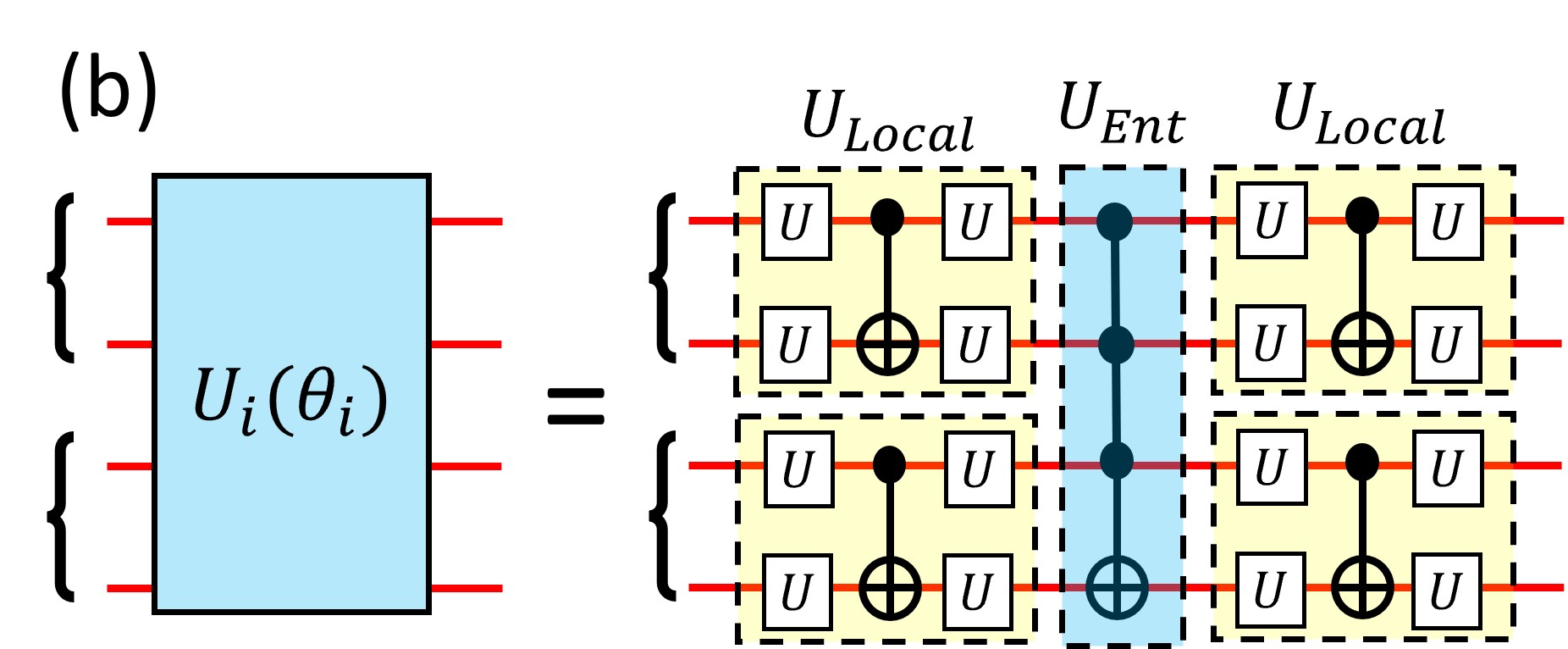}
    \caption{(a) A possible QLOQ quantum neural network (QNN) structure. (b) Circuit O as a QNN block, with a cost of 1 physical CNOT.}
    \label{fig:QNN_diagram}
\end{figure*}

QLOQ is not limited to the layer based ansatz structures analysed in our experiments. For example, entangling gates can be combined with parameterised local layers to create blocks, which can be arranged into a quantum neural net. An example of this is shown in Figure \ref{fig:QNN_diagram}, resembling the scheme presented in Fahri et al.\ \cite{farhi2018classification}. Each block in the circuit shown costs only a single two-level entangling gate, but entangles 4 qubits with significant expressibility. Any 4-qubit QLOQ circuit costing 1 physical CNOT can be used as a block in this map (max qubits-per-qudit $G=2$). The QLOQ ansatz Circuit O is used as the block in Figure \ref{fig:QNN_diagram} (b). Separating qubits mapped to the same qudit is possible but expensive in terms of CNOTs. 

\newpage

\section{VQE}

\subsection{QLOQ Optical VQE Circuit}
\label{sec:optical}
\begin{figure*}[!ht]
    \centering
    \includegraphics[width=0.95\textwidth]{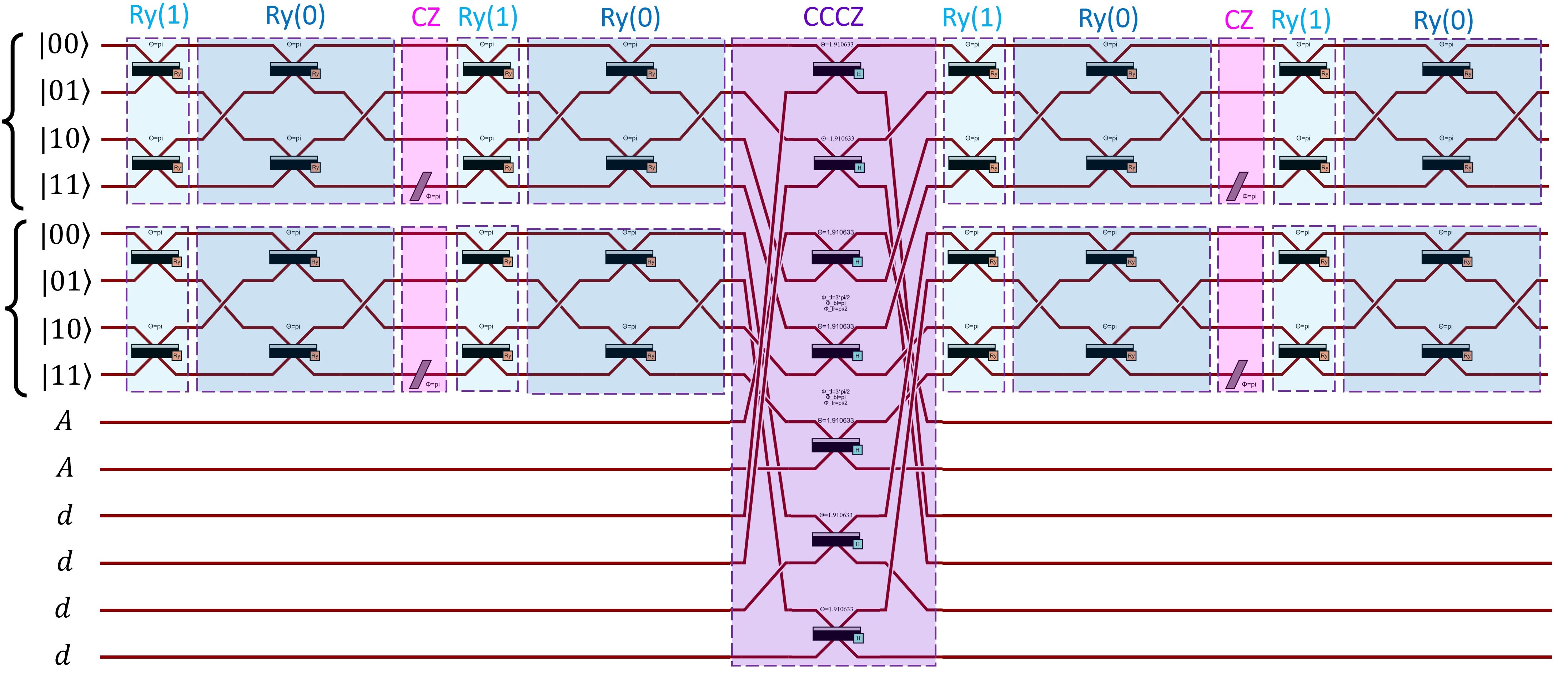}
    \caption{The pre-compilation optical component diagram of the QLOQ ansatz circuit used in the L-BFGS-B VQE simulations. $R_y$(0) and $R_y$(1) are qubit-logical Y-rotation gates applied to qubits 0 and 1 respectively. Modes marked `A' are ancillas for the Ralph et al.\ CZ, while modes marked `d' are dump ports used to balance the gate's success probability.  
    } 
    \label{fig:ascella}
\end{figure*}

Figure \ref{fig:ascella} shows the pre-compilation optical component diagram of the QLOQ ansatz circuit used in the L-BFGS-B VQE simulations.
In order to ensure that the success probability of the CCCZ gate doesn't change depending on its inputs, dump ports (marked `d' in the diagram) and additional beam splitters with reflectivity 1/3 were added. 
The success probability of a physical Ralph et al.\ CZ changes depending on the number of photons it receives as inputs. This isn't a problem in qubit encoding, but in QLOQ the Ralph et al.\ CZ acting as a CCCZ might receive 1 or 0 photons instead of 2. Dump ports (extra spatial modes) and beam splitters with particular reflectivity can reduce the success probability of the gate for different inputs by getting rid of some fraction of the photons, so that the overall success probability of the gate is constant. 
Due to hardware constraints, these dump ports were omitted in the ansatz used for the VQE QPU run shown in Figure \ref{fig:QPU_plot}. This creates a non-standard CCCZ gate with the correct classical truth table, but a success probability that varies depending on its inputs. The same QLOQ ansatz was used in the noise-free COBYLA simulation shown in Figure \ref{fig:time_plots}, with an average success probability of 0.448. 
During tests in which 20 noise-free COBYLA simulations were run with each ansatz, the mean GSEE value for the balanced QLOQ ansatz was -7.83937, compared to -7.825 for the unbalanced version. However the minimum value achieved was lower for the unbalanced version: -7.8504 compared to -7.84791 for the balanced version. 

\newpage
\subsection{VQE Results}

\begin{table*}[!ht]
    \centering
    \begin{tabular}{|l|l|l|l|l|l|l|l|l|}
    \hline
        Ansatz & QPU/ & Optimiser &  Entanglers & Minimum  & Mean  & GSEE  & Avg valid & Num  \\ 
        ~ & SIM & ~ &  ~ &  GSEE &  GSEE &  STD &  samples &  Runs \\ \hline
        Qubit  & Sim & L-BFGS-B & None & -7.82920 & -7.82920 & ~0 & inf & 100 \\ \hline
        Qubit  & Sim & L-BFGS-B & CZ Cascade & -7.85274 & -7.8526 & 0.00135 & inf & 100 \\ \hline
        \textbf{QLOQ} & \textbf{Sim} & \textbf{L-BFGS-B} & \textbf{CCCZ}  & \textbf{-7.86187} & \textbf{-7.86112} & \textbf{0.00366} & \textbf{inf} & \textbf{100} \\ \hline
        Qubit  & Sim & COBYLA & None & -7.82937 & -7.82224 & 0.02325 & 3000000 & 100 \\ \hline
        Qubit  & Sim & COBYLA & CZ Cascade & -7.85341 & -7.81474 & 0.03721 & 37,037 & 100 \\ \hline
        QLOQ & Sim & COBYLA & CCCZ & -7.85925 & -7.83886 & 0.01442 & 1344126.4 & 100 \\ \hline
        QLOQ & QPU & COBYLA & CCCZ & -7.84049 & -7.84049 & - & 39821.54 & 1 \\ \hline
        ~ & ~ & ~ & \textbf{Actual} & \textbf{-7.862} & ~ & ~ & ~ & ~ \\ 
        ~ & ~ & ~ & \textbf{GSEE} & ~ & ~ & ~ & ~ & ~ \\ \hline
    \end{tabular}
    \caption{\label{results table} The LiH VQE results. QLOQ L-BFGS-B in simulation (bold highlighted row) were the only runs to reach chemical accuracy. 
    The ``Mean GSEE" column reports the final ground state energy estimate averaged across all runs of a given type. The standard deviation of this mean is presented in the ``GSEE STD" column. The ``Avg valid samples" column reports the number of measurements taken for each Pauli group (of which there are 25 per iteration) on average. 
    }
\end{table*}

\subsection{VQE Error}
\label{vqe_error}
\begin{figure*}[!ht]
    \centering
    \includegraphics[width=0.75\textwidth]{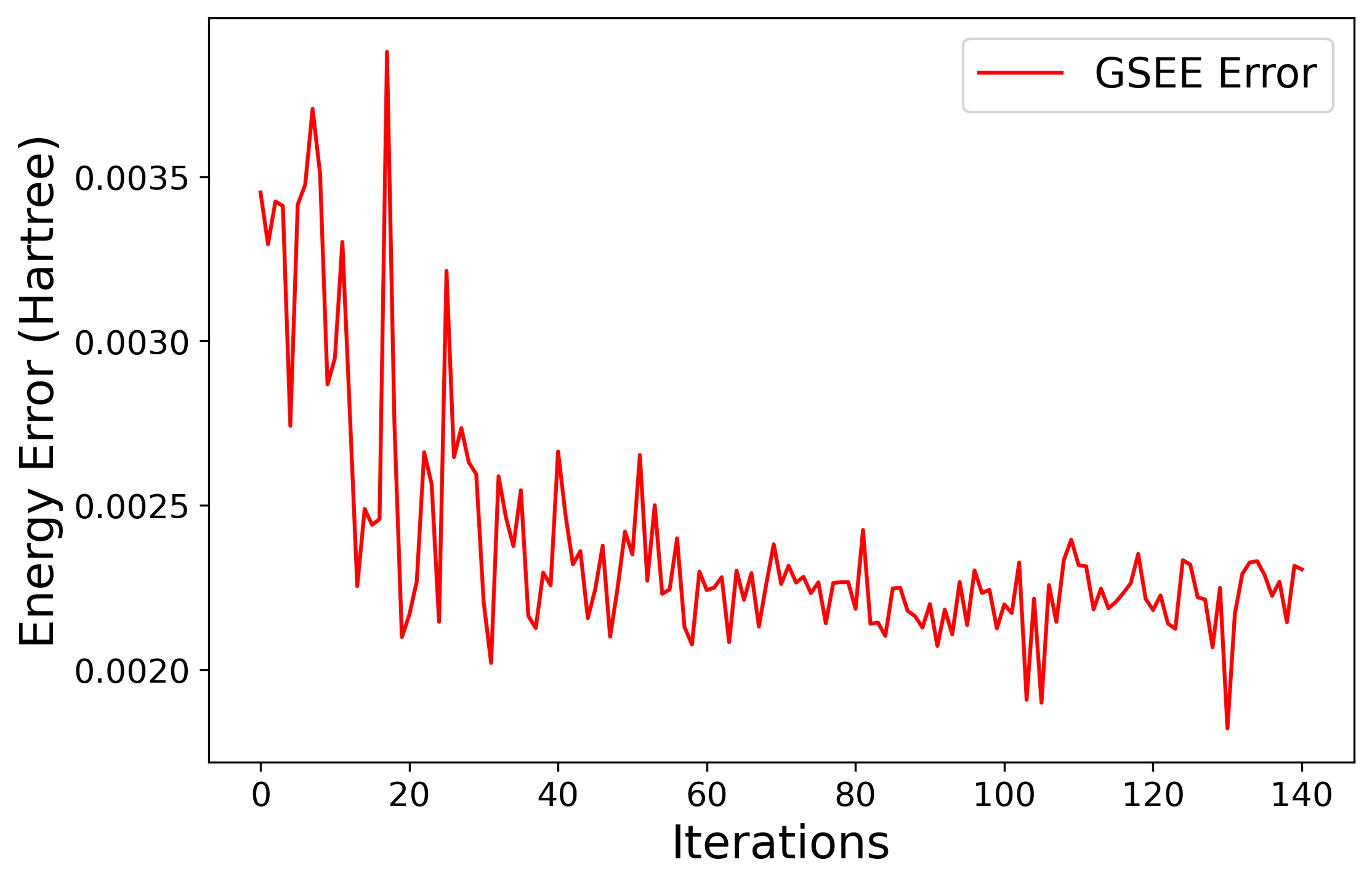}
    \caption{\centering The error estimate for the QLOQ LiH VQE run on Quandela's linear optical QPU. 
    }
    \label{fig:error_plots}
\end{figure*}

The error on the ground state energy estimate for the QPU VQE run was calculated using the method described in the Supplementary Information of Ref \cite{kandala2017hardware}. 
This error is shown for every iteration in Figure \ref{fig:error_plots}. 

\newpage

\newpage

\section{QSD Cost Comparison}
\label{sec:QSD_cost}
The CNOT cost $k_1$ of decomposing an $n$-qubit unitary with qubit-based QSD is:

\begin{equation}
k_1 = \tfrac{23}{48} \times 4^{n} - \tfrac{3}{2}\times2^{n} + \tfrac{4}{3} \; .
\end{equation}

\noindent The CNOT cost $k_2$ for QLOQ QSD using a 4L qudit with remapping is:

\begin{equation}
k_2 = \tfrac{3}{8}\times4^{n} - \tfrac{3}{2}\times2^{n} + 8 \; .
\end{equation}

\noindent The CNOT cost of 4L QLOQ QSD with remapping is less than that of qubit QSD where $k_2 < k_1$:

\begin{equation}
\tfrac{3}{8}\times4^{n} - \tfrac{3}{2}\times2^{n} + 8 < \tfrac{23}{48} \times 4^{n} - \tfrac{3}{2}\times2^{n} + \tfrac{4}{3} \; ,
\end{equation}

\noindent which can be expressed as: 

\begin{equation}
64 < 4^{n}  \; ,
\end{equation}

\noindent which is the case for all $n > 3$. 
\newline
\newline
\newline
\newline
\newline

\noindent The theoretical lower bound on the number of CNOTs required to decompose an $n$-qubit unitary in qubit encoding $k_L$ is: 

\begin{equation}
k_L = \tfrac{1}{4}\times4^{n} - \tfrac{3n}{4} - \tfrac{1}{4} \; .
\end{equation}

\noindent The CNOT cost $k_3$ for QLOQ QSD with an 8L qudit is:

\begin{equation}
k_3 = \tfrac{3}{16} \times 4^{n} - \tfrac{3}{2}\times2^{n} + 24 \; .
\end{equation}

\noindent So $k_3 < k_L$ where: 

\begin{equation}
\tfrac{3}{16}\times 4^{n} - \tfrac{3}{2}\times2^{n} + 24 < \tfrac{1}{4}\times 4^{n} - \tfrac{3n}{4} - \tfrac{1}{4} \; ,
\end{equation}

\noindent which can be expressed as: 

\begin{equation}
97 < 4^{n-1} + 6\times2^{n} - 3n\; ,
\end{equation}

\noindent which is the case for all $n > 3$.

\section{Previous Qudit-Based Unitary Decomposition Techniques}
\label{QSD_related_works}

The unitary decomposition technique presented in Li et al.\ relies on a multi-level entangling gate called the ``controlled-double-NOT" between 4-level qudits. 
Their paper assumes that this gate is available for the same resource cost as a qubit (two-level) CNOT, which is not generally the case (e.g.\ in LOQC and superconducting transmons \cite{litteken2023qompress}). 
The controlled-double-NOT applies a NOT operation to two pairs of levels on the target qudit if the control qudit is in either state $|2\rangle_{P}$ or $|3\rangle_{P}$. This operation can be decomposed into 4 standard two-level CNOTs.
The Li et al.\ decomposition implements its own form of remapping, and can therefore be inserted into qubit-based circuits if sufficient unused qudit levels are available. We include it in Table \ref{QSD_comparison} in terms of two-level CNOTs. QLOQ QSD outperforms the Li et al.\ decomposition for all $n$. 

Assuming that double-CNOTs are available, one could directly translate standard QSD into QLOQ. Even including the cost of remapping, a direct translation of 6-qubit QSD into QLOQ would only cost 114 double-CNOTs. The Li decomposition meanwhile requires 1212 double-CNOTs. Both are overly favourable given that the double-CNOT gate is not typically available on hardware.  

Ref \cite{di2013synthesis} decomposes multi-qudit unitaries using standard two-level CNOTs between qudits, but they do not implement remapping or qubit logic, so comparison is difficult. One could map 4 qubits to two 4-level qudits, and use their version of QSD to decompose any 4-qubit unitary into 108 CNOTs gates (compared to 72 CNOTs in QLOQ 4L-2L-2L). For a unitary between two 8L qudits (to which 6 qubits could be mapped), they required 1176 CNOTs. In QLOQ 8L-2L-2L-2L without remapping, 672 CNOTs are required.

%% file: main.bbl
\begin{thebibliography}{10}

\bibitem{reck1994experimental}
Michael Reck, Anton Zeilinger, Herbert~J Bernstein, and Philip Bertani.
\newblock ``Experimental realization of any discrete unitary operator''.
\newblock \href{https://dx.doi.org/10.1103/PhysRevLett.73.58}{Physical review
  letters {\bf 73}, 58}~(1994).

\bibitem{clements2016optimal}
William~R Clements, Peter~C Humphreys, Benjamin~J Metcalf, W~Steven Kolthammer,
  and Ian~A Walmsley.
\newblock ``Optimal design for universal multiport interferometers''.
\newblock \href{https://dx.doi.org/10.1364/OPTICA.3.001460}{Optica {\bf 3},
  1460--1465}~(2016).

\bibitem{da2021path}
Beatrice Da~Lio, Daniele Cozzolino, Nicola Biagi, Yunhong Ding, Karsten
  Rottwitt, Alessandro Zavatta, Davide Bacco, and Leif~K Oxenl{\o}we.
\newblock ``Path-encoded high-dimensional quantum communication over a 2-km
  multicore fiber''.
\newblock \href{https://dx.doi.org/10.1038/s41534-021-00398-y}{npj Quantum
  Information {\bf 7}, 63}~(2021).

\bibitem{lanyon2009simplifying}
Benjamin~P Lanyon, Marco Barbieri, Marcelo~P Almeida, Thomas Jennewein,
  Timothy~C Ralph, Kevin~J Resch, Geoff~J Pryde, Jeremy~L O’brien, Alexei
  Gilchrist, and Andrew~G White.
\newblock ``Simplifying quantum logic using higher-dimensional hilbert
  spaces''.
\newblock \href{https://dx.doi.org/10.1038/nphys1150}{Nature Physics {\bf 5},
  134--140}~(2009).

\bibitem{li2022chip}
Meng Li, Chu Li, Yang Chen, Lan-Tian Feng, Linyu Yan, Qian Zhang, Jueming Bao,
  Bi-Heng Liu, Xi-Feng Ren, Jianwei Wang, et~al.
\newblock ``On-chip path encoded photonic quantum toffoli gate''.
\newblock \href{https://dx.doi.org/10.1364/PRJ.452539}{Photonics Research {\bf
  10}, 1533--1542}~(2022).

\bibitem{galda2021implementing}
Alexey Galda, Michael Cubeddu, Naoki Kanazawa, Prineha Narang, and Nathan
  Earnest-Noble.
\newblock ``Implementing a ternary decomposition of the toffoli gate on
  fixed-frequencytransmon qutrits''~(2021).
\newblock  \href{http://arxiv.org/abs/2109.00558}{arXiv:2109.00558}.

\bibitem{ringbauer2022universal}
Martin Ringbauer, Michael Meth, Lukas Postler, Roman Stricker, Rainer Blatt,
  Philipp Schindler, and Thomas Monz.
\newblock ``A universal qudit quantum processor with trapped ions''.
\newblock \href{https://dx.doi.org/10.1038/s41567-022-01658-0}{Nature Physics
  {\bf 18}, 1053--1057}~(2022).

\bibitem{wang2020qudits}
Yuchen Wang, Zixuan Hu, Barry~C Sanders, and Sabre Kais.
\newblock ``Qudits and high-dimensional quantum computing''.
\newblock \href{https://dx.doi.org/10.3389/fphy.2020.589504}{Frontiers in
  Physics {\bf 8}, 589504}~(2020).

\bibitem{gao2022role}
Xiaoqin Gao, Paul Appel, Nicolai Friis, Martin Ringbauer, and Marcus Huber.
\newblock ``On the role of entanglement in qudit-based circuit
  compression''~(2022).
\newblock  \href{http://arxiv.org/abs/2209.14584}{arXiv:2209.14584}.

\bibitem{kiktenko2023realization}
Evgeniy~O Kiktenko, Anastasiia~S Nikolaeva, and Aleksey~K Fedorov.
\newblock ``Realization of quantum algorithms with qudits''~(2023).
\newblock  \href{http://arxiv.org/abs/2311.12003}{arXiv:2311.12003}.

\bibitem{litteken2023qompress}
Andrew Litteken, Lennart~Maximilian Seifert, Jason Chadwick, Natalia
  Nottingham, Frederic~T Chong, and Jonathan~M Baker.
\newblock ``Qompress: Efficient compilation for ququarts exploiting partial and
  mixed radix operations for communication reduction''.
\newblock In Proceedings of the 28th ACM International Conference on
  Architectural Support for Programming Languages and Operating Systems, Volume
  2.
\newblock \href{https://dx.doi.org/10.1145/3575693.3575726}{Pages 646--659}.
\newblock ~(2023).

\bibitem{mato2023compression}
Kevin Mato, Stefan Hillmich, and Robert Wille.
\newblock ``Compression of qubit circuits: Mapping to mixed-dimensional quantum
  systems''.
\newblock In 2023 IEEE International Conference on Quantum Software (QSW).
\newblock \href{https://dx.doi.org/10.1109/QSW59989.2023.00027}{Pages
  155--161}.
\newblock IEEE~(2023).

\bibitem{preskill2018quantum}
John Preskill.
\newblock ``Quantum computing in the nisq era and beyond''.
\newblock \href{https://dx.doi.org/10.22331/q-2018-08-06-79}{Quantum {\bf 2},
  79}~(2018).

\bibitem{zurek2003decoherence}
Wojciech~Hubert Zurek.
\newblock ``Decoherence and the transition from quantum to
  classical—revisited''.
\newblock In Quantum Decoherence: Poincar{\'e} Seminar 2005.
\newblock \href{https://dx.doi.org/10.1007/978-3-7643-7808-0_1}{Pages 1--31}.
\newblock Springer~(2003).

\bibitem{ralph2002linear}
Timothy~C Ralph, Nathan~K Langford, TB~Bell, and AG~White.
\newblock ``Linear optical controlled-not gate in the coincidence basis''.
\newblock \href{https://dx.doi.org/10.1103/PhysRevA.65.062324}{Physical Review
  A {\bf 65}, 062324}~(2002).

\bibitem{nikolaeva2024efficient}
Anastasiia~S Nikolaeva, Evgeniy~O Kiktenko, and Aleksey~K Fedorov.
\newblock ``Efficient realization of quantum algorithms with qudits''.
\newblock
  \href{https://dx.doi.org/https://doi.org/10.1140/epjqt/s40507-024-00250-0}{EPJ
  Quantum Technology {\bf 11}, 1--25}~(2024).

\bibitem{cerezo2021variational}
Marco Cerezo, Andrew Arrasmith, Ryan Babbush, Simon~C Benjamin, Suguru Endo,
  Keisuke Fujii, Jarrod~R McClean, Kosuke Mitarai, Xiao Yuan, Lukasz Cincio,
  et~al.
\newblock ``Variational quantum algorithms''.
\newblock \href{https://dx.doi.org/10.1038/s42254-021-00348-9}{Nature Reviews
  Physics {\bf 3}, 625--644}~(2021).

\bibitem{shende2005synthesis}
Vivek~V Shende, Stephen~S Bullock, and Igor~L Markov.
\newblock ``Synthesis of quantum logic circuits''.
\newblock In Proceedings of the 2005 Asia and South Pacific Design Automation
  Conference.
\newblock \href{https://dx.doi.org/10.1109/TCAD.2005.855930}{Pages 272--275}.
\newblock ~(2005).

\bibitem{krol2022efficient}
Anna~M Krol, Aritra Sarkar, Imran Ashraf, Zaid Al-Ars, and Koen Bertels.
\newblock ``Efficient decomposition of unitary matrices in quantum circuit
  compilers''.
\newblock \href{https://dx.doi.org/10.3390/app12020759}{Applied Sciences {\bf
  12}, 759}~(2022).

\bibitem{goubault2020methods}
Timothée Goubault~de Brugière.
\newblock ``Methods for optimizing the synthesis of quantum circuits''.
\newblock PhD thesis.
\newblock Université Paris-Saclay.
\newblock ~(2020).
\newblock
  url:~\href{https://inspirehep.net/files/2ac3fddc186c6f9af433f5e2387a8a86}{inspirehep.net/files/2ac3fddc186c6f9af433f5e2387a8a86}.

\bibitem{madden2022best}
Liam Madden and Andrea Simonetto.
\newblock ``Best approximate quantum compiling problems''.
\newblock \href{https://dx.doi.org/10.1145/3505181}{ACM Transactions on Quantum
  Computing {\bf 3}, 1--29}~(2022).

\bibitem{rakyta2022approaching}
P{\'e}ter Rakyta and Zolt{\'a}n Zimbor{\'a}s.
\newblock ``Approaching the theoretical limit in quantum gate decomposition''.
\newblock \href{https://dx.doi.org/10.22331/q-2022-05-11-710}{Quantum {\bf 6},
  710}~(2022).

\bibitem{di2013synthesis}
Yao-Min Di and Hai-Rui Wei.
\newblock ``Synthesis of multivalued quantum logic circuits by elementary
  gates''.
\newblock \href{https://dx.doi.org/10.1103/PhysRevA.87.012325}{Physical Review
  A {\bf 87}, 012325}~(2013).

\bibitem{li2013efficient}
Wen-Dong Li, Yong-Jian Gu, Kai Liu, Yuan-Harng Lee, and Yao-Zhong Zhang.
\newblock ``Efficient universal quantum computation with auxiliary hilbert
  space''.
\newblock \href{https://dx.doi.org/10.1103/PhysRevA.88.034303}{Physical Review
  A {\bf 88}, 034303}~(2013).

\bibitem{shende2004minimal}
Vivek~V Shende, Igor~L Markov, and Stephen~S Bullock.
\newblock ``Minimal universal two-qubit controlled-not-based circuits''.
\newblock \href{https://dx.doi.org/10.1103/PhysRevA.69.062321}{Physical Review
  A {\bf 69}, 062321}~(2004).

\bibitem{bremner2002practical}
Michael~J Bremner, Christopher~M Dawson, Jennifer~L Dodd, Alexei Gilchrist,
  Aram~W Harrow, Duncan Mortimer, Michael~A Nielsen, and Tobias~J Osborne.
\newblock ``Practical scheme for quantum computation with any two-qubit
  entangling gate''.
\newblock \href{https://dx.doi.org/10.1103/PhysRevLett.89.247902}{Physical
  review letters {\bf 89}, 247902}~(2002).

\bibitem{brylinski2002universal}
Jean-Luc Brylinski and Ranee Brylinski.
\newblock ``Universal quantum gates''.
\newblock \href{https://dx.doi.org/10.48550/arXiv.quant-ph/0108062}{Mathematics
  of quantum computation{\bf 79}}~(2002).

\bibitem{nakanishi2021quantum}
Ken~M Nakanishi, Takahiko Satoh, and Synge Todo.
\newblock ``Quantum-gate decomposer''~(2021).
\newblock  \href{http://arxiv.org/abs/2109.13223}{arXiv:2109.13223}.

\bibitem{shende2008cnot}
Vivek~V Shende and Igor~L Markov.
\newblock ``On the cnot-cost of toffoli gates''~(2008).
\newblock  \href{http://arxiv.org/abs/0803.2316}{arXiv:0803.2316}.

\bibitem{kiktenko2020scalable}
EO~Kiktenko, AS~Nikolaeva, Peng Xu, GV~Shlyapnikov, and AK~Fedorov.
\newblock ``Scalable quantum computing with qudits on a graph''.
\newblock \href{https://dx.doi.org/10.1103/PhysRevA.101.022304}{Physical Review
  A {\bf 101}, 022304}~(2020).

\bibitem{nikolaeva2022decomposing}
AS~Nikolaeva, EO~Kiktenko, and AK~Fedorov.
\newblock ``Decomposing the generalized toffoli gate with qutrits''.
\newblock \href{https://dx.doi.org/10.1103/PhysRevA.105.032621}{Physical Review
  A {\bf 105}, 032621}~(2022).

\bibitem{wang2011improved}
Yushi Wang and Marek Perkowski.
\newblock ``Improved complexity of quantum oracles for ternary grover algorithm
  for graph coloring''.
\newblock In 2011 41st IEEE International Symposium on Multiple-Valued Logic.
\newblock \href{https://dx.doi.org/10.1109/ISMVL.2011.42}{Pages 294--301}.
\newblock IEEE~(2011).

\bibitem{baker2020improved}
Jonathan~M Baker, Casey Duckering, Pranav Gokhale, Natalie~C Brown, Kenneth~R
  Brown, and Frederic~T Chong.
\newblock ``Improved quantum circuits via intermediate qutrits''.
\newblock \href{https://dx.doi.org/10.1145/3406309}{ACM Transactions on Quantum
  Computing {\bf 1}, 1--25}~(2020).

\bibitem{feynman2023quantum}
Richard~P Feynman and Tony Hey.
\newblock ``Quantum mechanical computers''.
\newblock In Feynman Lectures on Computation.
\newblock \href{https://dx.doi.org/10.1007/BF01886518}{Pages 169--192}.
\newblock CRC Press~(2023).

\bibitem{qiskit2024}
Ali Javadi-Abhari, Matthew Treinish, Kevin Krsulich, Christopher~J. Wood, Jake
  Lishman, Julien Gacon, Simon Martiel, Paul~D. Nation, Lev~S. Bishop,
  Andrew~W. Cross, Blake~R. Johnson, and Jay~M. Gambetta.
\newblock ``Quantum computing with {Q}iskit''~(2024).
\newblock  \href{http://arxiv.org/abs/2405.08810}{arXiv:2405.08810}.

\bibitem{nikolaeva2023generalized}
Anstasiia~S Nikolaeva, Evgeniy~O Kiktenko, and Aleksey~K Fedorov.
\newblock ``Generalized toffoli gate decomposition using ququints: Towards
  realizing grover’s algorithm with qudits''.
\newblock \href{https://dx.doi.org/10.3390/e25020387}{Entropy {\bf 25},
  387}~(2023).

\bibitem{molmer1999multiparticle}
Klaus M{\o}lmer and Anders S{\o}rensen.
\newblock ``Multiparticle entanglement of hot trapped ions''.
\newblock \href{https://dx.doi.org/10.1103/PhysRevLett.82.1835}{Physical Review
  Letters {\bf 82}, 1835}~(1999).

\bibitem{kandala2017hardware}
Abhinav Kandala, Antonio Mezzacapo, Kristan Temme, Maika Takita, Markus Brink,
  Jerry~M Chow, and Jay~M Gambetta.
\newblock ``Hardware-efficient variational quantum eigensolver for small
  molecules and quantum magnets''.
\newblock \href{https://dx.doi.org/10.1038/nature23879}{nature {\bf 549},
  242--246}~(2017).

\bibitem{havlivcek2019supervised}
Vojt{\v{e}}ch Havl{\'i}{\v{c}}ek, Antonio~D C{\'o}rcoles, Kristan Temme, Aram~W
  Harrow, Abhinav Kandala, Jerry~M Chow, and Jay~M Gambetta.
\newblock ``Supervised learning with quantum-enhanced feature spaces''.
\newblock \href{https://dx.doi.org/10.1038/s41586-019-0980-2}{Nature {\bf 567},
  209--212}~(2019).

\bibitem{sim2019expressibility}
Sukin Sim, Peter~D Johnson, and Al{\'a}n Aspuru-Guzik.
\newblock ``Expressibility and entangling capability of parameterized quantum
  circuits for hybrid quantum-classical algorithms''.
\newblock \href{https://dx.doi.org/10.1002/qute.201900070}{Advanced Quantum
  Technologies {\bf 2}, 1900070}~(2019).

\bibitem{holmes2022connecting}
Zo{\"e} Holmes, Kunal Sharma, Marco Cerezo, and Patrick~J Coles.
\newblock ``Connecting ansatz expressibility to gradient magnitudes and barren
  plateaus''.
\newblock \href{https://dx.doi.org/10.1103/PRXQuantum.3.010313}{PRX Quantum
  {\bf 3}, 010313}~(2022).

\bibitem{rasmussen2020reducing}
Stig~Elkj{\ae}r Rasmussen, Niels Jakob~S{\o}e Loft, Thomas B{\ae}kkegaard,
  Michael Kues, and Nikolaj~Thomas Zinner.
\newblock ``Reducing the amount of single-qubit rotations in vqe and related
  algorithms''.
\newblock \href{https://dx.doi.org/10.1002/qute.202000063}{Advanced Quantum
  Technologies {\bf 3}, 2000063}~(2020).

\bibitem{kullback1951information}
Solomon Kullback and Richard~A Leibler.
\newblock ``On information and sufficiency''.
\newblock \href{https://dx.doi.org/10.1214/aoms/1177729694}{The annals of
  mathematical statistics {\bf 22}, 79--86}~(1951).

\bibitem{meyer2002global}
David~A Meyer and Nolan~R Wallach.
\newblock ``Global entanglement in multiparticle systems''.
\newblock \href{https://dx.doi.org/10.1063/1.1497700}{Journal of Mathematical
  Physics {\bf 43}, 4273--4278}~(2002).

\bibitem{brennen2003observable}
Gavin~K Brennen.
\newblock ``An observable measure of entanglement for pure states of
  multi-qubit systems''~(2003).
\newblock
  \href{http://arxiv.org/abs/quant-ph/0305094v3}{arXiv:quant-ph/0305094v3}.

\bibitem{schuld2020circuit}
Maria Schuld, Alex Bocharov, Krysta~M Svore, and Nathan Wiebe.
\newblock ``Circuit-centric quantum classifiers''.
\newblock \href{https://dx.doi.org/10.1103/PhysRevA.101.032308}{Physical Review
  A {\bf 101}, 032308}~(2020).

\bibitem{romero2017quantum}
Jonathan Romero, Jonathan~P Olson, and Alan Aspuru-Guzik.
\newblock ``Quantum autoencoders for efficient compression of quantum data''.
\newblock \href{https://dx.doi.org/10.1088/2058-9565/aa8072}{Quantum Science
  and Technology {\bf 2}, 045001}~(2017).

\bibitem{de2024spin}
Gr{\'e}goire de~Gliniasty, Paul Hilaire, Pierre-Emmanuel Emeriau, Stephen~C
  Wein, Alexia Salavrakos, and Shane Mansfield.
\newblock ``A spin-optical quantum computing architecture''.
\newblock \href{https://dx.doi.org/10.22331/q-2024-07-24-1423}{Quantum {\bf 8},
  1423}~(2024).

\bibitem{hilaire2024enhanced}
Paul Hilaire, Th{\'e}o Dessertaine, Boris Bourdoncle, Aur{\'e}lie Denys,
  Gr{\'e}goire de~Gliniasty, Gerard Valent{\'\i}-Rojas, and Shane Mansfield.
\newblock ``Enhanced fault-tolerance in photonic quantum computing: Floquet
  code outperforms surface code in tailored architecture''~(2024).
\newblock  \href{http://arxiv.org/abs/2410.07065}{arXiv:2410.07065}.

\bibitem{Kok2007linear}
Pieter Kok, William~J Munro, Kae Nemoto, Timothy~C Ralph, Jonathan~P Dowling,
  and Gerard~J Milburn.
\newblock ``Linear optical quantum computing with photonic qubits''.
\newblock \href{https://dx.doi.org/10.1103/RevModPhys.79.135}{Reviews of modern
  physics {\bf 79}, 135}~(2007).

\bibitem{nielsen2004optical}
Michael~A Nielsen.
\newblock ``Optical quantum computation using cluster states''.
\newblock \href{https://dx.doi.org/10.1103/PhysRevLett.93.040503}{Physical
  review letters {\bf 93}, 040503}~(2004).

\bibitem{knill2001scheme}
Emanuel Knill, Raymond Laflamme, and Gerald~J Milburn.
\newblock ``A scheme for efficient quantum computation with linear optics''.
\newblock \href{https://dx.doi.org/10.1038/35051009}{nature {\bf 409},
  46--52}~(2001).

\bibitem{yoran2003deterministic}
Nadav Yoran and Benni Reznik.
\newblock ``Deterministic linear optics quantum computation with single photon
  qubits''.
\newblock \href{https://dx.doi.org/10.1103/PhysRevLett.91.037903}{Physical
  review letters {\bf 91}, 037903}~(2003).

\bibitem{browne2005resource}
Daniel~E Browne and Terry Rudolph.
\newblock ``Resource-efficient linear optical quantum computation''.
\newblock \href{https://dx.doi.org/10.1103/PhysRevLett.95.010501}{Physical
  Review Letters {\bf 95}, 010501}~(2005).

\bibitem{maring2024versatile}
Nicolas Maring, Andreas Fyrillas, Mathias Pont, Edouard Ivanov, Petr Stepanov,
  Nico Margaria, William Hease, Anton Pishchagin, Aristide Lema{\^\i}tre,
  Isabelle Sagnes, et~al.
\newblock ``A versatile single-photon-based quantum computing platform''.
\newblock \href{https://dx.doi.org/10.1038/s41566-024-01403-4}{Nature
  PhotonicsPages 1--7}~(2024).

\bibitem{knill2002quantum}
Emanuel Knill.
\newblock ``Quantum gates using linear optics and postselection''.
\newblock \href{https://dx.doi.org/10.1103/PhysRevA.66.052306}{Physical Review
  A {\bf 66}, 052306}~(2002).

\bibitem{peruzzo2014variational}
Alberto Peruzzo, Jarrod McClean, Peter Shadbolt, Man-Hong Yung, Xiao-Qi Zhou,
  Peter~J Love, Al{\'a}n Aspuru-Guzik, and Jeremy~L O’brien.
\newblock ``A variational eigenvalue solver on a photonic quantum processor''.
\newblock \href{https://dx.doi.org/10.1038/ncomms5213}{Nature communications
  {\bf 5}, 4213}~(2014).

\bibitem{tilly2022variational}
Jules Tilly, Hongxiang Chen, Shuxiang Cao, Dario Picozzi, Kanav Setia, Ying Li,
  Edward Grant, Leonard Wossnig, Ivan Rungger, George~H Booth, et~al.
\newblock ``The variational quantum eigensolver: a review of methods and best
  practices''.
\newblock \href{https://dx.doi.org/10.1016/j.physrep.2022.08.003}{Physics
  Reports {\bf 986}, 1--128}~(2022).

\bibitem{seeley2012bravyi}
Jacob~T Seeley, Martin~J Richard, and Peter~J Love.
\newblock ``The bravyi-kitaev transformation for quantum computation of
  electronic structure''.
\newblock \href{https://dx.doi.org/10.1063/1.4768229}{The Journal of chemical
  physics{\bf 137}}~(2012).

\bibitem{mcclean2020openfermion}
Jarrod~R McClean, Nicholas~C Rubin, Kevin~J Sung, Ian~D Kivlichan, Xavier
  Bonet-Monroig, Yudong Cao, Chengyu Dai, E~Schuyler Fried, Craig Gidney,
  Brendan Gimby, et~al.
\newblock ``Openfermion: the electronic structure package for quantum
  computers''.
\newblock \href{https://dx.doi.org/10.1088/2058-9565/ab8ebc}{Quantum Science
  and Technology {\bf 5}, 034014}~(2020).

\bibitem{zhu1997algorithm}
Ciyou Zhu, Richard~H Byrd, Peihuang Lu, and Jorge Nocedal.
\newblock ``Algorithm 778: L-bfgs-b: Fortran subroutines for large-scale
  bound-constrained optimization''.
\newblock \href{https://dx.doi.org/10.1145/279232.279236}{ACM Transactions on
  mathematical software (TOMS) {\bf 23}, 550--560}~(1997).

\bibitem{ralph2004scaling}
TC~Ralph.
\newblock ``Scaling of multiple postselected quantum gates in optics''.
\newblock \href{https://dx.doi.org/10.1103/PhysRevA.70.012312}{Physical Review
  A {\bf 70}, 012312}~(2004).

\bibitem{heurtel2023strong}
Nicolas Heurtel, Shane Mansfield, Jean Senellart, and Beno{\^\i}t Valiron.
\newblock ``Strong simulation of linear optical processes''.
\newblock \href{https://dx.doi.org/10.1016/j.cpc.2023.108848}{Computer Physics
  Communications {\bf 291}, 108848}~(2023).

\bibitem{heurtel2023perceval}
Nicolas Heurtel, Andreas Fyrillas, Gr{\'e}goire de~Gliniasty, Rapha{\"e}l
  Le~Bihan, S{\'e}bastien Malherbe, Marceau Pailhas, Eric Bertasi, Boris
  Bourdoncle, Pierre-Emmanuel Emeriau, Rawad Mezher, et~al.
\newblock ``Perceval: A software platform for discrete variable photonic
  quantum computing''.
\newblock \href{https://dx.doi.org/10.22331/q-2023-02-21-931}{Quantum {\bf 7},
  931}~(2023).

\bibitem{powell1994direct}
Michael~JD Powell.
\newblock ``A direct search optimization method that models the objective and
  constraint functions by linear interpolation''.
\newblock \href{https://dx.doi.org/10.1007/978-94-015-8330-5_4}{Springer}.
  ~(1994).

\bibitem{zhang2021mutual}
Zi-Jian Zhang, Thi~Ha Kyaw, Jakob~S Kottmann, Matthias Degroote, and Al{\'a}n
  Aspuru-Guzik.
\newblock ``Mutual information-assisted adaptive variational quantum
  eigensolver''.
\newblock \href{https://dx.doi.org/10.1088/2058-9565/abdca4}{Quantum Science
  and Technology {\bf 6}, 035001}~(2021).

\bibitem{choy2023molecular}
Boy Choy and David~J Wales.
\newblock ``Molecular energy landscapes of hardware-efficient ans\"atze in
  quantum computing''.
\newblock \href{https://dx.doi.org/10.1021/acs.jctc.2c01057}{Journal of
  Chemical Theory and Computation {\bf 19}, 1197--1206}~(2023).

\bibitem{peterson2012chemical}
Kirk~A Peterson, David Feller, and David~A Dixon.
\newblock ``Chemical accuracy in ab initio thermochemistry and spectroscopy:
  current strategies and future challenges''.
\newblock \href{https://dx.doi.org/10.1007/s00214-011-1079-5}{Theoretical
  Chemistry Accounts {\bf 131}, 1--20}~(2012).

\bibitem{sachdeva2024quantum}
Natasha Sachdeva, Gavin~S Harnett, Smarak Maity, Samuel Marsh, Yulun Wang, Adam
  Winick, Ryan Dougherty, Daniel Canuto, You~Quan Chong, Michael Hush, et~al.
\newblock ``Quantum optimization using a 127-qubit gate-model ibm quantum
  computer can outperform quantum annealers for nontrivial binary optimization
  problems''~(2024).
\newblock  \href{http://arxiv.org/abs/2406.01743}{arXiv:2406.01743}.

\bibitem{farhi2018classification}
Edward Farhi and Hartmut Neven.
\newblock ``Classification with quantum neural networks on near term
  processors''~(2018).
\newblock  \href{http://arxiv.org/abs/1802.06002}{arXiv:1802.06002}.

\end{thebibliography}
